\documentclass[12pt]{article} 

\usepackage[sectionbib]{natbib}
\usepackage{amsmath,amsthm}
\usepackage{natbib}
\usepackage{array,epsfig,fancyheadings,rotating}
\usepackage[]{hyperref}  
\usepackage{graphicx}
\usepackage{enumerate}
\usepackage{url} 
\usepackage{xcolor,colortbl}
\usepackage{bm}
\usepackage{amsmath}
\usepackage{amsfonts}   
\usepackage{graphicx}
\usepackage{array}
\usepackage{longtable}
\usepackage{multirow}
\usepackage{setspace}
\usepackage{makecell}
\usepackage{tablefootnote}
\usepackage{sectsty, secdot}
\sectionfont{\fontsize{12}{14pt plus.8pt minus .6pt}\selectfont}

\renewcommand{\theequation}{\thesection\arabic{equation}}
\subsectionfont{\fontsize{12}{14pt plus.8pt minus .6pt}\selectfont}
\newcommand*{\affaddr}[1]{#1} 
\newcommand*{\affmark}[1][*]{\textsuperscript{#1}}

\newcommand\notsotiny{\@setfontsize\notsotiny\@vipt\@viipt}
\usepackage{amsmath}
\usepackage{amssymb}
\usepackage{amsfonts}
\usepackage{multirow}
\usepackage{amsthm}

\newtheorem{theorem}{Theorem}
\newtheorem{assumption}{Assumption}

\theoremstyle{definition}
\newtheorem{definition}{Definition}

\newtheorem{remark}{Remark}
\def\Var{\operatorname{var}}

\def\E{\mbox{E}}

\begin{document}



\markright{ \hbox{\footnotesize\rm 
}\hfill\\[-13pt]
\hbox{\footnotesize\rm
}\hfill }

\markboth{\hfill{\footnotesize\rm You et al.} \hfill}
{\hfill {\footnotesize\rm Regularized nonlinear regression} \hfill}

\renewcommand{\thefootnote}{}
$\ $\par


\fontsize{12}{14pt plus.8pt minus .6pt}\selectfont \vspace{0.8pc}
\centerline{\large\bf 
Regularized Nonlinear Regression with Dependent Errors}
\vspace{2pt} 
\centerline{\large\bf 
		and its Application to a Biomechanical Model }
\vspace{.4cm}

\centerline{Hojun You\affmark[1], Kyubaek Yoon\affmark[3], Wei-Ying Wu\affmark[4], Jongeun Choi\affmark[3] and Chae Young Lim\affmark[2]} 
\vspace{.4cm} 
\centerline{\it {\affmark[1]University of Houston}\\\hspace{0.1cm}
\affaddr{\affmark[2]Seoul National University}\\\hspace{0.1cm}
\affaddr{\affmark[3]Yonsei University}\\\hspace{0.1cm}
\affaddr{\affmark[4]National Dong Hwa University}\\
}
 \vspace{.55cm} \fontsize{9}{11.5pt plus.8pt minus.6pt}\selectfont


\begin{quotation}
\noindent {\it Abstract:}
A biomechanical model often requires parameter estimation and selection in a known but complicated nonlinear function. Motivated by observing that the data from a head-neck position tracking system, one of biomechanical models, show multiplicative time dependent errors, we develop a modified penalized weighted least squares estimator. The proposed method can be also applied to a model with possible non-zero mean time dependent additive errors. Asymptotic properties of the proposed estimator are investigated under mild conditions on a weight matrix and the error process. A simulation study demonstrates that the proposed estimation works well in both parameter estimation and selection with time dependent error. The analysis and comparison with an existing method for head-neck position tracking data show better performance of the proposed method in terms of the variance accounted for (VAF). 

\vspace{9pt}
\par
\end{quotation}\par

\def\thefigure{\arabic{figure}}
\def\thetable{\arabic{table}}

\renewcommand{\theequation}{\thesection.\arabic{equation}}

\fontsize{12}{14pt plus.8pt minus .6pt}\selectfont

\section{Introduction}

\label{introduction}

A nonlinear regression model has been widely used to describe complicated relationships between variables \citep{wood2010nonlinear1, baker2011nonlinear3, paula2014nonlinear2, lim2014operator, salamh2021second}. In particular, various nonlinear problems are considered in the field of machinery and biomechanical engineering \citep{moon2016mechanic1, santos2017mechanic2}. Among such nonlinear problems,  a head-neck position tracking model with neurophysiological parameters in biomechanics motivated us to develop an estimation and selection method for a nonlinear regression model in this work. 

The head-neck position tracking application aims to figure out how characteristics of the vestibulocollic and cervicocollic reflexes (VCR and CCR) contribute to the head-neck system. The VCR activates neck muscles to stabilize the head-in-space and the CCR acts to hold the head on the trunk. A subject of the experiment follows a reference signal on a computer screen with his or her head and a head rotation angle is measured during the experiment. A reference signal is the input of the system and the measured head rotation angle is the output. The parameters related to VCR and CCR in this nonlinear system are of interest to understand the head-neck position tracking system. 

The head-neck position problem has been widely studied in the literature of biomechanics \citep{peng1996dynamical, chen2002modeling, forbes2013dependency, ramadan2018selecting, yoon2022regularized}. One of the prevalent issues in biomechanics is that a model suffers from a relatively large number of parameters and limited availability of data because the  subjects in the experiment cannot tolerate sufficient time without being fatigued. This leads to overfitting as well as non-identifiability of the parameters. To resolve this issue, selection approaches via a penalized regression method have been implemented to fix a subset of the parameters to the pre-specified values while the remaining parameters are estimated \citep{ramadan2018selecting, yoon2022regularized}. 

\begin{figure}[!ht]\small\center
	\includegraphics[width=0.5\textwidth]{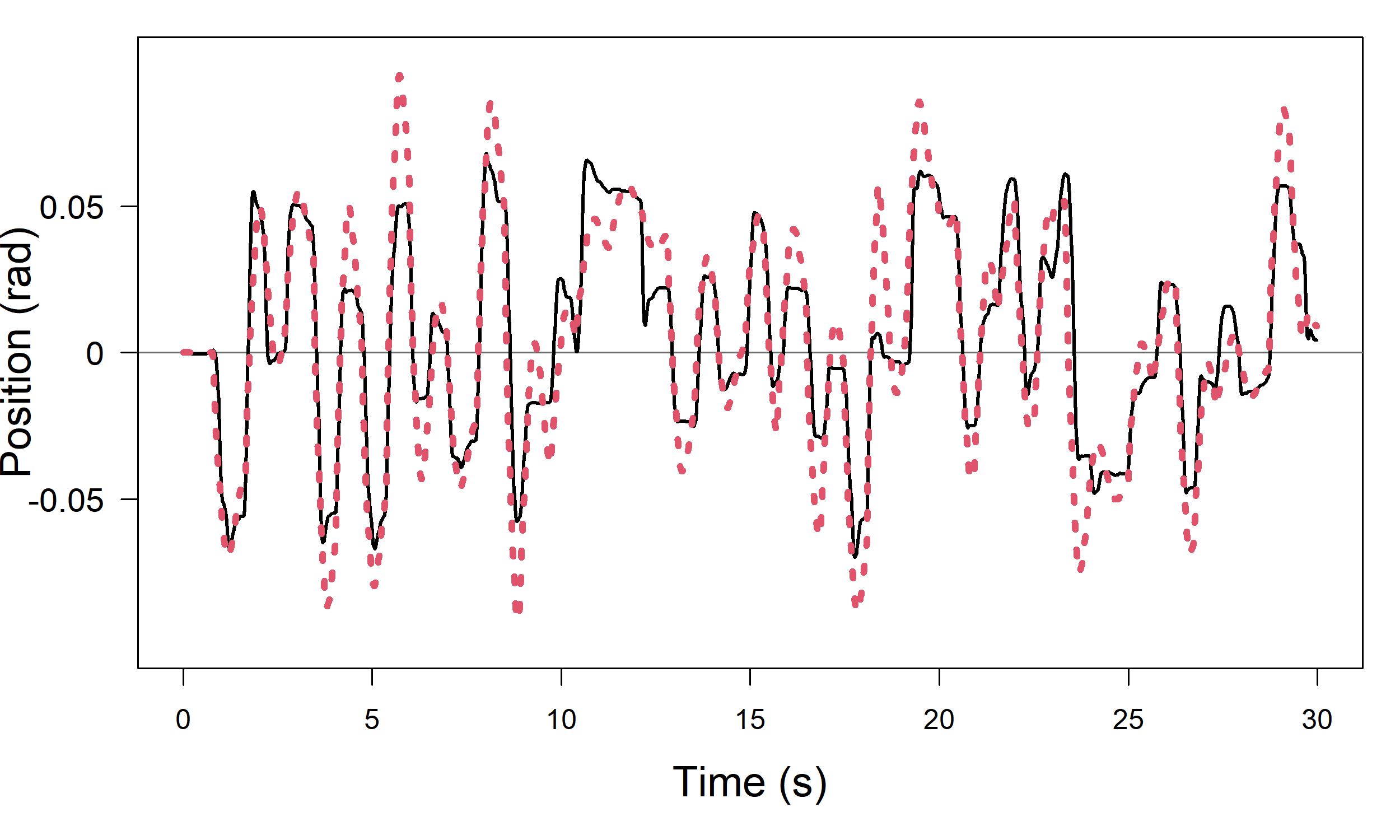}
	\caption{\footnotesize The black curve represents the measured responses (the observations) from the subject No. 8 in the head-neck position tracking experiment. The red dashed curve represents the estimated responses (the fitted values) from the nonlinear regression model with additive errors introduced in \cite{yoon2022regularized}.  }
	\label{fig:multiplicative}
	\includegraphics[width=0.45\textwidth]{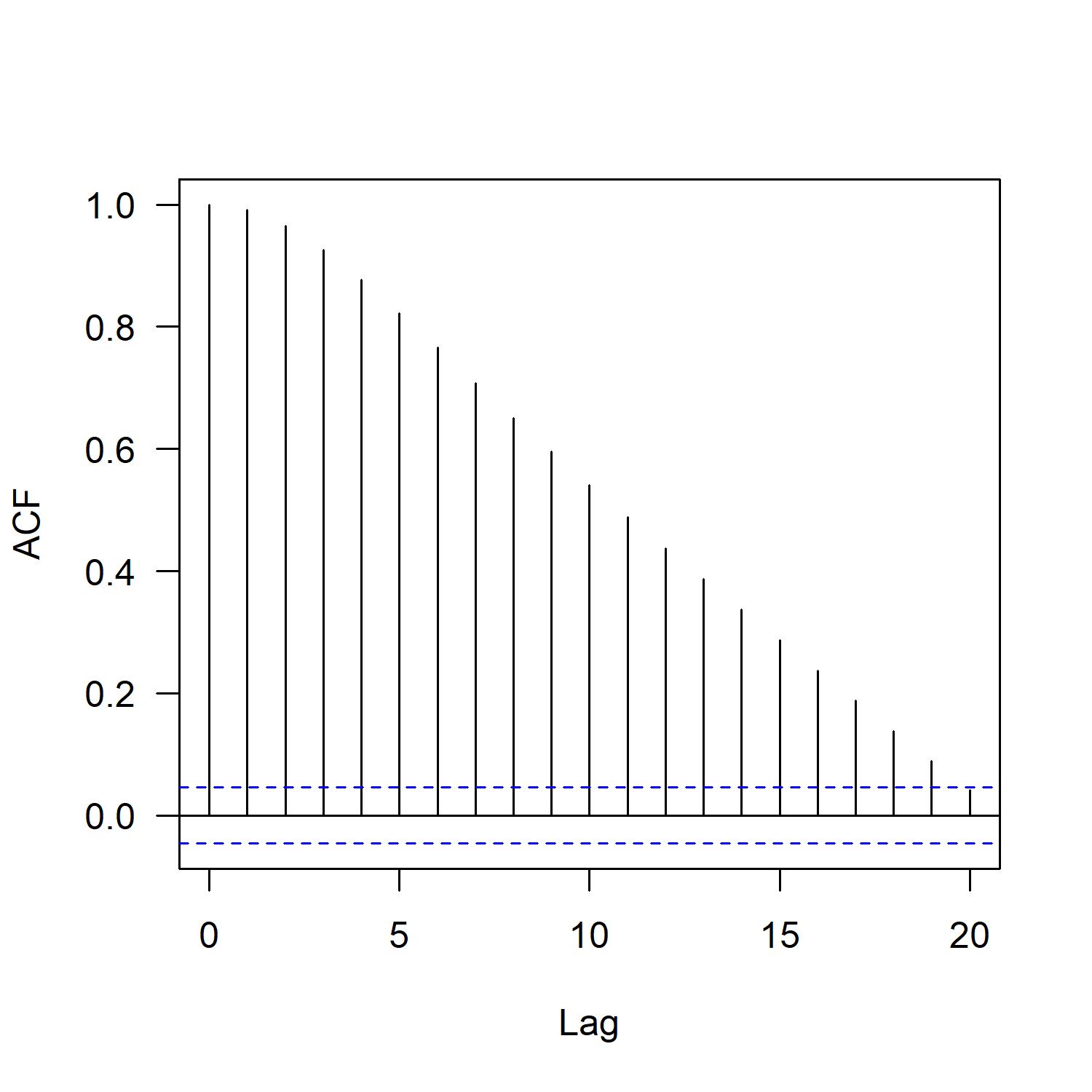}
	\includegraphics[width=0.45\textwidth]{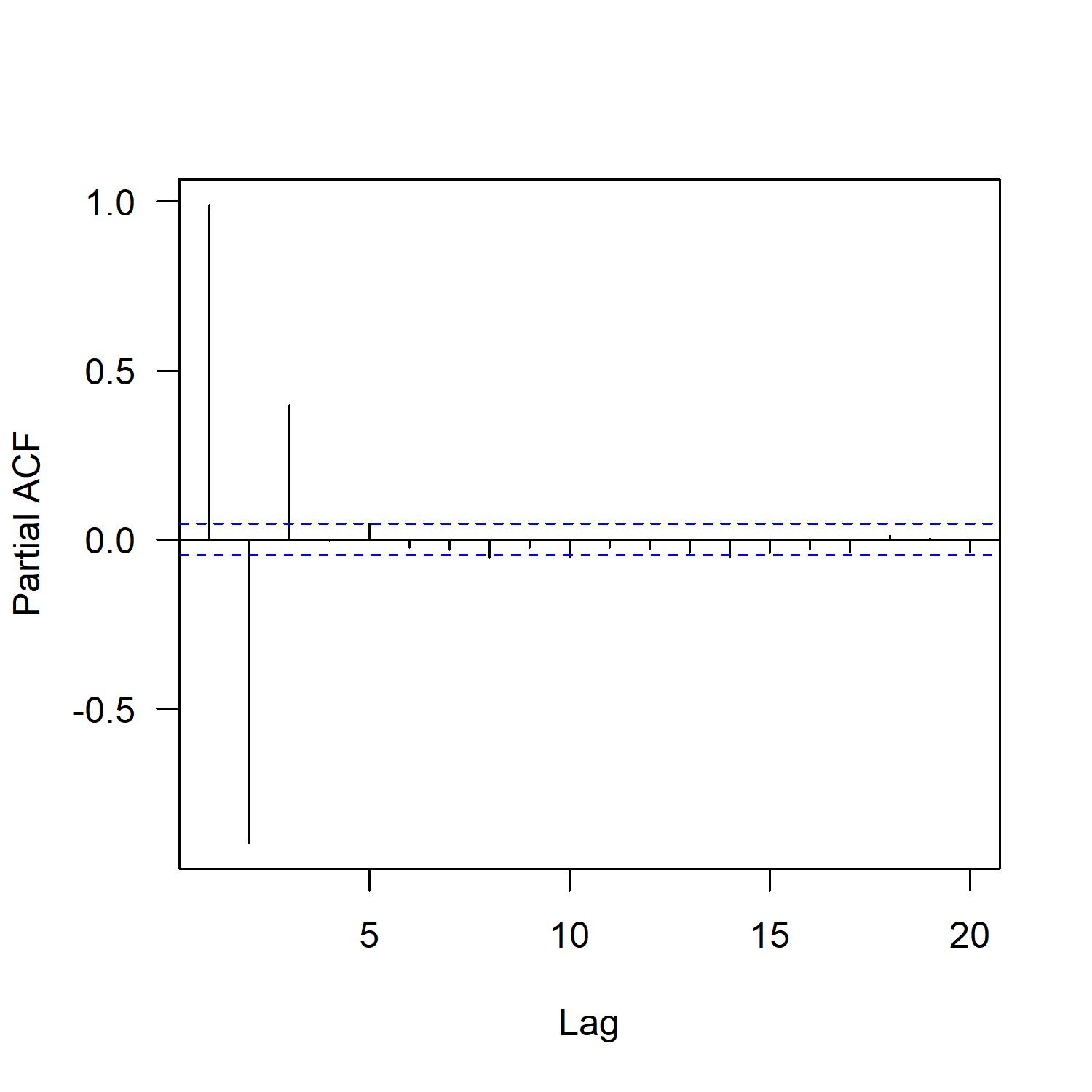}
	\caption{\footnotesize Sample autocorrelation (left) and sample partial autocorrelation (right) of the residuals $(\mbox{measured response}-\mbox{estimated response})$ for the subject No. 8 from the head-neck position tracking experiment. The estimated response is obtained from the method in \cite{yoon2022regularized}.} 
	\label{fig:additive_res}
\end{figure}

\indent The existing approaches, however, have some limitations. The fitted values from the penalized nonlinear regression with additive errors in \cite{yoon2022regularized} show larger discrepancy from the observed values when the head is turning in its direction. For example, Figure \ref{fig:multiplicative} shows the fitted values (the estimated responses) from the method in \cite{yoon2022regularized} and the observations (the measured responses) of the subject No. 8 in a head-neck position tracking experiment. The detailed description of the data from the experiment is given in Section \ref{section:application}.  We can observe that a larger measured response leads to a bigger gap between the measured response and the estimated response. 
Note that the largest measured responses occur when the head is turning in its direction to follow the reference signal's direction changes. It is more difficult to track the end positions correctly for some subjects, which can create more errors at each end. Hence, the additive error structure considered in \cite{yoon2022regularized} can not properly explain the data. Indeed, Figure \ref{fig:multiplicative} shows the model with additive errors did not successfully accommodate this characteristic.

Another point we pose in this study is temporal dependence of the data which previous studies did not also take into account while the experimental data exhibit temporal correlation. For example, Figure \ref{fig:additive_res} shows sample autocorrelation function and sample partial autocorrelation function of the residuals from the fitted model for the subject No. 8 by the method in \cite{yoon2022regularized}. The residuals clearly show temporal dependence while 
\cite{yoon2022regularized} worked under independent error assumption. Lastly, \cite{ramadan2018selecting} and \cite{yoon2022regularized} restrict the number of sensitive parameters to five, where the sensitive parameters refer to the parameters whose estimates are not shrunk to the pre-specified values. Not only may this restriction increase computational instability but also reduce estimation and prediction performances. Provided the head-neck position tracking task already suffers from computational challenges, additional computational issues should be avoided. 

To resolve the above-mentioned issues, we consider a nonlinear regression model with multiplicative errors for the head-neck position tracking system, which can be written as 
\begin{equation}
z_{t}=g(\bm x_t;\bm\theta)\times\varsigma_{t}, \label{eq_multi}
\end{equation}
where $z_t$ is a measured response, $g(\bm x;\bm \theta)$ is a known nonlinear function with an input $\bm x$. $\bm \theta$ is a set of parameters in $g$ and $\varsigma_{t}$ is a multiplicative error. The details for $g$ and $\bm{\theta}$ for the head-neck position tracking system are described in \cite{ramadan2018selecting} and \cite{yoon2022regularized}. {The multiplicative error structure can better explain the data than the previous studies in that the error may increase as the signal increases. It is of importance that we choose suitable regression model for data because \cite{bhattacharyya1992inconsistent} showed that an ordinary least squares estimator for a nonlinear regression model with additive errors may not possess strong consistency when the true underlying model has multiplicative errors. 

A typical approach for the multiplicative error model is to take the logarithm in both sides of \eqref{eq_multi} so that the resulting model becomes a nonlinear model with additive errors:}
\begin{equation}
y_{t}=f(\bm x_t;\bm\theta) + \epsilon_{t}, \label{eq_additive}
\end{equation}
with $y_{t}=\log(z_t)$, $f(\bm x_t;\bm\theta)=\log(g(\bm x_t;\bm\theta))$ and $\epsilon_{t} = \log(\varsigma_t)$ by assuming all components are positive. Note that the difference from the classical additive regression model in \cite{yoon2022regularized} is that the additive error $\epsilon_t$ in \eqref{eq_additive} may have non-zero mean due to the log transformation. By the relationship $f=\log (g)$, the conditions imposed on $g$ are inherited from the conditions on $f$. Therefore, the degree of flexibility allowed for $g$ is linked to the flexibility allowed for $f$ based on the assumptions outlined in Section \ref{section:method}.

Motivated by the head-neck position tracking application, we propose a parameter estimation method for a nonlinear model with multiplicative error in \eqref{eq_multi} by applying a modified weighted least squares estimation method to \eqref{eq_additive} which accommodate temporal dependence as well as non-zero mean errors. Given by the structure, the proposed estimation can also handle the nonlinear regression model with possible non-zero mean additive errors in \eqref{eq_additive}. The asymptotic properties of the estimator obtained from the proposed method are studied under the assumption of temporally correlated errors as we observed temporal dependence in the head-neck position tracking system data. Introducing an additional intercept for non-zero mean error could be a simple solution for handling \eqref{eq_additive}. However, this approach introduces an extra parameter to the original model, which could lead to inefficiency in the estimation process. Indeed,  as detailed in Section \ref{section:simulation}, our proposed approach shows better performance in the simulation study  for most cases and the difference is apparent, in particular, when the non-zero mean is large or temporal dependence is stronger. Furthermore, if the function $f(\bm x_t; \bm \theta)$ already includes an intercept term, the two intercept parameters are not identifiable.

A penalized estimator and its asymptotic properties are also investigated. In the application of the head-neck position tracking system, a set of parameters needs to be shrunk to the pre-specified values instead of the zero-values. Thus, we use the penalized estimation approach to shrink estimates to the pre-specified values. By doing so, we not only resolve the non-identifiability issue but also keep all estimates meaningful in the head-neck position tracking system. For a penalty function, we allow a general penalty function with mild conditions for the asymptotic properties of the penalized least squares estimators. In the numerical study, least absolute shrinkage and selection operator \citep[LASSO,][]{tibshirani1996regression} and smoothly clipped absolute deviation \citep[SCAD,][]{fan2001scad} are considered for results.

Numerous studies have explored the properties of least squares estimators in nonlinear regression models. \cite{jennrich1969asymp} first proved the strong consistency and asymptotic normality of the nonlinear least squares estimator with independent errors. \cite{wu1981asymptotic} provided necessary conditions for the existence of any kind of weakly consistent estimators and extended the results in \cite{jennrich1969asymp} under weaker conditions. \cite{pollard2006empirical} established asymptotic properties for least squares estimators under second-moment assumptions for errors. Later, several studies \citep{wang2008second, kim2012efficiency, ivanov2015estimation, radchenko2015high, salamh2021second, wang2021least} delved further into the properties of the least squares estimator in nonlinear regression.

There are several studies on the nonlinear regression with multiplicative errors. \cite{xu2000multiplicative} studied least squares estimation for nonlinear multiplicative noise models with independent errors, but their proposed estimator induces a bias and needs correction. \cite{lim2014operator} also investigated the nonlinear multiplicative noise models with independent errors by the log transformation and proposed the modified least square estimation by including a sample mean component in the objective function. \cite{chen2010least, chen2016least} proposed least absolute relative error (LARE) and least product relative error (LPRE), respectively, as alternatives to least squares for multiplicative regression. \cite{zhang2022maximum} further devised maximum nonparametric kernel likelihood estimation for multiplicative regression. However, proposed methods in \cite{chen2010least, chen2016least} and \cite{zhang2022maximum} can only accommodate exponential nonlinear function, which highly limit the applicability of the methods. Our study follows a similar approach to \cite{lim2014operator} but least squares estimation is enhanced with a weight matrix and temporally dependent errors are taken into account in addition to the penalization.

To address temporal dependence in the errors, we consider mixing conditions, including strong mixing ($\alpha$-mixing), $\phi$-mixing, and $\rho$-mixing, which are well-established techniques for handling temporal data dependencies. Prior research has explored various regression models involving mixing errors. 
\cite{jhang2012strongmixingsemi} studied a semi-parametric regression model with strong mixing errors and established the asymptotic normality of a weighted least squares estimator. Subsequently, \cite{guo2019regression} devised wavelet regression estimators under strong mixing data and \cite{ullah2022nonlinear} investigated the asymptotic properties of nonlinear modal estimator with strong mixing errors. For more articles that accommodate mixing conditions in regression problems, we refer to \cite{el2017local, almanjahie2022nonparametric, kurisu2022nonparametric, mokhtari2022consistency}. In our study, the asymptotic properties of the proposed estimator are established under various mixing conditions.

In Section \ref{section:method}, we demonstrate the proposed estimation method for a nonlinear regression model and establish the asymptotic properties of the proposed estimators. In Section \ref{section:simulation}, several simulation studies are conducted with various settings.  The analysis on head-neck position tracking data with the  proposed method is introduced in Section \ref{section:application}. At last, we provide a discussion in Section \ref{section:discussion}. The technical proofs for the theorems and additional results of the simulation studies  are presented in a supplementary material.

\section{Methods}
\label{section:method}

\subsection{Modified Weighted Least Squares}\label{section:MLS}

We consider a following nonlinear regression model
$$y_{t}=f(\bm x_t;\bm\theta)+ \epsilon_{t},$$
for $t=1, \cdots, n$, where $\bm x_t\in D\subset R^d$ is a fixed covariate vector and $f({\bm x};\bm\theta)$ is a known nonlinear function on $\bm x$, which also depends on the parameter vector $ \bm{\theta}:=(\theta_1,\theta_2,...,\theta_p)^{T} \in \Theta$. ${A}^{T}$ is the transpose of a matrix $A$. $\epsilon_{t}$ is temporally correlated and its mean, $E(\epsilon_t)=\mu$, may not be zero. The assumption on a possible non-zero mean of the error comes from a nonlinear regression model with multiplicative errors introduced in \eqref{eq_multi}. We further assume that only a few entries of the true parameter are non-zero. Without loss of generality, we let the first $s$ entries of the true $\bm\theta_0$ be non-zero. That is,   $\bm\theta_0=(\theta_{01},\theta_{02},...,\theta_{0s},\theta_{0s+1},...,\theta_{0p})^{T}$ and $\theta_{0t}\neq 0$ for $1\leq t\leq s$ and $\theta_{0t}= 0$ for $s+1 \leq t\leq p$. It should be noted that the dimension of $\bm x_t,\ d,$ can be different from the dimension of $\bm \theta$, $p$, as we allow $f$ to have a complicated nonlinear structure. For example, our method can handle  $f=\sin(\theta_1 x)/(1+\exp(-\theta_2 x))$ where $x\in \mathcal{R}$ and $\bm \theta = (\theta_1, \theta_2)^T$.

We begin with a modified least squares method using
\begin{align*}
\displaystyle \bm S_n^{(ind)}(\bm\theta)&= \sum^{n}_{t=1}\left(y_t-f({\bm{x}_t};\bm\theta)-\frac{1}{n}\sum_{t'=1}^{n}(y_{t'}-f(\bm{x}_{t'};\bm{\theta}))\right)^2, \\
&=\left(\bm{y}-\bm f(\bm{x}, \bm{\theta})\right)^{T}\bm \Sigma_n\left(\bm{y}-\bm f(\bm{x}, \bm{\theta})\right),
\end{align*}
where $\bm y=(y_1,\ldots, y_n)^T,\ \bm f(\bm x,\bm\theta)=(f(\bm x_1;\bm\theta),\ldots, f(\bm x_n;\bm\theta))^T$, and $\bm \Sigma_n = \bm I_n-n^{-1}\bm 1\bm 1^T$. $\bm I_n$ is the identity matrix and $\bm 1$ is the column vector of one's. 
This objective function is different from a typical least squares expression in that the sample mean of the errors is subtracted from the error at each data point. This is motivated by taking into account possible non-zero mean of the errors \citep{lim2014operator}.

Since we consider temporally dependent data, we introduce a temporal weight matrix $\bm W$ to account for the temporal dependence so that the modified objective function is 
\begin{align}
	\bm S_n(\bm{\theta}) & = \left(\bm{y}-\bm f(\bm{x}, \bm{\theta})\right)^{T}\bm W^T\bm \Sigma_n\bm W\left(\bm{y}-\bm f(\bm{x}, \bm{\theta})\right), \nonumber \\
	 & = \left(\bm y - \bm f(\bm x,\bm\theta)\right)^T\bm\Sigma_w \left(\bm y - \bm f(\bm x,\bm\theta)\right) \label{eq:mainS},
\end{align}
where $\bm\Sigma_w = \bm W^T\bm \Sigma_n\bm W$.  If we know the true temporal dependence model of the error process, a natural choice of the weight matrix is from the  true covariance matrix of the error process. However, we allow the weight matrix more flexible since we do not want to assume a specific temporal dependence model for the error process.   Conditions for $\bm W$ will be introduced in the next section.

We can add a penalty function $p_{\tau_n}(\cdot)$ when our interest is to detect the relevant parameters. Then, the penalized estimator is obtained by minimizing
\begin{equation}\label{eq:main}
    \bm Q_n(\bm\theta)=\bm S_n(\bm\theta) 
+n\sum^{p}_{i=1}p_{\tau_n}(|\theta_i|).
\end{equation}
To investigate theoretical properties of the proposed estimators obtained by minimizing $\bm S_n(\bm{\theta})$ and $\bm Q_n(\bm{\theta})$, we
introduce notations and assumptions in the next section.

\subsection{Notations and Assumptions}\label{section:notations}

We start with three mixing conditions for temporal dependence: $\alpha$-mixing, $\phi$-mixing, and $\rho$-mixing. 
\begin{definition}\label{def:mixings}
\begin{enumerate}[(a)]
Consider a sequence of random variables, $\{\xi_i, i \geq 1 \}$ and let $\mathcal{F}_{i}^{j}$ denote the $\sigma$-field generated by $\{\xi_i, \ldots, \xi_j \}$. Then, $\{\xi_i, i \geq 1 \}$ is said to be 
    \item strong mixing or $\alpha$-mixing if $\alpha(m)\rightarrow 0$ as $m \rightarrow \infty$, where 
    \begin{align*}
        \alpha(m) &= \displaystyle \underset{n}{\sup} ~\alpha(\mathcal{F}_{1}^n, \mathcal{F}_{n+m}^{\infty}) \;
        \hbox{with}~~\alpha(\mathcal{F}, \mathcal{G}) & = \displaystyle \underset{A\in\mathcal{F},B\in \mathcal{G}}{\sup}~|P(AB)-P(A)P(B)|,
    \end{align*}
    \item $\phi$-mixing if $\phi(m)\rightarrow 0$ as $m \rightarrow \infty$, where 
    \begin{align*}
        \phi(m) &= \displaystyle \underset{n}{\sup} ~\phi(\mathcal{F}_{1}^n, \mathcal{F}_{n+m}^{\infty})\;
        \hbox{with}~~ \phi(\mathcal{F}, \mathcal{G}) &=\displaystyle \sup_{A\in\mathcal{F},B\in\mathcal{G},P(A)>0}|P(B|A)-P(B)|, 
    \end{align*}
    \item $\rho$-mixing if $\rho(m)\rightarrow 0$ as $m \rightarrow \infty$, where 
    \begin{align*}
        \rho(m) &= \displaystyle \underset{n}{\sup} ~\rho(\mathcal{F}_{1}^n, \mathcal{F}_{n+m}^{\infty}),\;
        \hbox{with}~~ \rho(\mathcal{F}, \mathcal{G})&=\displaystyle \sup_{f \in \mathcal{L}^2_{real}(\mathcal{F}),\ g \in \mathcal{L}^2_{real}(\mathcal{G})}|corr(f,g)|. 
    \end{align*}
    Here, $L^2_{real}(\mathcal{F})$ denotes the space of square-integrable, $\mathcal{F}$-measurable real-valued random variables \citep{bradley2005mixing}.
\end{enumerate}

\end{definition}

These mixing conditions have been widely adopted to explain dependence of a random sequence in the literature \citep{machkouri2017strongmixinglocallinear, geller2018strongmixinglocalpolynomial}. It is well-known that $\phi$-mixing implies $\rho$-mixing, $\rho$-mixing implies $\alpha$-mixing, and the strong mixing condition is one of the weakest conditions among many mixing conditions \citep{peligrad1997central, bradley2005mixing}. We assume an appropriate mixing condition for our temporal data and derive the asymptotic properties of the proposed estimators under such condition. The details appear in Assumption \ref{assumption:general}.

Next, we introduce notations and assumptions for theoretical results. Define $d_t(\bm\theta, \bm\theta^{\prime})=f({\bm{x}}_t; \bm\theta)-f({\bm{x}}_t;\bm\theta^{\prime})$ and $\bm{d}=(d_1(\bm\theta, \bm\theta^{\prime}), d_2(\bm\theta, \bm\theta^{\prime}), \ldots, d_n(\bm\theta, \bm\theta^{\prime}))^{T}$. When $f$ is twice differentiable with respect to $\bm\theta$, let $\bm{f}_k=\left(\frac{\partial{f(\bm x_1,\bm\theta)}}{\partial\theta_k},\cdots,	\frac{\partial f(\bm x_n,\bm \theta)}{\partial\theta_k}\right)^{T}$ and $\bm{f}_{kl}=\left(\frac{\partial^2 \bm f(\bm x_1,\bm \theta)}{\partial\theta_k\partial\theta_l},..., \frac{\partial^2 \bm f(\bm x_n,\bm\theta)}{\partial\theta_k\partial\theta_l}\right)^{T}$. Using $\bm{f}_k$ and $\bm{f}_{kl}$, we define ${\bm{\dot{F}}}(\bm\theta)=(\bm{f}_1,...,\bm{f}_p)$ and $\bm{{\ddot{F}}}(\bm\theta)$=Block($\bm{f}_{kl}$) so that $\bm{\dot{F}}(\bm\theta)$ is $n \times p $ matrix whose $k$th column is $\bm{f}_k$ and $\bm{{\ddot{F}}}$ is a $pn\times p$ block matrix whose $(k,l)$th block is $\bm{f}_{kl}$. $\E (\bm \epsilon)=\mu\bm 1 $ and $\Var(\bm \epsilon)=\bm\Sigma_\epsilon$. Let $\lambda_w$ and $\lambda_\epsilon$ denote the maximum eigenvalues of $\bm\Sigma_w$ and $\bm\Sigma_\epsilon$, respectively. We consider a temporal weight matrix satisfying $\bm W\bm 1 = \bm1$, i.e. the row-sums are 1's.   This condition is to handle the nonzero mean of the errors. Let $\|\cdot \|$ for a vector denote a euclidean norm and $\|\cdot\|_1$ and $\|\cdot\|_{\infty}$ for a matrix denote 1-norm and infinity norm, respectively.

The assumptions on the nonlinear function $f$, the errors $\epsilon_i$, the weight matrix $\bm W$ and the penalty function $p_{\tau_n}(\cdot)$ to investigate asymptotic properties are now introduced.    

\begin{assumption}\label{assumption:general}
    \begin{enumerate}[(1)]
        \item The nonlinear function $f\in C^2$ on the compact set $\mathcal{D} \times \Theta$ where $C^2$ is the set of twice continuously differentiable functions. 
        \item As $\| \bm \theta - \bm \theta_0 \| \rightarrow 0 $, $\left({\bm{\dot{F}}}(\bm\theta_0)^{T}\bm\Sigma_w {\bm{\dot{F}}}(\bm\theta_0)\right)^{-1} {\bm{\dot{F}}}(\bm\theta)^{T}\bm\Sigma_w {\bm{\dot{F}}}(\bm\theta)\rightarrow I_p$, elementwisely and uniformly in $\bm\theta$.
        \item There exist symmetric positive definite matrices $\bm\Gamma$ and $\bm\Gamma_\epsilon$ such that
        \begin{align*}
            \frac{1}{n\lambda_w}\bm{\dot{F}}(\bm\theta_0)^{T}\bm\Sigma_w \bm{\dot{F}}(\bm\theta_0) & \rightarrow \bm\Gamma \\
            \frac{1}{n\lambda_\epsilon\lambda_w^2}{\bm{\dot{F}}}(\bm\theta_0)^{T}\bm\Sigma_w\bm\Sigma_\epsilon\bm\Sigma_w {\bm{\dot{F}}}(\bm\theta_0) & \rightarrow \bm\Gamma_\epsilon.
        \end{align*}
        \item $\frac{\|\bm W\|_1\cdot\|\bm W\|_{\infty}}{\|\bm W^T \bm \Sigma_n \bm W\|_2} = o(n^{1/2}\lambda_\epsilon^{1/2})$.
        \item $O(1) \leq \lambda_\epsilon \leq o(n)$ and $\lambda_w \geq O(1)$.
        \item $\{\epsilon_i^2\}$ is uniformly integrable.
        \item One of the following conditions is satisfied for $\epsilon_i$.
        \begin{itemize}
            \item [$(a)$] $\{\epsilon_i\}$ is a $\phi$-mixing.
            \item [$(b)$] $\{\epsilon_i\}$ is a $\rho$-mixing and $\sum_{j \in \mathcal{N}} \rho(2^j)<\infty$.
            \item [$(c)$] For $\delta >0$, $\{\epsilon_i\}$ is a $\alpha$-mixing, $\{|\epsilon_i|^{2+\delta}\}$ is uniformly integrable, and $\sum_{j \in \mathcal{N}}n^{2/\delta}\alpha(n) < \infty$. 
        \end{itemize}
    \end{enumerate}
\end{assumption}

\begin{assumption}\label{assumption:penalty}
The first derivative of a penalty function $p_{\tau_n}(\cdot)$ denoted by $ q_{\tau_n}(\cdot)$, has the following properties:
\begin{enumerate}[(1)]

    \item $c_n=\max_{i\in\{1, \ldots, s\}}\left\{| q_{\tau_n}(|\theta_{0i}|)|\right\} = O\left(\left( \lambda_\epsilon/n\right)^{1/2}\right)$
    \item $ q_{\tau_n}(\cdot)$ is Lipschitz continuous given $\tau_n$
    \item $n^{1/2}\lambda_\epsilon^{-1/2}\lambda_w^{-1}\tau_n \rightarrow \infty$
    \item For any $C>0$, $\displaystyle\liminf_{n\rightarrow \infty}\inf_{\theta \in \left(0, C(\lambda_\epsilon/n)^{1/2}\right)} \tau_n^{-1} q_{\tau_n}(\theta)>0 $
\end{enumerate}
\end{assumption}

Assumption \ref{assumption:general} imposes mild conditions on the nonlinear function, its domain, the weight matrix, and the error process. Assumption \ref{assumption:general}-(1), (2), and (3) introduce reasonably weak conditions for the nonlinear function and the domain of data and parameters. The first condition in Assumption \ref{assumption:general}-(3) is a modified version of Grenander condition for our objective function \citep{grenander1954estimation, wang2009spatial, lim2014operator}. The second condition in Assumption \ref{assumption:general}-(3) is required to derive the variance of the asymptotic distribution. Remark \ref{remark:convergence} explains the plausibility of these conditions by addressing that slightly weaker conditions can be easily satisfied. We impose a weak condition on the temporal weight matrix in Assumption \ref{assumption:general}-(4) so that flexible weight matrices are allowed. Remark \ref{remark:weight} further discusses on the condition for the temporal weight matrix. In Assumption \ref{assumption:general}-(5), a lower bound for $\lambda_\epsilon$ can be attained if the error process is stationary with a bounded spectral density. In addition, $\lambda_\epsilon$, the maximum eigenvalue of the covariance matrix from a stationary process cannot exceed the order of $O(n)$ \citep[equation (11) in section 5.2,][]{grenander1958toeplitz}. This indicates that we allow for the wide range of the error process in Assumption \ref{assumption:general}-(5) for the asymptotic properties.

Assumption \ref{assumption:general} contains additional conditions for the asymptotic normality of the unpenalized estimator from $\bm S_n(\bm\theta)$. Assumptions \ref{assumption:general}-(6) and (7) refer to \cite{peligrad1997central}, which studied central limit theorems for linear processes. Assumption \ref{assumption:general}-(6) implies uniform boundedness of the second moment for the errors. Assumption \ref{assumption:general}-(7) provides weak conditions for temporal dependence of the errors. The detailed discussion on Assumption \ref{assumption:general}-(7) is given in Remark \ref{remark:mixing}. 

Assumption \ref{assumption:penalty} demonstrates typical conditions for a penalty function. The first two conditions guarantee that the penalized least squares estimator possess consistency with the same order as the modified weighted least squares estimator. The other two conditions contribute to the sparsity of the penalized estimator. The conditions in Assumption \ref{assumption:penalty} are similar to those in \cite{fan2001scad} and \cite{wang2009spatial}. Typically, LASSO and SCAD penalty functions are considered. The former satisfies only the first two conditions in Assumption \ref{assumption:penalty} while the latter satisfies all conditions in Assumption \ref{assumption:penalty} with properly chosen $\tau_n$. This means LASSO fails to correctly identify significant parameters while an estimator using the SCAD penalty function possesses selection consistency as well as estimation consistency. 

\begin{remark}\label{remark:convergence}
We discuss the positive definiteness of $\bm\Gamma$ and $\bm \Gamma_\epsilon$ and the boundedness of the matrix sequence. First, $\bm\Sigma_w$ is a symmetric and semi-positive definite matrix since $\bm\Sigma_w = \bm W^T\bm\Sigma_n \bm W$ and $\bm\Sigma_n$ has rank of $n-1$. Despite the rank deficiency, the sequences of the matrices are $p\times p$ matrices with $p<n$, so we believe that the limits of the sequences are likely to acquire positive definiteness. Next, the first sequence of the matrices in Assumption \ref{assumption:general}-(3) are clearly bounded above. Since $\bm \Sigma_w$ is a semi-positive definite matrix, $(n\lambda_w)^{-1}\bm{\dot{F}}(\bm\theta_0)^{T}\bm\Sigma_w \bm{\dot{F}}(\bm\theta_0) \leq n^{-1}\bm{\dot{F}}(\bm\theta_0)^{T}\bm{\dot{F}}(\bm\theta_0)=O(1)$ by the compactness of the domain (Assumption \ref{assumption:general}-(1)). With $\lambda_{max}(\bm\Sigma_w\bm\Sigma_\epsilon\bm\Sigma_w)\leq \lambda_\epsilon\lambda_w^2$, we obtain the same result for the second sequence. 
\end{remark}

\begin{remark}\label{remark:weight}
We give detailed justification for assumptions on the temporal weight matrix. By H$\ddot{\mbox{o}}$lder's inequality, $\|\bm W\|_2^2 \leq \|\bm W\|_1\|\bm W\|_{\infty}$. Hence, with $\lambda_w = \|\bm W^T \bm\Sigma_n \bm W\|_2$ Assumption \ref{assumption:general}-(4) leads to $\|\bm W\|_2 \leq o(n^{1/4}\lambda_\epsilon^{1/4}\lambda_w^{1/2})$. Recall $O(1) \leq \lambda_\epsilon \leq o(n)$ and $\lambda_w \geq O(1)$ from Assumption \ref{assumption:general}-(5). Thus, the upper bound is sufficiently large for $\|\bm W\|_2$ to allow flexible $\bm W$. In addition, since product matrices $\bm A\bm B$ and $\bm B\bm A$ share their eigenvalues, $\lambda_w = \lambda_{max}(\bm W^T \bm\Sigma_n \bm W) = \lambda_{max}(\bm \Sigma_n \bm W \bm W^T) \leq \|\bm W\|_2^2$. In summary, we obtain $O(1) \leq \| \bm W\|_2^2 \leq o(n^{1/2}\lambda_\epsilon^{1/2} \lambda_w)$, so Assumptions \ref{assumption:general}-(4) and (5) together provide a flexible upper bound and lower bound for $\bm W$. The flexible structure of $\bm W$ will expand the applicability of our method and help practitioners to obtain their desirable results without excessive concerns about the choice of $\bm W$. This advantage can be observed from our simulation results presented in Tables \ref{table:add_scad_5_estimation}-\ref{table:add_scad_8_4_estimation}. Our proposed approach with a weight matrix outperforms  the other method as well as the approach without a weight matrix for most cases in the estimation results.  Moreover, the choice of $\bm W$ does not significantly affect the results.
\end{remark}

\begin{remark}\label{remark:mixing}
There exist many familiar time series processes that satisfy Assumption \ref{assumption:general}-(7). Autoregressive (AR) processes and moving average (MA) processes are strongly mixing under mild conditions \citep{athreya1986arma}. \cite{athreya1986arma} also mentions that finite order autoregressive moving average (ARMA) processes are $\phi$-mixing under mild conditions. Furthermore, ARMA processes and bilinear processes are strong mixing with $\alpha(n) = O(e^{-n\rho})$ with some $\rho >0$ \citep{roussas1992strongmixing}.
\end{remark}

\subsection{Theoretical results}\label{section:theorems}

First, we construct the existence and the consistency of the penalized least squares estimator. We have the existence and the consistency results of the modified weighted least squares estimator, but provide them to the Supplementary Material as Lemma 1, as our final goal lies in providing the asymptotic properties of the penalized estimators.

\begin{theorem}\label{thm:consistency-penalty}
For any $\varepsilon>0$ and $b_n = (\lambda_\epsilon/n)^{1/2}+c_n$, under Assumptions 1-(1), (2), (3), (5) and \ref{assumption:penalty}-(1),(2), there exists a positive constant $C$ such that
    $$P\left(\inf_{\|\bm v\|=C} \bm Q_n(\bm \theta_0+b_n\bm v) - \bm Q_n(\bm\theta_0) >0 \right) > 1-\varepsilon$$
    for large enough $n$. Therefore, with probability tending to 1, there exists a local minimizer of $\bm Q_n (\bm \theta)$, say $\hat{\bm\theta}$, in the ball centered at $\bm\theta_0$ with the radius $b_n \bm v$. By Assumptions \ref{assumption:general}-(5) and \ref{assumption:penalty}-(1), $b_n=o(1)$, which leads to the consistency of $\hat{\bm\theta}$.
\end{theorem}

The next theorem establishes the oracle property of the penalized least squares estimator. We also obtained the asymptotic normality of the modified weighted least squares estimator without penalization. The result is provided as Lemma 3 in the supplementary materials.

\begin{theorem}[Oracle property]\label{thm:sparsity}
With $\hat{\bm \theta}$, a consistent estimator introduced in Theorem \ref{thm:consistency-penalty} using $\bm Q_n(\bm\theta)$, if Assumptions \ref{assumption:general} and \ref{assumption:penalty} are satisfied,
\begin{enumerate}[(i)]
    \item $P\left(\hat{\theta}_i=0\right) \rightarrow 1,$ for $i \in \{s+1, \ldots, p\}$.
    \item Also,
    $$\left(\frac{n}{\lambda_\epsilon}\right)^{1/2}\left(\hat{\bm\theta}_1-\bm\theta_{01}+\left((2\lambda_w\bm\Gamma)^{-1}\right)_{11}\bm \beta_{n,s}\right)~\overset{d}{\longrightarrow}~N\left(0,\left( \bm\Gamma^{-1}\bm\Gamma_\epsilon\bm\Gamma^{-1}\right)_{11}\right),$$
    where $\hat{\bm \theta}_1=(\hat\theta_1, \ldots, \hat\theta_s)^T,\ \bm\theta_{01} = (\theta_{01}, \ldots, \theta_{0s})^T,\ \bm\beta_{n,s}=(q_{\tau_n}({|\theta_{01}|)sgn(\theta}_{01}),\ldots,$ $q_{\tau_n}(|\theta_{0s}|)sgn(\theta_{0s}))^{T}$ and $\bm A_{11}$ is the $s\times s$ upper-left matrix of $\bm A$.
\end{enumerate}
\end{theorem}

Note that the estimators from Lemma 1 in the supplementary material and Theorem \ref{thm:consistency-penalty} have convergence orders of $a_n$ and $b_n$, respectively. $a_n$ and $b_n$ eventually have the same order of $(\lambda_\epsilon/n)^{1/2}$ by Assumption \ref{assumption:penalty}-(1). One may think  $\lambda_w$, information of $\bm W$, makes no contribution to both theorems, even though we have $\bm W$ in the objective functions. Recall that $\lambda_w$ contributes to $\tau_n$ via Assumption \ref{assumption:penalty}-(2) and (4), and $c_n$, which appears in $b_n$, is related to  $\tau_n$ by Assumption \ref{assumption:penalty}-(1). This is where $\lambda_w$ implicitly comes into the theorems. We could impose different conditions to make $\lambda_w$ explicitly appear in the theorems. However, such conditions restrict the flexibility of $\bm \Sigma_\epsilon$ so that the applicability of the proposed methods becomes limited. Instead, we decide to keep the current assumptions to allow flexible $\bm\Sigma_\epsilon$ and attain the implicit involvement of $\lambda_w$ in the theorems. The proofs for Theorem \ref{thm:sparsity} and necessary lemmas are given in the supplementary material. Note that the asymptotic variances of $\hat{\bm\theta}^{(s)}$ and $ \hat{\bm\theta}$ do not have closed forms since we do not assume any specific form on the dependence structure of the error process other than mixing conditions.  Hence, estimation of asymptotic variance goes beyond the scope of this work.

\section{Simulation Study}
\label{section:simulation}

We investigate performance of the proposed estimator, in particular, the penalized version with two different penalty functions, LASSO and SCAD using simulated data sets. First, we consider the data generated from a nonlinear additive error model: 
\begin{equation}
\label{eq:additive}
\displaystyle y_t=\frac{1}{1+\exp(-\bm x_t^{T}\bm\theta_0)}+\epsilon_t,
\end{equation}
where $\bm\theta_0=(\theta_{01}, \theta_{02}, \ldots, \theta_{0,20})^{T}$ with $\theta_{01}=1, \theta_{02}=1.2, \theta_{03}=0.6$, and the others being zero.  The first component of the covariate $\bm x$ comes from $U[-1,1]$, a uniform distribution on $[-1,1]$, and the other components of $\bm x$ are simulated from a joint normal distribution with the zero mean, the variance being 0.6 and pairwise covariance being 0.1. This is a slight modification of the covariates setting used in \cite{jiang2012nonlinearwcqr}. For $\epsilon_t$, the AR(1) and ARMA(1,1) with the non-zero mean are considered since these processes not only represent typical time series processes but also possess the strong mixing property. The choices of the AR(1) coefficient ($\rho$) are 0.5 and 0.9. For the ARMA process, the parameters for the AR and MA parts are fixed as 0.8 ($\rho$) and 0.4 ($\phi$), respectively. For the non-zero mean, $\mu$, the choices are 0.1 and 0.5 and the same for the standard deviation, $\sigma$. We only show the results when $\sigma=0.5$ to highlight findings as it is more difficult settings due to a larger variance. We consider three sample sizes; $n= 50, 100$ and $200$ to investigate improvements according to the increasing sample sizes.

A coordinate descent (CD) algorithm, in particular, a cyclic CD algorithm  \citep{breheny2011cdscad}, was implemented to calculate the minimizer of the objective function given in \eqref{eq:main}. Although \cite{breheny2011cdscad} considered the convergence of the CD algorithm in a linear model, the cyclic CD algorithm worked well for our penalized nonlinear regression problem as well.

To select the tuning parameter, $\tau_n$, of the penalty function, we use a following BIC-type criterion which is similar to the BIC-type criterion in  \cite{chu2011penalized} for spatially correlated data. $\mbox{BIC} = \log(\hat{\sigma}^2) + \log(n)\cdot\widehat{df}/n,$ where $\widehat{df}$ is the number of significant estimates, $\hat{\sigma}^2=\overline{r^2}-\bar{r}^2$ (variance of the residuals) with $\bar{r}=n^{-1}\sum_{t=1}^{n}r_t$ and $\overline{r^2}=n^{-1}\sum_{t=1}^{n}r_t^2$. Here, $ r_t = y_t - f(\bm x_t; \bm{\hat{\theta}})$. The tuning parameter $a$ in the SCAD penalty was fixed at 3.7 as recommended in \cite{fan2001scad}.

Our proposed method is denoted as penalized modified weighted least squares (PMWLS). For the simulation study, the square roots of the inverse of the covariance matrices from the AR(1) process with the AR coefficient $\rho=0.5, 0.9$ and the ARMA(1,1) process with AR and MA coefficients $(\rho,\phi)=(0.8,0.4)$ are considered for the weight matrices after scaling to have the row-sums 1. We compare the results of the proposed PMWLS method with the results from a penalized  least squares with a weight matrix (PWLS). For fair comparison with the proposed method, a temporal weight matrix is also  considered for the PWLS method. In the PWLS method, we introduce an intercept term to account for a possible non-zero mean of the error in the equation \eqref{eq:additive}, which is a straightforward way to handle non-zero mean of the error. That is, PWLS minimizes 
$$\left(\bm y-\beta_0-\bm f(\bm x;\bm\theta)\right)^T\widetilde{\bm W} (\bm y-\beta_0-\bm f(\bm x;\bm\theta))+n\sum_{i=1}^{p} p_{\tau_n}(|\theta_i|),$$
where $\beta_0$ is the intercept term and  $\widetilde{\bm W}$ is a weight matrix. For weight matrices, we used the inverse of the covariance matrices from the same models considered for the PMWLS method such as  AR(1) and ARMA(1,1) processes.

Tables \ref{table:add_scad_5_estimation}-\ref{table:add_scad_8_4_estimation} report the values of mean squared error (MSE) with standard deviation of squared error (SD) in parenthesis for the estimates with a SCAD penalty from 100 repetitions of data. MSE and SD are calculated as 
\begin{align*}
    MSE & =\frac{1}{R\,p}\sum_{j=1}^{R}\left\|\hat{\bm\theta}^{(j)}-\bm\theta_0\right\|^2, \\
    SD & = \displaystyle \sqrt{\frac{1}{R-1} \sum_{j=1}^{R} \left\|\hat{\bm\theta}^{(j)}-{\bm{\bar\theta}}\right\|^2},
\end{align*}
where $\hat{\bm\theta}^{(j)}$ stands for the estimate from the $j$-th repetition and ${\bm {\bar\theta}}$ is the sample mean of $\hat{\bm\theta}^{(j)}$ for $j=1,\ldots, 100$. 
The results for the estimates with a LASSO penalty are provided in the supplementary material. 

The MSE and SD results in Tables \ref{table:add_scad_5_estimation}-\ref{table:add_scad_8_4_estimation} show good performances in estimating the true parameters. Overall, the estimation results are robust over various weight matrices. While the results with the LASSO penalty for both PMWLS and PWLS  are comparable  (Tables S1-S3 in the supplementary material), the proposed method (PMWLS) outperforms the PWLS method for most cases with the SCAD penalty. In particular, the improvement by the PMWLS method with the SCAD penalty is more apparent when the error process has stronger dependence (Tables \ref{table:add_scad_9_estimation} and  \ref{table:add_scad_8_4_estimation}). This performance improvement would be from estimation efficiency in finite sample since the PMWLS method estimates one less number of parameters, an intercept, compared to the PWLS method. Weight matrices for both PMWLS and PWLS methods help to reduce MSE and the improvement is more clear when the dependence is strong. On the other hand, the choice of the weight matrices do not make much difference except that the performance is better when the weight matrix is introduced.

\begin{table}%
\footnotesize
    \centering
    \caption{\small{Estimation results with SCAD from the equation \eqref{eq:additive} when the error process is AR(1) with $\rho=0.5$. Mean squared error values are presented with standard deviation in the parenthesis. The rows without $\rho$ or $(\rho,\phi)$ indicate that no weight matrix is used.}}
    \begin{tabular}{clrrr}
    \Xhline{3\arrayrulewidth}
    \multicolumn{5}{c}{AR(1) with $\rho=0.5$} \\ \hline
     $(\mu,\sigma)$ & Methods & \multicolumn{1}{c}{$n=50$} & \multicolumn{1}{c}{$n=100$} & \multicolumn{1}{c}{$n=200$} \\ \hline
    
    \multirow{8}{*}{$(0.1,0.5)$} & PMWLS & 15.34 (1.67) & 4.98 (0.97) & 2.38 (0.66) \\ \cline{2-2}
    & $\mbox{PMWLS } {\tiny (\rho=0.5)}$ & 11.36 (1.38) & 4.17 (0.85) & 2.36 (0.64) \\ \cline{2-2}
    & $\mbox{PMWLS } {\tiny (\rho=0.9)}$ & 10.69 (1.30) & 4.22 (0.84) & 2.22 (0.61) \\ \cline{2-2}
    & $\mbox{PMWLS } {\tiny (\rho=0.8,\phi=0.4)}$ & 10.94 (1.29) & 4.42 (0.86) & 2.49 (0.64) \\ \cline{2-2}
    & PWLS & 15.00 (1.64) & 4.89 (0.96) & 2.51 (0.67) \\ \cline{2-2}
    & $\mbox{PWLS } {\tiny (\rho=0.5)}$ & 11.88 (1.40) & 4.22 (0.85) & 2.29 (0.63) \\ \cline{2-2}
    & $\mbox{PWLS } {\tiny (\rho=0.9)}$ & 11.15 (1.34) & 4.09 (0.83) & 2.42 (0.63) \\ \cline{2-2}
    & $\mbox{PWLS } {\tiny (\rho=0.8,\phi=0.4)}$ & 11.07 (1.33) & 4.53 (0.86) & 2.38 (0.63) \\\hline
    \multirow{8}{*}{$(0.5,0.5)$} & PMWLS & 9.44 (1.32) & 5.15 (0.98) & 2.60 (0.70) \\ \cline{2-2}
    & $\mbox{PMWLS } {\tiny (\rho=0.5)}$ & 9.84 (1.31) & 4.19 (0.84) & 2.02 (0.59) \\ \cline{2-2}
    & $\mbox{PMWLS } {\tiny (\rho=0.9)}$ & 9.58 (1.27) & 4.07 (0.82) & 1.97 (0.58) \\ \cline{2-2}
    & $\mbox{PMWLS } {\tiny (\rho=0.8,\phi=0.4)}$ & 11.71 (1.39) & 4.73 (0.87) & 2.24 (0.62) \\ \cline{2-2}
    & PWLS & 10.17 (1.34) & 6.78 (1.10) & 3.81 (0.83) \\ \cline{2-2}
    & $\mbox{PWLS } {\tiny (\rho=0.5)}$ & 9.17 (1.24) & 4.20 (0.84) & 2.17 (0.60) \\ \cline{2-2}
    & $\mbox{PWLS } {\tiny (\rho=0.9)}$ & 10.66 (1.34) & 4.37 (0.83) & 2.16 (0.59) \\ \cline{2-2}
    & $\mbox{PWLS } {\tiny (\rho=0.8,\phi=0.4)}$ & 12.17 (1.43) & 5.10 (0.90) & 2.67 (0.66) \\
    \Xhline{3\arrayrulewidth}
    \multicolumn{5}{l}{\footnotesize{$\ast$ The actual MSE values are $0.01 \times$ the reported values.}}
    \end{tabular}
    \label{table:add_scad_5_estimation}
\end{table}

\begin{table}
\footnotesize
    \centering
    \caption{\small{Estimation results with SCAD from the equation \eqref{eq:additive} when the error process is AR(1) with $\rho=0.9$. The other configurations are identical to Table \ref{table:add_scad_5_estimation}.}}
    \begin{tabular}{clrrr}
    \Xhline{3\arrayrulewidth}
    \multicolumn{5}{c}{AR(1) with $\rho=0.9$} \\ \hline
    $(\mu,\sigma)$ & Methods & \multicolumn{1}{c}{$n=50$} & \multicolumn{1}{c}{$n=100$} & \multicolumn{1}{c}{$n=200$} \\ \hline
    
    \multirow{8}{*}{$(0.1,0.5)$} & PMWLS & 9.42 (1.31) & 4.14 (0.85) & 2.49 (0.68) \\ \cline{2-2}
    & $\mbox{PMWLS } {\tiny (\rho=0.5)}$ & 4.90 (0.83) & 3.65 (0.69) & 1.37 (0.47) \\ \cline{2-2}
    & $\mbox{PMWLS } {\tiny (\rho=0.9)}$ & 4.96 (0.79) & 3.76 (0.67) & 1.49 (0.47) \\ \cline{2-2}
    & $\mbox{PMWLS } {\tiny (\rho=0.8,\phi=0.4)}$ & 5.29 (0.82) & 3.76 (0.69) & 1.42 (0.46) \\ \cline{2-2}
    & PWLS & 14.59 (1.64) & 4.58 (0.89) & 2.47 (0.68) \\ \cline{2-2}
    & $\mbox{PWLS } {\tiny (\rho=0.5)}$ & 5.14 (0.86) & 2.93 (0.66) & 1.41 (0.48) \\ \cline{2-2}
    & $\mbox{PWLS } {\tiny (\rho=0.9)}$ & 4.95 (0.77) & 3.56 (0.65) & 1.41 (0.46) \\ \cline{2-2}
    & $\mbox{PWLS } {\tiny (\rho=0.8,\phi=0.4)}$ & 5.08 (0.80) & 3.48 (0.65) & 1.44 (0.47) \\\hline
    \multirow{8}{*}{$(0.5,0.5)$} & PMWLS & 9.38 (1.35) & 4.51 (0.91) & 2.33 (0.67) \\ \cline{2-2}
    & $\mbox{PMWLS } {\tiny (\rho=0.5)}$ & 5.13 (0.84) & 2.95 (0.65) & 1.42 (0.47) \\ \cline{2-2}
    & $\mbox{PMWLS } {\tiny (\rho=0.9)}$ & 5.19 (0.82) & 2.84 (0.63) & 1.41 (0.47) \\ \cline{2-2}
    & $\mbox{PMWLS } {\tiny (\rho=0.8,\phi=0.4)}$ & 5.76 (0.84) & 3.22 (0.67) & 1.32 (0.45) \\ \cline{2-2}
    & PWLS & 12.02 (1.53) & 7.19 (1.17) & 2.93 (0.72) \\ \cline{2-2}
    & $\mbox{PWLS } {\tiny (\rho=0.5)}$ & 5.53 (0.90) & 3.31 (0.68) & 1.65 (0.50)  \\ \cline{2-2}
    & $\mbox{PWLS } {\tiny (\rho=0.9)}$ & 5.25 (0.82) & 3.36 (0.67) & 1.69 (0.50) \\ \cline{2-2}
    & $\mbox{PWLS } {\tiny (\rho=0.8,\phi=0.4)}$ & 5.75 (0.83) & 3.38 (0.67) & 1.60 (0.48) \\
    \Xhline{3\arrayrulewidth}
    \multicolumn{5}{l}{\footnotesize{$\ast$ The actual MSE values are $0.01 \times$ the reported values.}}
    \end{tabular}
    \label{table:add_scad_9_estimation}
\vskip0.4cm
\footnotesize
    \caption{\small{Estimation results with SCAD from the equation \eqref{eq:additive} when the error process is ARMA(1,1) with $(\rho,\phi)=(0.8, 0.4)$. The other configurations are identical to Table \ref{table:add_scad_5_estimation}.}}
    \begin{tabular}{clrrr}
    \Xhline{3\arrayrulewidth}
    \multicolumn{5}{c}{ARMA(1,1) with $\rho=0.8,\phi=0.4$} \\ \hline
     $(\mu,\sigma)$ & Methods & \multicolumn{1}{c}{$n=50$} & \multicolumn{1}{c}{$n=100$} & \multicolumn{1}{c}{$n=200$} \\ \hline
    
    \multirow{8}{*}{$(0.1,0.5)$} & PMWLS & 5.83 (1.07) & 2.62 (0.71) & 0.83 (0.41) \\ \cline{2-2}
    & $\mbox{PMWLS } {\tiny (\rho=0.5)}$ & 3.53 (0.78) & 1.94 (0.57) & 0.50 (0.31) \\ \cline{2-2}
    & $\mbox{PMWLS } {\tiny (\rho=0.9)}$ & 3.38 (0.75) & 1.93 (0.55) & 0.50 (0.30) \\ \cline{2-2}
    & $\mbox{PMWLS } {\tiny (\rho=0.8,\phi=0.4)}$ & 3.40 (0.74) & 1.78 (0.54) & 0.46 (0.29) \\ \cline{2-2}
    & PWLS & 5.73 (1.05) & 2.66 (0.71) & 0.90 (0.43) \\ \cline{2-2}
    & $\mbox{PWLS } {\tiny (\rho=0.5)}$ & 3.52 (0.78) & 1.95 (0.57) & 0.43 (0.29) \\ \cline{2-2}
    & $\mbox{PWLS } {\tiny (\rho=0.9)}$ & 3.22 (0.73) & 1.93 (0.56) & 0.59 (0.33) \\ \cline{2-2}
    & $\mbox{PWLS } {\tiny (\rho=0.8,\phi=0.4)}$ & 3.60 (0.75) & 1.89 (0.55) & 0.51 (0.31) \\\hline
    \multirow{8}{*}{$(0.5,0.5)$} & PMWLS & 4.86 (0.91) & 2.35 (0.67) & 0.85 (0.40) \\ \cline{2-2}
    & $\mbox{PMWLS } {\tiny (\rho=0.5)}$ & 4.09 (0.76) & 1.49 (0.50) & 0.58 (0.32) \\ \cline{2-2}
    & $\mbox{PMWLS } {\tiny (\rho=0.9)}$ & 4.16 (0.74) & 1.53 (0.50) & 0.58 (0.31) \\ \cline{2-2}
    & $\mbox{PMWLS } {\tiny (\rho=0.8,\phi=0.4)}$ & 4.46 (0.77) & 1.58 (0.51) & 0.55 (0.30) \\ \cline{2-2}
    & PWLS & 9.08 (1.30) & 3.07 (0.77) & 1.01 (0.44) \\ \cline{2-2}
    & $\mbox{PWLS } {\tiny (\rho=0.5)}$ & 4.68 (0.80) & 1.86 (0.55) & 0.88 (0.37) \\ \cline{2-2}
    & $\mbox{PWLS } {\tiny (\rho=0.9)}$ & 4.29 (0.73) & 1.81 (0.51) & 0.79 (0.35) \\ \cline{2-2}
    & $\mbox{PWLS } {\tiny (\rho=0.8,\phi=0.4)}$ & 4.56 (0.75) & 1.94 (0.55) & 0.79 (0.35) \\
    \Xhline{3\arrayrulewidth}
    \multicolumn{5}{l}{\footnotesize{$\ast$ The actual MSE values are $0.01 \times$ the reported values.}}
    \end{tabular}
    \label{table:add_scad_8_4_estimation}
\end{table}

Tables \ref{table:add_scad_5_selection}-\ref{table:add_scad_8_4_selection} demonstrate selection results of PMWLS and PWLS methods with the SCAD penalty. The results for the LASSO penalty are provided in Tables S4-S6 in the supplementary material. True positive (TP) counts the number of significant estimates among the significant true parameters and true negative (TN) counts the number of insignificant estimates among the insignificant true parameters. As the sample size increases, the values of TP approaches the true value. The performance in terms of TN is better for SCAD compared to LASSO. These results correspond to  Theorem \ref{thm:sparsity} since LASSO does not fulfill all conditions in Assumption \ref{assumption:penalty} as discussed. Lastly, selection performance between PMWLS and PWLS methods are comparable. Different from the estimation performance, the results are comparable over the choice of weight matrices including no weight matrix.

\begin{table}
\footnotesize
    \centering
    \caption{\small{Selection results with SCAD from the equation \eqref{eq:additive} when the error process is AR(1) with $\rho=0.5$. TP counts the number of significant estimates among the significant true parameters and TN counts the number of insignificant estimates among the insignificant true parameters. }}
    \begin{tabular}{clrrrrrrr}
    \Xhline{3\arrayrulewidth}
    \multicolumn{9}{c}{AR(1) with $\rho=0.5$} \\ \hline
    \multirow{2}{*}{$(\mu,\sigma)$} & \multirow{2}{*}{Methods} & \multicolumn{3}{c}{TP} && \multicolumn{3}{c}{TN} \\ \cline{3-5} \cline{7-9} 
    & & \multicolumn{1}{c}{50} & \multicolumn{1}{c}{100} & \multicolumn{1}{c}{200} && \multicolumn{1}{c}{50} & \multicolumn{1}{c}{100} & \multicolumn{1}{c}{200}\\ \hline
    \multirow{8}{*}{$(0.1,0.5)$} & PMWLS & 1.35 & 2.04 & 2.40 & & 16.99 & 17.00 & 17.00 \\ \cline{2-2}
    & $\mbox{PMWLS } {\tiny (\rho=0.5)}$ & 1.31 & 2.02 & 2.33 & & 17.00 & 17.00 & 17.00 \\ \cline{2-2}
    & $\mbox{PMWLS } {\tiny (\rho=0.9)}$ & 1.26 & 1.97 & 2.35 & & 16.97 & 17.00 & 17.00 \\ \cline{2-2}
    & $\mbox{PMWLS } {\tiny (\rho=0.8,\phi=0.4)}$ & 1.11 & 1.93 & 2.33 & & 17.00 & 17.00 & 17.00 \\ \cline{2-2}
    & PWLS & 1.27 & 2.07 & 2.38 & & 17.00 & 17.00 & 17.00 \\ \cline{2-2}
    & $\mbox{PWLS } {\tiny (\rho=0.5)}$ & 1.28 & 1.98 & 2.35 & & 16.98 & 17.00 & 17.00 \\ \cline{2-2}
    & $\mbox{PWLS } {\tiny (\rho=0.9)}$ & 1.26 & 2.00 & 2.31 & & 17.00 & 17.00 & 17.00 \\ \cline{2-2}
    & $\mbox{PWLS } {\tiny (\rho=0.8,\phi=0.4)}$ & 1.17 & 1.92 & 2.36 & & 16.98 & 17.00 & 17.00 \\\hline
    \multirow{8}{*}{$(0.5,0.5)$} & PMWLS & 1.51 & 2.03 & 2.39 & & 16.99 & 17.00 & 17.00 \\ \cline{2-2}
    & $\mbox{PMWLS } {\tiny (\rho=0.5)}$ & 1.38 & 1.99 & 2.39 & & 17.00 & 16.99 & 17.00 \\ \cline{2-2}
    & $\mbox{PMWLS } {\tiny (\rho=0.9)}$ & 1.36 & 2.00 & 2.41 & & 16.96 & 17.00 & 16.99 \\ \cline{2-2}
    & $\mbox{PMWLS } {\tiny (\rho=0.8,\phi=0.4)}$ & 1.19 & 1.90 & 2.39 & & 17.00 & 17.00 & 16.99 \\ \cline{2-2}
    & PWLS & 1.35 & 1.81 & 2.24 & & 17.00 & 17.00 & 17.00 \\ \cline{2-2}
    & $\mbox{PWLS } {\tiny (\rho=0.5)}$ & 1.35 & 1.98 & 2.32 & & 17.00 & 17.00 & 17.00 \\ \cline{2-2}
    & $\mbox{PWLS } {\tiny (\rho=0.9)}$ & 1.25 & 1.90 & 2.32 & & 17.00 & 17.00 & 17.00 \\ \cline{2-2}
    & $\mbox{PWLS } {\tiny (\rho=0.8,\phi=0.4)}$ & 1.20 & 1.80 & 2.25 & & 17.00 & 17.00 & 17.00 \\
    \Xhline{3\arrayrulewidth}
    \end{tabular}
    \label{table:add_scad_5_selection}
\vskip0.4cm 
\footnotesize
    \centering
    \caption{\small{Selection results with SCAD from the equation \eqref{eq:additive} when the error process is AR(1) with $\rho=0.9$. The other configurations are identical to Table \ref{table:add_scad_5_selection}. }}
    \begin{tabular}{clrrrrrrr}
    \Xhline{3\arrayrulewidth}
    \multicolumn{9}{c}{AR(1) with $\rho=0.9$} \\ \hline
    \multirow{2}{*}{$(\mu,\sigma)$} & \multirow{2}{*}{Methods} & \multicolumn{3}{c}{TP} && \multicolumn{3}{c}{TN} \\ \cline{3-5} \cline{7-9} 
    & & \multicolumn{1}{c}{50} & \multicolumn{1}{c}{100} & \multicolumn{1}{c}{200} && \multicolumn{1}{c}{50} & \multicolumn{1}{c}{100} & \multicolumn{1}{c}{200}\\ \hline
    \multirow{8}{*}{$(0.1,0.5)$} & PMWLS & 1.69 & 2.04 & 2.50 & & 16.97 & 17.00 & 16.99 \\ \cline{2-2}
    & $\mbox{PMWLS } {\tiny (\rho=0.5)}$ & 1.69 & 1.91 & 2.50 & & 17.00 & 17.00 & 17.00 \\ \cline{2-2}
    & $\mbox{PMWLS } {\tiny (\rho=0.9)}$ & 1.64 & 1.88 & 2.46 & & 17.00 & 17.00 & 17.00 \\ \cline{2-2}
    & $\mbox{PMWLS } {\tiny (\rho=0.8,\phi=0.4)}$ & 1.58 & 1.89 & 2.47 & & 17.00 & 17.00 & 17.00 \\ \cline{2-2}
    & PWLS & 1.62 & 1.99 & 2.49 & & 16.98 & 17.00 & 17.00 \\ \cline{2-2}
    & $\mbox{PWLS } {\tiny (\rho=0.5)}$ & 1.64 & 1.93 & 2.49 & & 17.00 & 17.00 & 17.00 \\ \cline{2-2}
    & $\mbox{PWLS } {\tiny (\rho=0.9)}$ & 1.62 & 1.92 & 2.47 & & 17.00 & 17.00 & 17.00 \\ \cline{2-2}
    & $\mbox{PWLS } {\tiny (\rho=0.8,\phi=0.4)}$ & 1.62 & 1.93 & 2.49 & & 17.00 & 17.00 & 17.00 \\\hline
    \multirow{8}{*}{$(0.5,0.5)$} & PMWLS & 1.83 & 2.05 & 2.49 & & 16.98 & 17.00 & 16.99 \\ \cline{2-2}
    & $\mbox{PMWLS } {\tiny (\rho=0.5)}$ & 1.71 & 2.09 & 2.50 & & 17.00 & 17.00 & 17.00 \\ \cline{2-2}
    & $\mbox{PMWLS } {\tiny (\rho=0.9)}$ & 1.65 & 2.09 & 2.49 & & 16.98 & 17.00 & 17.00 \\ \cline{2-2}
    & $\mbox{PMWLS } {\tiny (\rho=0.8,\phi=0.4)}$ & 1.54 & 2.01 & 2.51 & & 16.98 & 17.00 & 17.00 \\ \cline{2-2}
    & PWLS & 1.68 & 1.98 & 2.35 & & 16.95 & 17.00 & 17.00 \\ \cline{2-2}
    & $\mbox{PWLS } {\tiny (\rho=0.5)}$ & 1.66 & 1.98 & 2.41 & & 17.00 & 17.00 & 17.00 \\ \cline{2-2}
    & $\mbox{PWLS } {\tiny (\rho=0.9)}$ & 1.62 & 1.96 & 2.39 & & 17.00 & 17.00 & 17.00 \\ \cline{2-2}
    & $\mbox{PWLS } {\tiny (\rho=0.8,\phi=0.4)}$ & 1.52 & 1.95 & 2.40 & & 17.00 & 17.00 & 17.00 \\
    \Xhline{3\arrayrulewidth}
    \end{tabular}
    \label{table:add_scad_9_selection}
\end{table}

\begin{table}
\footnotesize
    \centering
    \caption{\small{Selection results with SCAD from the equation \eqref{eq:additive} when the error process is ARMA(1) with $(\rho,\phi)=(0.8, 0.4)$. The other configurations are identical to Table \ref{table:add_scad_5_selection}.}}
    \begin{tabular}{clrrrrrrr}
    \Xhline{3\arrayrulewidth}
    \multicolumn{9}{c}{ARMA(1,1) with $\rho=0.8$, $\phi=0.4$} \\ \hline
    \multirow{2}{*}{$(\mu,\sigma)$} & \multirow{2}{*}{Methods} & \multicolumn{3}{c}{TP} && \multicolumn{3}{c}{TN} \\ \cline{3-5} \cline{7-9} 
    & & \multicolumn{1}{c}{50} & \multicolumn{1}{c}{100} & \multicolumn{1}{c}{200} && \multicolumn{1}{c}{50} & \multicolumn{1}{c}{100} & \multicolumn{1}{c}{200}\\ \hline
    \multirow{8}{*}{$(0.1,0.5)$} & PMWLS & 2.15 & 2.43 & 2.85 & & 17.00 & 16.98 & 16.98 \\ \cline{2-2}
    & $\mbox{PMWLS } {\tiny (\rho=0.5)}$ & 2.15 & 2.40 & 2.83 & & 16.99 & 17.00 & 17.00 \\ \cline{2-2}
    & $\mbox{PMWLS } {\tiny (\rho=0.9)}$ & 2.17 & 2.39 & 2.82 & & 16.99 & 17.00 & 17.00 \\ \cline{2-2}
    & $\mbox{PMWLS } {\tiny (\rho=0.8,\phi=0.4)}$ & 2.15 & 2.44 & 2.84 & & 17.00 & 17.00 & 17.00 \\ \cline{2-2}
    & PWLS & 2.14 & 2.44 & 2.85 & & 16.98 & 17.00 & 17.00 \\ \cline{2-2}
    & $\mbox{PWLS } {\tiny (\rho=0.5)}$ & 2.16 & 2.40 & 2.85 & & 17.00 & 17.00 & 17.00 \\ \cline{2-2}
    & $\mbox{PWLS } {\tiny (\rho=0.9)}$ & 2.16 & 2.40 & 2.79 & & 17.00 & 17.00 & 17.00 \\ \cline{2-2}
    & $\mbox{PWLS } {\tiny (\rho=0.8,\phi=0.4)}$ & 2.09 & 2.42 & 2.82 & & 17.00 & 17.00 & 17.00 \\\hline
    \multirow{8}{*}{$(0.5,0.5)$} & PMWLS & 1.93 & 2.53 & 2.79 & & 16.98 & 16.98 & 17.00 \\ \cline{2-2}
    & $\mbox{PMWLS } {\tiny (\rho=0.5)}$ & 1.88 & 2.50 & 2.76 & & 16.99 & 17.00 & 17.00 \\ \cline{2-2}
    & $\mbox{PMWLS } {\tiny (\rho=0.9)}$ & 1.85 & 2.48 & 2.75 & & 17.00 & 17.00 & 17.00 \\ \cline{2-2}
    & $\mbox{PMWLS } {\tiny (\rho=0.8,\phi=0.4)}$ & 1.82 & 2.49 & 2.77 & & 17.00 & 17.00 & 17.00 \\ \cline{2-2}
    & PWLS & 1.80 & 2.47 & 2.73 & & 17.00 & 17.00 & 17.00 \\ \cline{2-2}
    & $\mbox{PWLS } {\tiny (\rho=0.5)}$ & 1.74 & 2.39 & 2.62 & & 17.00 & 17.00 & 17.00 \\ \cline{2-2}
    & $\mbox{PWLS } {\tiny (\rho=0.9)}$ & 1.76 & 2.34 & 2.65 & & 17.00 & 17.00 & 17.00 \\ \cline{2-2}
    & $\mbox{PWLS } {\tiny (\rho=0.8,\phi=0.4)}$ & 1.72 & 2.36 & 2.66 & & 17.00 & 17.00 & 17.00 \\
    \Xhline{3\arrayrulewidth}
    \end{tabular}
    \label{table:add_scad_8_4_selection}
\end{table}

Next, we consider a following nonlinear multiplicative model: 
\begin{equation}
\label{eq:multiplicative}
\displaystyle y_t=\frac{1}{1+\exp(-\bm x_t^{T}\bm\theta_0)}\times\epsilon_t.
\end{equation}
For $\epsilon_t$, the exponentiated AR processes or an ARMA process are considered since the $\epsilon_{t}$'s in the equation~\eqref{eq:multiplicative} are allowed to have only positive values. The AR and ARMA coefficients and the parameter setting of $\bm\theta$ are the same as the one in the model \eqref{eq:additive}. We transformed the model in the log scale and apply our approach. Then, we compare the results with the PWLS method and an `additive' method, where the estimator from the additive method is calculated as if the data are from a nonlinear additive model without log transformation. For this simulation, we provide the results using the SCAD penalty. 

\indent Tables \ref{table:multi_scad_5_estimation}-\ref{table:multi_scad_8_4_estimation} show estimation performances of PMWLS and PWLS methods for the model given in \eqref{eq:multiplicative} and Tables \ref{table:multi_scad_5_selection}-\ref{table:multi_scad_8_4_selection} describe selection performances. For most data  generation settings, our proposed method (PMWLS) shows better results. In particular, when $(\mu, \sigma) = (0.5, 0.5)$ and the sample size is small, the difference in performance between PMWLS and PWLS methods becomes more evident. Hence, we argue that PMWLS is preferred over PWLS  in practice since PMWLS shows better finite sample performance. In terms of choice of weight matrices, the results are similar to those in the first simulation study. That is, estimation performance is better when we use a weight matrix for both approaches while the selection performances are comparable with and without a weight matrix.

\begin{table}
\footnotesize
    \centering
    \caption{\small{Estimation results with SCAD from the equation \eqref{eq:multiplicative} when the error process is the exponentiated AR(1) with $\rho=0.5$. Mean squared error values are presented with standard deviation in the parenthesis. }}
    \begin{tabular}{clrrr}
    \Xhline{3\arrayrulewidth}
    \multicolumn{5}{c}{AR(1) with $\rho=0.5$} \\ \hline
     $(\mu,\sigma)$ & Methods & \multicolumn{1}{c}{50} & \multicolumn{1}{c}{100} & \multicolumn{1}{c}{200} \\ \hline
    
    \multirow{8}{*}{$(0.1,0.5)$} & PMWLS & 0.88 (0.42) & 0.32 (0.26) & 0.15 (0.17) \\ \cline{2-2}
    & $\mbox{PMWLS } {\tiny (\rho=0.5)}$ & 0.79 (0.39) & 0.20 (0.20) & 0.08 (0.12) \\ \cline{2-2}
    & $\mbox{PMWLS } {\tiny (\rho=0.9)}$ & 0.85 (0.41) & 0.20 (0.20) & 0.08 (0.13) \\ \cline{2-2}
    & $\mbox{PMWLS } {\tiny (\rho=0.8,\phi=0.4)}$ & 0.87 (0.41) & 0.24 (0.22) & 0.08 (0.13) \\ \cline{2-2}
    & PWLS & 0.91 (0.42) & 0.32 (0.25) & 0.13 (0.15) \\ \cline{2-2}
    & $\mbox{PWLS } {\tiny (\rho=0.5)}$ & 0.73 (0.38) & 0.18 (0.19) & 0.07 (0.12) \\ \cline{2-2}
    & $\mbox{PWLS } {\tiny (\rho=0.9)}$ & 0.77 (0.39) & 0.21 (0.20) & 0.07 (0.12) \\ \cline{2-2}
    & $\mbox{PWLS } {\tiny (\rho=0.8,\phi=0.4)}$ & 0.83 (0.40) & 0.23 (0.21) & 0.08 (0.13) \\\hline
    \multirow{8}{*}{$(0.5,0.5)$} & PMWLS & 0.68 (0.37) & 0.30 (0.25) & 0.14 (0.16) \\ \cline{2-2}
    & $\mbox{PMWLS } {\tiny (\rho=0.5)}$ & 0.56 (0.33) & 0.19 (0.19) & 0.08 (0.12) \\ \cline{2-2}
    & $\mbox{PMWLS } {\tiny (\rho=0.9)}$ & 0.64 (0.36) & 0.18 (0.19) & 0.09 (0.14) \\ \cline{2-2}
    & $\mbox{PMWLS } {\tiny (\rho=0.8,\phi=0.4)}$ & 0.75 (0.38) & 0.22 (0.21) & 0.11 (0.15) \\ \cline{2-2}
    & PWLS & 0.78 (0.39) & 0.30 (0.24) & 0.14 (0.17) \\ \cline{2-2}
    & $\mbox{PWLS } {\tiny (\rho=0.5)}$ & 0.71 (0.36) & 0.25 (0.21) & 0.10 (0.13) \\ \cline{2-2}
    & $\mbox{PWLS } {\tiny (\rho=0.9)}$ & 0.74 (0.37) & 0.25 (0.22) & 0.08 (0.13) \\ \cline{2-2}
    & $\mbox{PWLS } {\tiny (\rho=0.8,\phi=0.4)}$ & 0.93 (0.41) & 0.26 (0.23) & 0.12 (0.15) \\
    \Xhline{3\arrayrulewidth}
    \end{tabular}
    \label{table:multi_scad_5_estimation}
\vskip0.4cm
\footnotesize
    \centering
    \caption{\small{Estimation results with SCAD from the equation \eqref{eq:multiplicative} when the error process is exponentiated AR(1) with $\rho=0.9$. The other configurations are identical to Table \ref{table:multi_scad_5_estimation}. }}
    \begin{tabular}{clrrr}
    \Xhline{3\arrayrulewidth}
    \multicolumn{5}{c}{AR(1) with $\rho=0.9$} \\ \hline
     $(\mu,\sigma)$ & Methods & \multicolumn{1}{c}{50} & \multicolumn{1}{c}{100} & \multicolumn{1}{c}{200} \\ \hline
    
    \multirow{8}{*}{$(0.1,0.5)$} & PMWLS & 0.61 (0.35) & 0.24 (0.22) & 0.13 (0.16) \\ \cline{2-2}
    & $\mbox{PMWLS } {\tiny (\rho=0.5)}$ & 0.25 (0.22) & 0.03 (0.08) & 0.02 (0.06) \\ \cline{2-2}
    & $\mbox{PMWLS } {\tiny (\rho=0.9)}$ & 0.32 (0.25) & 0.09 (0.14) & 0.01 (0.05) \\ \cline{2-2}
    & $\mbox{PMWLS } {\tiny (\rho=0.8,\phi=0.4)}$ & 0.40 (0.28) & 0.03 (0.08) & 0.01 (0.05) \\ \cline{2-2}
    & PWLS & 0.76 (0.39) & 0.22 (0.21) & 0.12 (0.16) \\ \cline{2-2}
    & $\mbox{PWLS } {\tiny (\rho=0.5)}$ & 0.27 (0.23) & 0.11 (0.15) & 0.02 (0.06) \\ \cline{2-2}
    & $\mbox{PWLS } {\tiny (\rho=0.9)}$ & 0.39 (0.27) & 0.09 (0.14) & 0.01 (0.05) \\ \cline{2-2}
    & $\mbox{PWLS } {\tiny (\rho=0.8,\phi=0.4)}$ & 0.40 (0.28) & 0.10 (0.14) & 0.01 (0.05) \\\hline
    \multirow{8}{*}{$(0.5,0.5)$} & PMWLS & 0.55 (0.33) & 0.28 (0.24) & 0.13 (0.16) \\ \cline{2-2}
    & $\mbox{PMWLS } {\tiny (\rho=0.5)}$ & 0.16 (0.18) & 0.08 (0.13) & 0.02 (0.06) \\ \cline{2-2}
    & $\mbox{PMWLS } {\tiny (\rho=0.9)}$ & 0.17 (0.18) & 0.07 (0.12) & 0.01 (0.05) \\ \cline{2-2}
    & $\mbox{PMWLS } {\tiny (\rho=0.8,\phi=0.4)}$ & 0.19 (0.19) & 0.05 (0.10) & 0.02 (0.06) \\ \cline{2-2}
    & PWLS & 0.47 (0.31) & 0.27 (0.23) & 0.14 (0.17) \\ \cline{2-2}
    & $\mbox{PWLS } {\tiny (\rho=0.5)}$ & 0.22 (0.20) & 0.10 (0.13) & 0.03 (0.06)  \\ \cline{2-2}
    & $\mbox{PWLS } {\tiny (\rho=0.9)}$ & 0.23 (0.21) & 0.11 (0.14) & 0.01 (0.05) \\ \cline{2-2}
    & $\mbox{PWLS } {\tiny (\rho=0.8,\phi=0.4)}$ & 0.22 (0.21) & 0.07 (0.12) & 0.02 (0.06) \\
    \Xhline{3\arrayrulewidth}
    \end{tabular}
    \label{table:multi_scad_9_estimation}
\end{table}
	
\begin{table}
\footnotesize
    \centering
    \caption{\small{Estimation results with SCAD from the equation \eqref{eq:multiplicative} when the error process is exponentiated ARMA(1,1) with $(\rho, \phi)=(0.8,0.4)$. The other configurations are identical to Table \ref{table:multi_scad_5_estimation}. }}
    \begin{tabular}{clrrr}
    \Xhline{3\arrayrulewidth}
    \multicolumn{5}{c}{ARMA(1,1) with $\rho=0.8, \phi=0.4$} \\ \hline
     $(\mu,\sigma)$ & Methods & \multicolumn{1}{c}{50} & \multicolumn{1}{c}{100} & \multicolumn{1}{c}{200} \\ \hline
    \multirow{8}{*}{$(0.1,0.5)$} & PMWLS & 0.37 (0.27) & 0.14 (0.17) & 0.09 (0.14) \\ \cline{2-2}
    & $\mbox{PMWLS } {\tiny (\rho=0.5)}$ & 0.18 (0.19) & 0.04 (0.08) & 0.02 (0.06) \\ \cline{2-2}
    & $\mbox{PMWLS } {\tiny (\rho=0.9)}$ & 0.17 (0.19) & 0.03 (0.08) & 0.01 (0.05) \\ \cline{2-2}
    & $\mbox{PMWLS } {\tiny (\rho=0.8,\phi=0.4)}$ & 0.12 (0.15) & 0.04 (0.09) & 0.01 (0.05) \\ \cline{2-2}
    & PWLS & 0.29 (0.24) & 0.16 (0.17) & 0.09 (0.13) \\ \cline{2-2}
    & $\mbox{PWLS } {\tiny (\rho=0.5)}$ & 0.18 (0.19) & 0.04 (0.09) & 0.02 (0.06) \\ \cline{2-2}
    & $\mbox{PWLS } {\tiny (\rho=0.9)}$ & 0.17 (0.19) & 0.03 (0.08) & 0.01 (0.05) \\ \cline{2-2}
    & $\mbox{PWLS } {\tiny (\rho=0.8,\phi=0.4)}$ & 0.12 (0.16) & 0.04 (0.09) & 0.01 (0.05) \\\hline
    \multirow{8}{*}{$(0.5,0.5)$} & PMWLS & 0.44 (0.30) & 0.15 (0.18) & 0.07 (0.12) \\ \cline{2-2}
    & $\mbox{PMWLS } {\tiny (\rho=0.5)}$ & 0.19 (0.19) & 0.04 (0.09) & 0.02 (0.06) \\ \cline{2-2}
    & $\mbox{PMWLS } {\tiny (\rho=0.9)}$ & 0.18 (0.19) & 0.03 (0.08) & 0.01 (0.05) \\ \cline{2-2}
    & $\mbox{PMWLS } {\tiny (\rho=0.8,\phi=0.4)}$ & 0.21 (0.20) & 0.04 (0.08) & 0.02 (0.06) \\ \cline{2-2}
    & PWLS & 0.42 (0.29) & 0.15 (0.17) & 0.06 (0.11) \\ \cline{2-2}
    & $\mbox{PWLS } {\tiny (\rho=0.5)}$ & 0.25 (0.21) & 0.06 (0.10) & 0.04 (0.08) \\ \cline{2-2}
    & $\mbox{PWLS } {\tiny (\rho=0.9)}$ & 0.23 (0.21) & 0.05 (0.10) & 0.02 (0.06) \\ \cline{2-2}
    & $\mbox{PWLS } {\tiny (\rho=0.8,\phi=0.4)}$ & 0.32 (0.25) & 0.03 (0.08) & 0.02 (0.06) \\
    \Xhline{3\arrayrulewidth}
    \end{tabular}
    \label{table:multi_scad_8_4_estimation}
\vskip0.4cm
\footnotesize
    \centering
    \caption{\small{Selection results with SCAD from the equation \eqref{eq:multiplicative} when the error process is the exponentiated AR(1) with $\rho=0.5$. TP counts the number of significant estimates among the significant true parameters and TN counts the number of insignificant estimates among the insignificant true parameters. }}
    \begin{tabular}{clrrrrrrr}
    \Xhline{3\arrayrulewidth}
    \multicolumn{9}{c}{AR(1) with $\rho=0.5$} \\ \hline
    \multirow{2}{*}{$(\mu,\sigma)$} & \multirow{2}{*}{Methods} & \multicolumn{3}{c}{TP} && \multicolumn{3}{c}{TN} \\ \cline{3-5} \cline{7-9} 
    & & \multicolumn{1}{c}{50} & \multicolumn{1}{c}{100} & \multicolumn{1}{c}{200} && \multicolumn{1}{c}{50} & \multicolumn{1}{c}{100} & \multicolumn{1}{c}{200}\\ \hline
    \multirow{8}{*}{$(0.1,0.5)$} & PMWLS & 2.88 & 2.99 & 3.00 & & 16.83 & 16.92 & 16.94 \\ \cline{2-2}
    & $\mbox{PMWLS } {\tiny (\rho=0.5)}$ & 2.85 & 2.99 & 3.00 & & 16.90 & 16.93 & 16.99 \\ \cline{2-2}
    & $\mbox{PMWLS } {\tiny (\rho=0.9)}$ & 2.83 & 2.98 & 3.00 & & 16.89 & 16.99 & 16.99 \\ \cline{2-2}
    & $\mbox{PMWLS } {\tiny (\rho=0.8,\phi=0.4)}$ & 2.84 & 2.98 & 3.00 & & 16.95 & 16.98 & 16.99 \\ \cline{2-2}
    & PWLS & 2.87 & 2.99 & 3.00 & & 16.84 & 16.94 & 17.00 \\ \cline{2-2}
    & $\mbox{PWLS } {\tiny (\rho=0.5)}$ & 2.87 & 2.99 & 3.00 & & 16.89 & 16.98 & 17.00 \\ \cline{2-2}
    & $\mbox{PWLS } {\tiny (\rho=0.9)}$ & 2.84 & 2.98 & 3.00 & & 16.98 & 17.00 & 17.00 \\ \cline{2-2}
    & $\mbox{PWLS } {\tiny (\rho=0.8,\phi=0.4)}$ & 2.84 & 2.98 & 3.00 & & 17.00 & 17.00 & 17.00 \\\hline
    \multirow{8}{*}{$(0.5,0.5)$} & PMWLS & 2.94 & 2.99 & 3.00 & & 16.84 & 16.94 & 17.00 \\ \cline{2-2}
    & $\mbox{PMWLS } {\tiny (\rho=0.5)}$ & 2.92 & 2.99 & 3.00 & & 16.85 & 16.97 & 17.00 \\ \cline{2-2}
    & $\mbox{PMWLS } {\tiny (\rho=0.9)}$ & 2.91 & 2.99 & 2.99 & & 16.93 & 16.97 & 17.00 \\ \cline{2-2}
    & $\mbox{PMWLS } {\tiny (\rho=0.8,\phi=0.4)}$ & 2.89 & 2.99 & 2.99 & & 16.85 & 16.97 & 16.99 \\ \cline{2-2}
    & PWLS & 2.92 & 2.99 & 3.00 & & 16.82 & 16.94 & 16.96 \\ \cline{2-2}
    & $\mbox{PWLS } {\tiny (\rho=0.5)}$ & 2.82 & 2.96 & 2.99 & & 16.91 & 16.98 & 16.99 \\ \cline{2-2}
    & $\mbox{PWLS } {\tiny (\rho=0.9)}$ & 2.79 & 2.95 & 3.00 & & 16.95 & 17.00 & 16.99 \\ \cline{2-2}
    & $\mbox{PWLS } {\tiny (\rho=0.8,\phi=0.4)}$ & 2.76 & 2.96 & 2.99 & & 17.00 & 16.99 & 16.99 \\
    \Xhline{3\arrayrulewidth}
    \end{tabular}
    \label{table:multi_scad_5_selection}
\end{table}

\begin{table}
\footnotesize
    \centering
    \caption{\small{Selection results with SCAD from the equation \eqref{eq:multiplicative} when the error process is exponentiated AR(1) with $\rho=0.9$. The other configurations are identical to Table \ref{table:multi_scad_5_selection}.}}
    \begin{tabular}{clrrrrrrr}
    \Xhline{3\arrayrulewidth}
    \multicolumn{9}{c}{AR(1) with $\rho=0.9$} \\ \hline
    \multirow{2}{*}{$(\mu,\sigma)$} & \multirow{2}{*}{Methods} & \multicolumn{3}{c}{TP} && \multicolumn{3}{c}{TN} \\ \cline{3-5} \cline{7-9} 
    & & \multicolumn{1}{c}{50} & \multicolumn{1}{c}{100} & \multicolumn{1}{c}{200} && \multicolumn{1}{c}{50} & \multicolumn{1}{c}{100} & \multicolumn{1}{c}{200}\\ \hline
    \multirow{8}{*}{$(0.1,0.5)$} & PMWLS & 2.94 & 3.00 & 3.00 & & 16.85 & 16.92 & 16.96 \\ \cline{2-2}
    & $\mbox{PMWLS } {\tiny (\rho=0.5)}$ & 2.93 & 3.00 & 3.00 & & 16.94 & 16.99 & 17.00 \\ \cline{2-2}
    & $\mbox{PMWLS } {\tiny (\rho=0.9)}$ & 2.90 & 2.98 & 3.00 & & 16.99 & 17.00 & 17.00 \\ \cline{2-2}
    & $\mbox{PMWLS } {\tiny (\rho=0.8,\phi=0.4)}$ & 2.88 & 3.00 & 3.00 & & 17.00 & 17.00 & 17.00 \\ \cline{2-2}
    & PWLS & 2.91 & 3.00 & 3.00 & & 16.75 & 16.98 & 16.97 \\ \cline{2-2}
    & $\mbox{PWLS } {\tiny (\rho=0.5)}$ & 2.93 & 2.98 & 3.00 & & 16.98 & 17.00 & 17.00 \\ \cline{2-2}
    & $\mbox{PWLS } {\tiny (\rho=0.9)}$ & 2.88 & 2.98 & 3.00 & & 17.00 & 17.00 & 17.00 \\ \cline{2-2}
    & $\mbox{PWLS } {\tiny (\rho=0.8,\phi=0.4)}$ & 2.88 & 2.98 & 3.00 & & 17.00 & 17.00 & 17.00 \\\hline
    \multirow{8}{*}{$(0.5,0.5)$} & PMWLS & 2.97 & 2.99 & 3.00 & & 16.74 & 16.90 & 16.97 \\ \cline{2-2}
    & $\mbox{PMWLS } {\tiny (\rho=0.5)}$ & 2.96 & 2.98 & 3.00 & & 16.99 & 16.98 & 17.00 \\ \cline{2-2}
    & $\mbox{PMWLS } {\tiny (\rho=0.9)}$ & 2.95 & 2.98 & 3.00 & & 16.97 & 17.00 & 17.00 \\ \cline{2-2}
    & $\mbox{PMWLS } {\tiny (\rho=0.8,\phi=0.4)}$ & 2.95 & 2.99 & 3.00 & & 16.96 & 17.00 & 17.00 \\ \cline{2-2}
    & PWLS & 2.97 & 2.99 & 3.00 & & 16.88 & 16.97 & 16.96 \\ \cline{2-2}
    & $\mbox{PWLS } {\tiny (\rho=0.5)}$ & 2.94 & 2.98 & 3.00 & & 16.97 & 17.00 & 17.00 \\ \cline{2-2}
    & $\mbox{PWLS } {\tiny (\rho=0.9)}$ & 2.92 & 2.96 & 3.00 & & 17.00 & 17.00 & 17.00 \\ \cline{2-2}
    & $\mbox{PWLS } {\tiny (\rho=0.8,\phi=0.4)}$ & 2.93 & 2.98 & 3.00 & & 17.00 & 17.00 & 17.00 \\
    \Xhline{3\arrayrulewidth}
    \end{tabular}
    \label{table:multi_scad_9_selection}
\vskip0.4cm
\footnotesize
    \centering
    \caption{\small{Selection results with SCAD from the equation \eqref{eq:multiplicative} when the error process is exponentiated ARMA(1) with $(\rho,\phi)=(0.8,0.4)$. The other configurations are identical to Table \ref{table:multi_scad_5_selection}.}}
    \begin{tabular}{clrrrrrrr}
    \Xhline{3\arrayrulewidth}
    \multicolumn{9}{c}{ARMA(1,1) with $\rho=0.8, \phi=0.4$} \\ \hline
    \multirow{2}{*}{$(\mu,\sigma)$} & \multirow{2}{*}{Methods} & \multicolumn{3}{c}{TP} && \multicolumn{3}{c}{TN} \\ \cline{3-5} \cline{7-9} 
    & & \multicolumn{1}{c}{50} & \multicolumn{1}{c}{100} & \multicolumn{1}{c}{200} && \multicolumn{1}{c}{50} & \multicolumn{1}{c}{100} & \multicolumn{1}{c}{200}\\ \hline
    \multirow{8}{*}{$(0.1,0.5)$} & PMWLS & 2.99 & 3.00 & 3.00 & & 16.81 & 16.96 & 16.92 \\ \cline{2-2}
    & $\mbox{PMWLS } {\tiny (\rho=0.5)}$ & 2.96 & 3.00 & 3.00 & & 16.94 & 16.99 & 16.98 \\ \cline{2-2}
    & $\mbox{PMWLS } {\tiny (\rho=0.9)}$ & 2.96 & 3.00 & 3.00 & & 17.00 & 16.99 & 16.99 \\ \cline{2-2}
    & $\mbox{PMWLS } {\tiny (\rho=0.8,\phi=0.4)}$ & 2.98 & 3.00 & 3.00 & & 17.00 & 17.00 & 16.99 \\ \cline{2-2}
    & PWLS & 2.99 & 3.00 & 3.00 & & 16.93 & 16.98 & 16.92 \\ \cline{2-2}
    & $\mbox{PWLS } {\tiny (\rho=0.5)}$ & 2.96 & 3.00 & 3.00 & & 16.98 & 16.99 & 16.99 \\ \cline{2-2}
    & $\mbox{PWLS } {\tiny (\rho=0.9)}$ & 2.96 & 3.00 & 3.00 & & 17.00 & 17.00 & 17.00 \\ \cline{2-2}
    & $\mbox{PWLS } {\tiny (\rho=0.8,\phi=0.4)}$ & 2.98 & 3.00 & 3.00 & & 17.00 & 17.00 & 17.00 \\\hline
    \multirow{8}{*}{$(0.5,0.5)$} & PMWLS & 2.98 & 3.00 & 3.00 & & 16.70 & 16.94 & 16.95 \\ \cline{2-2}
    & $\mbox{PMWLS } {\tiny (\rho=0.5)}$ & 2.95 & 3.00 & 3.00 & & 16.98 & 16.98 & 17.00 \\ \cline{2-2}
    & $\mbox{PMWLS } {\tiny (\rho=0.9)}$ & 2.95 & 3.00 & 3.00 & & 16.99 & 16.99 & 17.00 \\ \cline{2-2}
    & $\mbox{PMWLS } {\tiny (\rho=0.8,\phi=0.4)}$ & 2.94 & 3.00 & 3.00 & & 17.00 & 16.99 & 16.99 \\ \cline{2-2}
    & PWLS & 2.97 & 3.00 & 3.00 & & 16.76 & 16.95 & 16.99 \\ \cline{2-2}
    & $\mbox{PWLS } {\tiny (\rho=0.5)}$ & 2.93 & 3.00 & 3.00 & & 16.91 & 16.98 & 16.99 \\ \cline{2-2}
    & $\mbox{PWLS } {\tiny (\rho=0.9)}$ & 2.92 & 2.99 & 3.00 & & 16.97 & 17.00 & 17.00 \\ \cline{2-2}
    & $\mbox{PWLS } {\tiny (\rho=0.8,\phi=0.4)}$ & 2.90 & 3.00 & 3.00 & & 16.97 & 17.00 & 17.00 \\
    \Xhline{3\arrayrulewidth}
    \end{tabular}
    \label{table:multi_scad_8_4_selection}
\end{table}

For the comparison between  PMWLS and the additive method,  where  the estimator of the additive method is calculated as if the data are from a nonlinear additive model without log transformation, the MSE values using the additive method are large. The detailed results are provided in Tables S7 and S8 in the supplementary material. This has been pointed out by \cite{bhattacharyya1992inconsistent} that a mis-specified additive nonlinear model while the true model is a multiplicative nonlinear model may lead to an inconsistent estimator. On the other hand, both PMWLS and the additive method successfully discriminate significant and insignificant parameters but PMWLS outperformed the additive method in terms of TP.  

\section{Application to a head-neck position tracking system}\label{section:application}

In this section, we apply our PMWLS method to the parametric nonlinear model of the head-neck position tracking task which is introduced in detail in \cite{ramadan2018selecting}. The nonlinear function considered in this application is extremely complicated to express in a closed form equation. Therefore, for the readers interested in gaining a deeper understanding, we recommend referring to a diagram provided at Figure 1 in \cite{ramadan2018selecting} or Figure 2 in \cite{yoon2022regularized}. The penalization is done with the SCAD penalty. The weight matrix is chosen by fitting the residuals from the PMWLS method without the weight matrix to the ARMA(1,1) process. Note that, from the simulation study, we found the results are better when we include a weight matrix even if it does not reflect the true dependence structure. Hence, we use the weight matrix constructed by assuming some dependence structure for the real data, although it may not describe the true dependent structure.

The weight matrix for the PWLS method is also constructed in a similar way.
The choice of the weight matrix in real data analysis can be flexible, but fitting residuals from the unweighted method to ARMA(1,1) worked well even for the data with slowly decreasing autocorrelation. 

There are ten subjects in total, and each subject participated in three trials of an experiment. In each trial, the subject's head-neck movement as an angle (radian) for 30 seconds was collected with measurement frequency of 60Hz, i.e. 1800 observations per each trial. Reference signals as an input guided the subjects to follow with their eyes. Since the neurophysiological parameters are different by the subjects, the model is separately fitted to each subject.

\begin{table}[t]
    \centering
    \footnotesize
    \caption{\small{The neurophysiological parameters of the head-neck position tracking model. The notation and description are adopted from \cite{ramadan2018selecting}. Max and Min are the range of the parameter values and the units of the parameters are the values in brackets.}}
    \begin{tabular}{c c c l }
    \hline
    {\textbf{Parameters}}& {\textbf{Max}}&
    {\textbf{Min}}&
    {\textbf{Description}}\\ 
    \hline
    $K_{vis}\left[\frac{Nm}{rad}\right]$&$10^{3}$&50&Visual feedback gain\\
    $K_{vcr}\left[\frac{Nms^2}{rad}\right]$&$10^{4}$&500&Vestibular feedback gain\\
    $K_{ccr}\left[\frac{Nm}{rad}\right]$&300&1&Proprioceptive feedback gain\\
    $\tau[s]$&0.4&0.1&Visual feedback delay\\
    $\tau_{1A}[s]$&0.2&0.01&Lead time constant of the irregular vestibular afferent neurons\\
    $\tau_{CNS1}[s]$&1&0.05&Lead time constant of the central nervous system\\
    $\tau_{C}[s]$&5&0.1&Lag time constant of the irregular vestibular afferent neurons\\
    $\tau_{CNS2}[s]$&60&5&Lag time constant of the central nervous system\\
    $\tau_{MS1}[s]$&1&0.01&First lead time constant of the neck muscle spindle\\
    $\tau_{MS2}[s]$&1&0.01&Second lead time constant of the neck muscle spindle\\
    $B\left[\frac{Nms}{rad}\right]$&5&0.1&Intrinsic damping\\
    $K\left[\frac{Nm}{rad}\right]$&5&0.1&Intrinsic stiffness\\\hline
    \end{tabular}
    \label{table:parameter}
\end{table}

The parametric model of the head-neck position tracking system involves a highly nonlinear structure with 12 parameters to be estimated. Table \ref{table:parameter} shows a list of the parameters for the head-neck position tracking system. Each parameter measures a neurophysiological function such as visual feedback gain, vestibular feedback gain, and so on. The parametric model of the head-neck position tracking task suffers from its complicated structure and limited data availability, which could bring overfitting and non-identifiability. 

 A penalized method has been considered to stabilize parameter estimation and build a sparse model for identifiability in head-neck position tracking system  \citep{ramadan2018selecting, yoon2022regularized}. The penalized method shrinks the parameter values to the pre-specified values, called the typical values, instead of the zero value since the parameters in the model have their physical meanings. For the determination of the typical values, we pre-estimated the parameters 10 times from 10 random initial values and set the average as the typical values, $\bm{\tilde\theta}$. When overfitting is highly concerned, the resulting estimates tend to possess non-ignorable gaps from the true optimum. This means even slight variations in the initial conditions can result in significant deviations in the resulting estimates. To reduce this instability, we averaged multiple overfitted estimates, which is expected to yield a new estimate closer to the true optimum than each estimate. Therefore, we chose the average of pre-estimated values as our typical values. Given that the final estimates from the penalization methods are ultimately shrunk towards these typical values, it is likely that certain elements in the final estimates will match with the corresponding elements in the typical values. As a result, the typical values play a crucial role in shaping the overall quality of both estimation and prediction performance for the penalization methods including our approach as well as the other approach.

As described in the introduction, the fitted values from the additive error model with the penalized ordinary least square method used in \cite{yoon2022regularized} indicate a possibility of the multiplicative errors and their residuals still exhibit temporal dependence. Thus, we apply our approach in estimating the neurophysiological parameters, $\bm\theta$, of the head-neck position tracking model. Then we compare our PMWLS method  to an additive approach without log transformation \citep{yoon2022regularized} and PWLS method. Note that both PMWLS and PWLS methods are applied after log transformation by assuming multiplicative errors. For the comparison, we consider variance accounted for (VAF), which is defined as 
${\text{VAF}}(\hat{\bm\theta})(\%) = \left [1 - \frac{\sum^{n}_{t=1}(y_t-\hat{y}_t)^{2}}{\sum^{n}_{t=1}y_t^{2}} \right ] \times 100, $
where $\hat{y}_t$ is the $t$-th component of $\bm f(\bm x, \hat{\bm\theta})$. VAF has been frequently used to assess the fit from the obtained estimates in biomechanics \citep{van2013vaf, ramadan2018selecting, yoon2022regularized}. As observed in the expression of VAF, the estimates with higher VAF values are translated into the estimates with lower MSE values. 

\begin{table}[!ht]
\small
\centering
\caption{\small{VAFs for 10 subjects with `No.' refering to the subject number.  `Additive' refers to the additive method studied in \cite{yoon2022regularized}, The Train column is for the train set and the Test column is for the test set. }}
\begin{tabular}{cccc}
    \hline
    \multirow{2}{*}{No.} & \multicolumn{3}{c}{Train} \\ \cline{2-4}
    & Additive & PWLS & PMWLS \\\hline
    1 & 8453 & 8472 & 8471 \\
    2 & 6896 & 6927 & 7139 \\
    3 & 8229 & 8229 & 8229 \\
    4 & 8476 & 8464 & 8459 \\
    5 & 8424 & 8516 & 8647 \\
    6 & 8824 & 8795 & 8846 \\
    7 & 9308 & 9308 & 9308 \\
    8 & 7900 & 8624 & 8771 \\
    9 & 8857 & 8858 & 8842 \\
    10 & 7672 & 7672 & 7668 \\\hline
    Average & 8304 & 8386 & 8438\\\hline 
    & \\\hline
    \multirow{2}{*}{No.} & \multicolumn{3}{c}{Test} \\ \cline{2-4}
    & Additive & PWLS & PMWLS \\\hline
    1 & 8815  & 8810  & 8823 \\
    2 & 5785 & 5827  & 6079 \\
    3 & 7680 & 7681  & 7680 \\
    4 & 8064 & 8058  & 8066 \\
    5 & 8350 & 8417  & 8502 \\
    6 & 8821 & 8816  & 8819 \\
    7 & 9114 & 9114  & 9114 \\
    8 & 8187 & 8975 & 9089 \\
    9 & 8276 & 8248  & 8327 \\
    10 & 7842 & 7842  & 7837 \\\hline
    Average & 8093 & 8179  & 8234 \\\hline
\multicolumn{4}{l}{\footnotesize{$\ast$ The actual VAF values are $10^{-4} \times$ the reported values.}}
\end{tabular}
\label{tab:VAF}
\end{table}

Recall that there are three sets of measurements (three trials) per subject. We used the measurements from one trial (train set) to fit the model and the measurements from two other trials (test set) were used to test the fitted model. Thus, we have one VAF value using the train set and two VAF values using the test set. We repeat this for each trial as a train set and report the average VAF values. 

Table \ref{tab:VAF} shows VAF values of all ten subjects for the additive approach, PWLS method and PMWLS method. The VAF values among different methods for some subjects (No. 1, 3, 4, 7, 9 and 10) are similar and the differences are small. On the other hand, the VAF values from our PMWLS method are higher than the VAF values from the other methods for the subjects No. 2, 5, 6 and 8.  The difference among the three approaches is more clear when VAF values are averaged over all subjects. The averages of VAF values over all subjects when all the parameter values were set to the typical values, i.e. no penalized estimation method, are 0.8149 for the train set and 0.7928 for the test set. Hence, the improvement by our approach is larger compared to the improvements by the other methods as well. We believe these results are originated from the fact that multiplicative error assumption is valid and our PMWLS method has successfully captured the error structure underneath the data.

\begin{figure}[!ht]
\centering
\includegraphics[width=.48\textwidth]{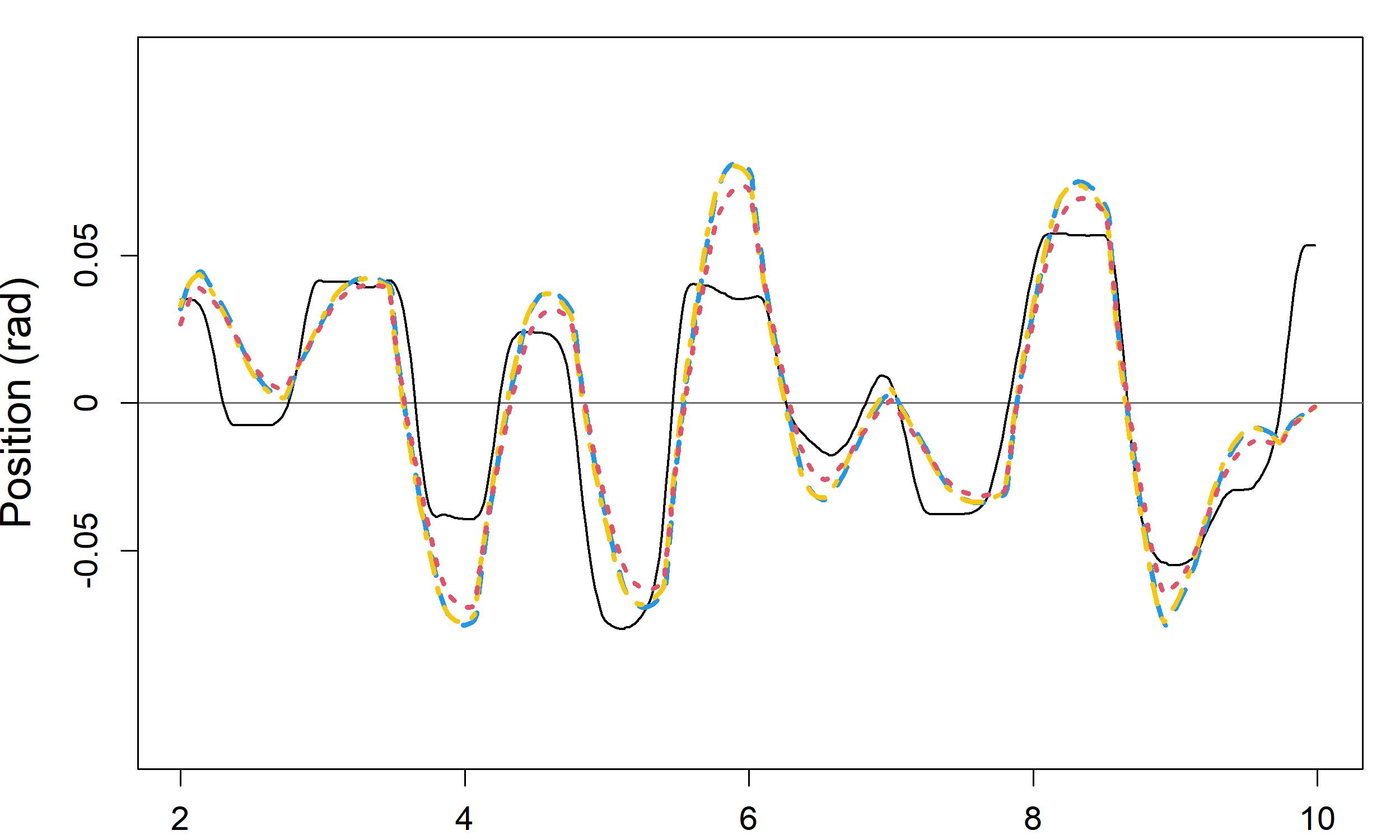}
\includegraphics[width=.48\textwidth]{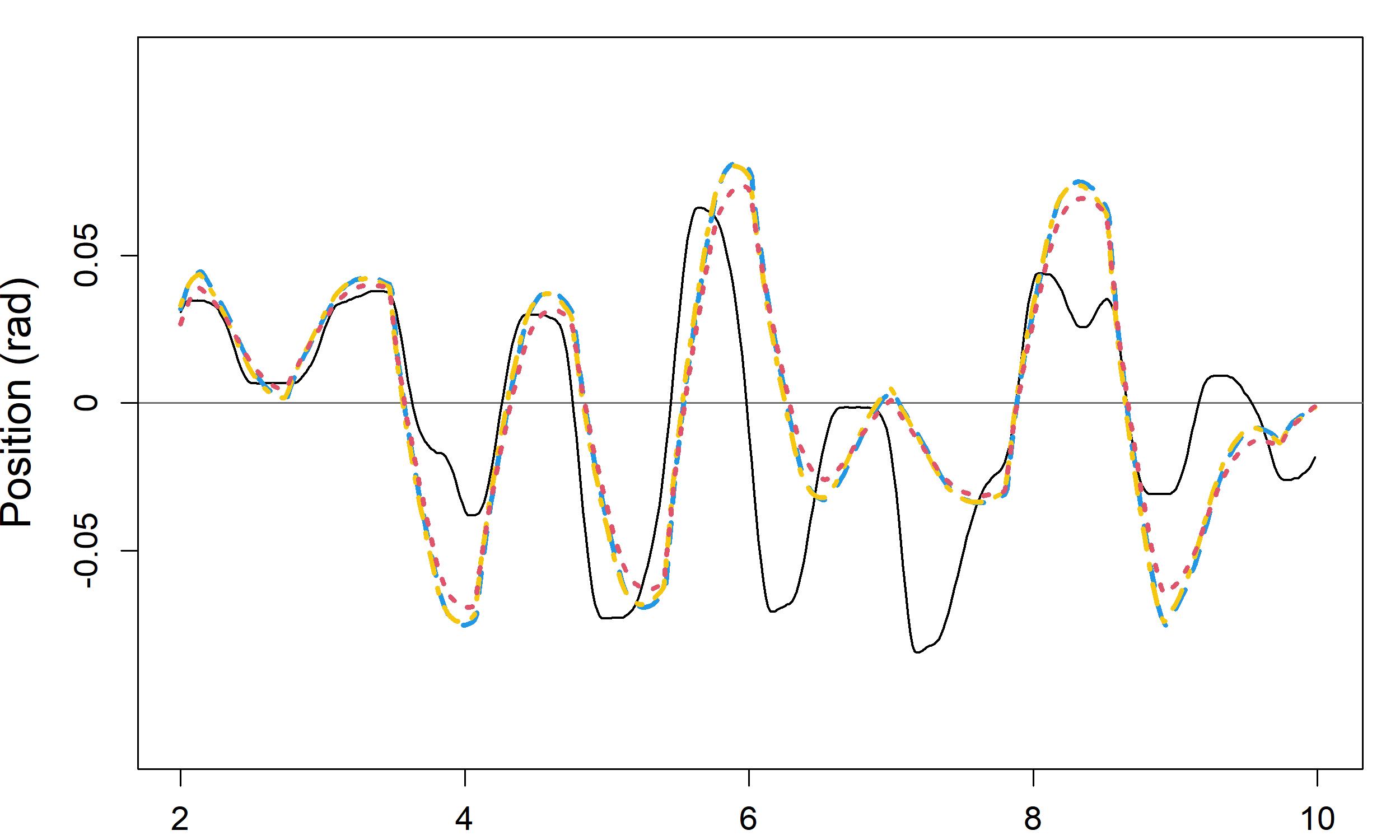}
\caption{\small Estimation (left) and prediction (right) results for the subject No.2. Measured responses (black line) and fitted values from the additive approach in  \cite{yoon2022regularized} (blue dashed line, $--$), PWLS (yellow dot-dashed line, $\cdot -\cdot$), and PMWLS (red dotted line, $\cdots$) are exhibited.}
\label{fig:real_results_2}
\end{figure}

\begin{figure}[!ht]
\centering
\includegraphics[width=.48\textwidth]{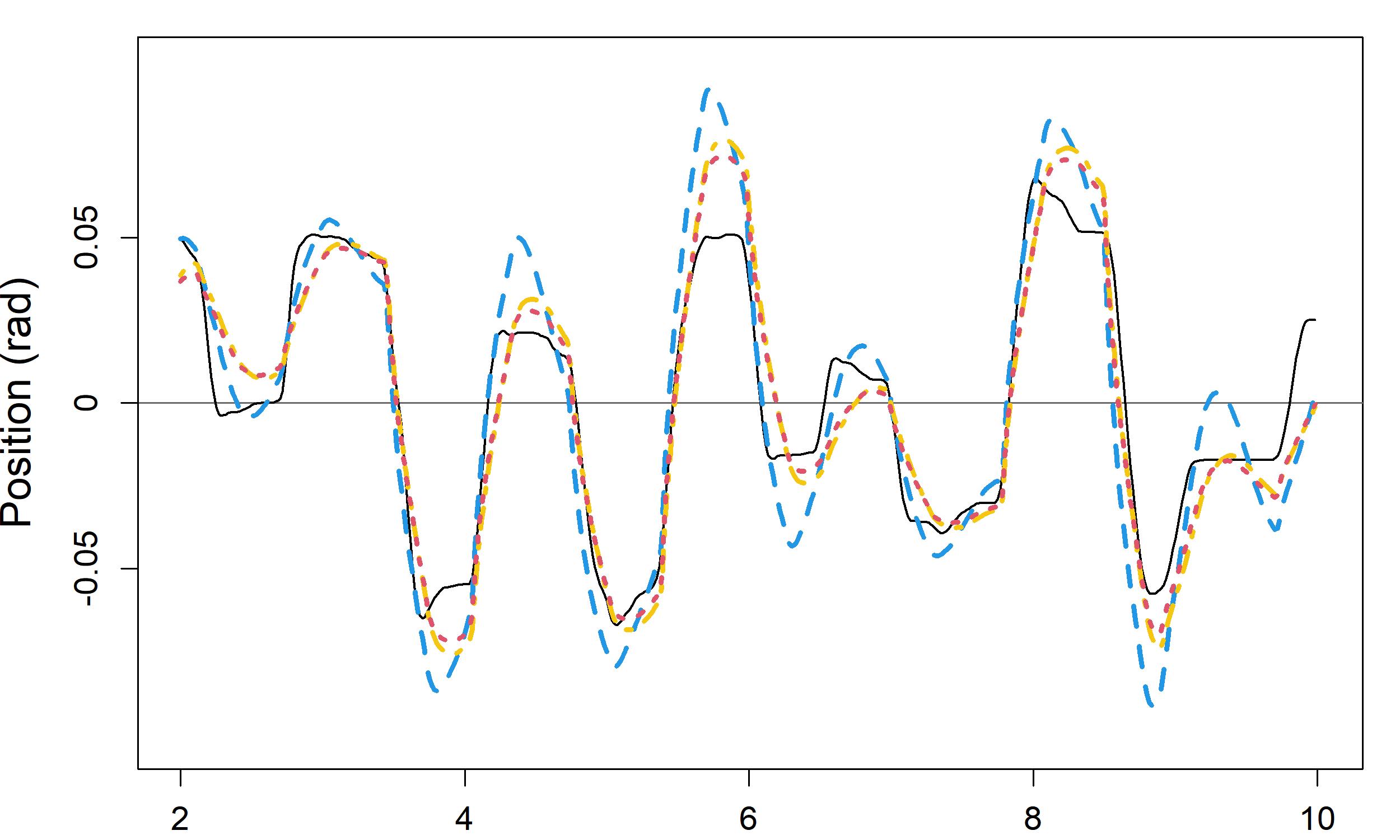}
\includegraphics[width=.48\textwidth]{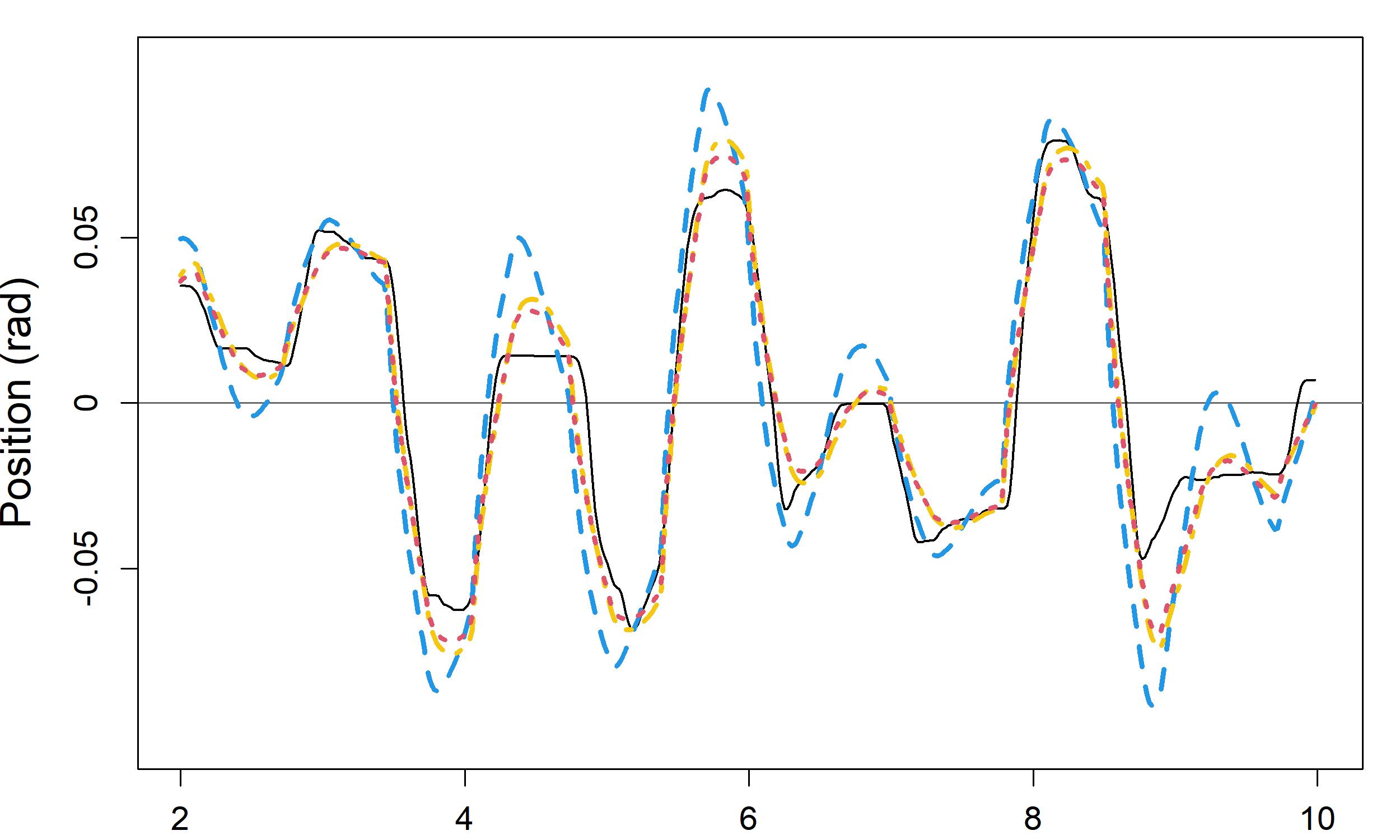}
\caption{\small Same configurations as in Figure \ref{fig:real_results_2}, but with the subject No. 8.}
\label{fig:real_results_8}
\end{figure}

The estimation and prediction results for the subjects No.2 and No.8 are provided in Figures \ref{fig:real_results_2} and \ref{fig:real_results_8}, respectively.
The left plots in both Figures \ref{fig:real_results_2} and \ref{fig:real_results_8} exhibit estimation results from one trial out of three trials per subject as an example. The right plots in both Figures \ref{fig:real_results_2} and \ref{fig:real_results_8} show prediction results for another trial using the fitted model. In both plots, our PMWLS method (the red dotted line) is better at capturing peaks of the measured response than the additive method (the blue dashed line), which motivated this study at the beginning. We believe that the reason our approach outperforms the additive approach is that the data implicitly have multiplicative structure. In addition, the PMWLS method also slightly excels the PWLS method in both estimation and prediction. Therefore, one can benefit from adopting our proposed method for data with multiplicative structure. The number of selected parameters, i.e. sensitive parameters, on average were 2.33 (additive), 3.23 ($\mbox{PWLS}$) and 3.63 ($\mbox{PMWLS}$) per subject. This might imply that a smaller number of selected parameters in the additive approach causes poor performance in estimation and prediction.

\section{Conclusion}
\label{section:discussion}

We proposed an estimation and selection method for the parameters in a nonlinear model when the errors have possible non-zero mean and temporal dependence. Our approach can also handle the multiplicative error model as shown in the simulation study and the real data example. One can consider simply adding an intercept term to a nonlinear model and estimating the parameters using the least squares to handle possible non-zero mean. Simulation results show that both approaches are overall comparable but our approach performs better compared to the approach with an intercept term when non-zero mean is larger and a sample size is rather small. We provided asymptotic properties of the proposed estimator and numerical studies supported our theoretical results. 

Introducing a weight matrix to reflect the temporal dependence improves estimation. One could assume a parametric model for temporal dependence of the error process and construct a weight matrix from the covariance matrix of the error process. However, theoretical results would be limited to the specific parametric model, resulting in limited applicability. In addition, simulation results show there is not much difference in estimation performance from a choice of the weight matrix. Hence, we allow various weight matrices and do not assume the exact temporal dependence model. 

In Table \ref{table:add_scad_5_selection}, PMWLS without any weight matrix slightly outperforms PMWLS with weight matrices in terms of TP and TN even when the weight matrix is obtained from the true error process. Although this difference is observed in a very limited manner and may not be significant, one possible reason for this can be found from weight normalization process, described in section \ref{section:notations}. The normalization process changes the structure of the weight matrices and potentially influences the parameter selection performance. Furthermore, this may bring up the need for further investigation into the tuning parameter selection criteria as it plays a key role in the selection performance yet remains not fully explored in depth. The topic of tuning parameter selection criteria presents a promising research topic for future investigation.

The proposed method assumes a fixed number of parameters. Hence, it is natural to think about an extension to an increasing number of parameters as the sample size grows, which we leave as a  future study. Since we shrink the estimates toward the typical values, obtaining good typical values can play a critical role in the final results. Therefore, a better way to locate the typical values, though the task is very challenging due to the complicated structure of the nonlinear function, would contribute greatly to solving the problem in the head-neck position tracking task. In our result, we used the same typical values as \cite{ramadan2018selecting} and \cite{yoon2022regularized} so that the comparison was carried out fairly. 

In overall, the proposed method was successfully applied to the head-neck position tracking model and produced the state-of-the-art performance in both estimation and prediction. Since the multiplicative structure with temporally correlated errors is frequently observed in other fields such as finance, signal processing and image processing \citep{vilcu2011geometric, wang2018two}, the use of proposed method is not limited to the application we considered in this study. 

\section*{Supplementary Materials}
The supplementary materials contain
proofs of theoretical results discussed in the text and additional simulation studies. 
\par
\section*{Acknowledgements}
Wu was supported by Ministry of Science and Technology
					of Taiwan under grants (MOST 111-2118-M-259 -002). The work of Lim was supported by National Research Foundation of Korea (NRF) grant funded by the Korea government (MSIT) (NRF-2019R1A2C1002213 and 2020R1A4A1018207).
The  work of Yoon and Choi was supported by the National Research Foundation of Korea (NRF) grants funded by the Korea government (MSIT) (No.RS-2023-00221762 and No. 2021R1A2B5B01002620).

\par


\bibhang=1.7pc
\bibsep=2pt
\fontsize{9}{14pt plus.8pt minus .6pt}\selectfont
\renewcommand\bibname{\large \bf References}
\expandafter\ifx\csname
natexlab\endcsname\relax\def\natexlab#1{#1}\fi
\expandafter\ifx\csname url\endcsname\relax
  \def\url#1{\texttt{#1}}\fi
\expandafter\ifx\csname urlprefix\endcsname\relax\def\urlprefix{URL}\fi



\bibliographystyle{chicago}
\bibliography{main}

\end{document}



\renewcommand{\baselinestretch}{1}

\markright{ \hbox{\footnotesize\rm 
}\hfill\\[-13pt]
\hbox{\footnotesize\rm
}\hfill }

\markboth{\hfill{\footnotesize\rm You et al.} \hfill}
{\hfill {\footnotesize\rm Regularized nonlinear regression} \hfill}

\renewcommand{\thefootnote}{}
$\ $\par \fontsize{12}{14pt plus.8pt minus .6pt}\selectfont


 \centerline{\large\bf Supplementary Material for }
\vspace{2pt}
\centerline{\large\bf Regularized Nonlinear Regression with Dependent Errors}
\vspace{2pt} 
\centerline{\large\bf 
		and its Application to a Biomechanical Model }
\vspace{.25cm}
\centerline{Hojun You\affmark[1], Kyubaek Yoon\affmark[3], Wei-Ying Wu\affmark[4], Jongeun Choi\affmark[3] and Chae Young Lim\affmark[2]} 
\vspace{.4cm} 
\centerline{\it {\affmark[1]University of Houston}\\\hspace{0.1cm}
\affaddr{\affmark[2]Seoul National University}\\\hspace{0.1cm}
\affaddr{\affmark[3]Yonsei University}\\\hspace{0.1cm}
\affaddr{\affmark[4]National Dong Hwa University}\\
}
\vspace{1.55cm}
\noindent
The proof of theoretical results and additional simulation studies are discussed in this supplementary material.\\
\par

\setcounter{section}{0}
\setcounter{equation}{0}
\def\theequation{S\arabic{section}.\arabic{equation}}
\def\thesection{S\arabic{section}}

\fontsize{12}{14pt plus.8pt minus .6pt}\selectfont

\section {Proof of Theorems}
For the completeness of the proofs, we state assumptions and theorems again and the proofs are followed.

\begin{assumption}\label{assumption:general}
    \begin{enumerate}[(1)]
        \item The nonlinear function $f\in C^2$ on the compact set $\mathcal{D} \times \Theta$ where $C^2$ is the set of twice continuously differentiable functions. 
        \item As $\| \bm \theta - \bm \theta_0 \| \rightarrow 0 $, $\left({\bm{\dot{F}}}(\bm\theta_0)^{T}\bm\Sigma_w {\bm{\dot{F}}}(\bm\theta_0)\right)^{-1} {\bm{\dot{F}}}(\bm\theta)^{T}\bm\Sigma_w {\bm{\dot{F}}}(\bm\theta)\rightarrow I_p$, elementwisely and uniformly in $\bm\theta$.
        \item There exist symmetric positive definite matrices $\bm\Gamma$ and $\bm\Gamma_\epsilon$ such that
        \begin{align*}
            \frac{1}{n\lambda_w}\bm{\dot{F}}(\bm\theta_0)^{T}\bm\Sigma_w \bm{\dot{F}}(\bm\theta_0) & \rightarrow \bm\Gamma \\
            \frac{1}{n\lambda_\epsilon\lambda_w^2}{\bm{\dot{F}}}(\bm\theta_0)^{T}\bm\Sigma_w\bm\Sigma_\epsilon\bm\Sigma_w {\bm{\dot{F}}}(\bm\theta_0) & \rightarrow \bm\Gamma_\epsilon.
        \end{align*}
        \item $\frac{\|\bm W\|_1\cdot\|\bm W\|_{\infty}}{\|\bm W^T \bm \Sigma_n \bm W\|_2} = o(n^{1/2}\lambda_\epsilon^{1/2})$.
        \item $O(1) \leq \lambda_\epsilon \leq o(n)$ and $\lambda_w \geq O(1)$.
        \item $\{\epsilon_i^2\}$ is uniformly integrable.
        \item One of the following conditions is satisfied for $\epsilon_i$.
        \begin{itemize}
            \item [$(a)$] $\{\epsilon_i\}$ is a $\phi$-mixing.
            \item [$(b)$] $\{\epsilon_i\}$ is a $\rho$-mixing and $\sum_{j \in \mathcal{N}} \rho(2^j)<\infty$.
            \item [$(c)$] For $\delta >0$, $\{\epsilon_i\}$ is a $\alpha$-mixing, $\{|\epsilon_i|^{2+\delta}\}$ is uniformly integrable, and $\sum_{j \in \mathcal{N}}n^{2/\delta}\alpha(n) < \infty$. 
        \end{itemize}
    \end{enumerate}
\end{assumption}

\begin{assumption}\label{assumption:penalty}
The first derivative of a penalty function $p_{\tau_n}(\cdot)$ denoted by $ q_{\tau_n}(\cdot)$, has the following properties:
\begin{enumerate}[(1)]

    \item $c_n=\max_{i\in\{1, \ldots, s\}}\left\{| q_{\tau_n}(|\theta_{0i}|)|\right\} = O\left(\left( \lambda_\epsilon/n\right)^{1/2}\right)$
    \item $ q_{\tau_n}(\cdot)$ is Lipschitz continuous given $\tau_n$
    \item $n^{1/2}\lambda_\epsilon^{-1/2}\lambda_w^{-1}\tau_n \rightarrow \infty$
    \item For any $C>0$, $\displaystyle\liminf_{n\rightarrow \infty}\inf_{\theta \in \left(0, C(\lambda_\epsilon/n)^{1/2}\right)} \tau_n^{-1} q_{\tau_n}(\theta)>0 $
\end{enumerate}
\end{assumption}

\newpage
\begin{lemma}\label{thm:consistency}
For any $\varepsilon>0$ and $a_n = (\lambda_\epsilon/n)^{1/2}$, under Assumption \ref{assumption:general}-(1), (2), (3), and (5) there exists a positive constant $C$ such that
    $$P\left(\inf_{\|\bm v\|=C} S_n(\bm \theta_0+a_n\bm v) - S_n(\bm \theta_0) >0 \right) > 1-\varepsilon$$
    for large enough $n$. Therefore, with probability tending to 1, there exists a local minimizer of $S_n (\bm \theta)$, denoted by $\hat{\bm \theta}^{(s)}$, in the ball centered at $\bm \theta_0$ with the radius $a_n \bm v$. 
    Since $a_n = o(1)$ by Assumption \ref{assumption:general}-(5), we have the consistency of $\hat{\bm\theta}^{(s)}$.
\end{lemma}

\begin{proof}
	\noindent By Taylor's theorem,
    \begin{align*}
    S_n(\bm \theta_0+a_n \bm v) -S_n(\bm\theta_0) &= a_n\bm v^T \nabla S_n(\bm\theta_0)
    +\frac{1}{2}a_n^2\bm v^{T}\nabla^2 S_n(\bm{\theta}_0+a_n\bm vt) \bm v\\
    &:=\mathbb{A}+ \mathbb{B}, \hspace{1cm}\mbox{where}\ t \in (0,1).
    \end{align*}
    For the term $\mathbb{A}$, since $\bm\Sigma_w \bm 1 =\bm 0$
    \begin{align*}
        \mathbb{A} & = -2a_n \bm v^T \dot{\bm F}(\bm \theta_0)^T\bm\Sigma_w \bm \epsilon \\
        & = -2a_n \bm v^T \dot{\bm F}(\bm \theta_0)^T\bm\Sigma_w \bm \eta,
    \end{align*}
    where $\bm \eta = \bm \epsilon-\mu \bm 1$.
    \begin{align*}
        \Var(\mathbb{A}) & = 4a_n^2\bm v^T \dot{\bm F}(\bm \theta_0)^T\bm\Sigma_w \bm\Sigma_\epsilon\bm\Sigma_w \dot{\bm F}(\bm \theta_0)\bm v \\
        & \leq 4a_n^2\lambda_\epsilon \bm v^T \dot{\bm F}(\bm \theta_0)^T\bm\Sigma_w^2 \dot{\bm F}(\bm \theta_0)\bm v \\
        & \leq 4a_n^2 \lambda_\epsilon \lambda_w^2 \|\dot{\bm F}(\bm \theta_0)\bm v\|^2.
    \end{align*}
    By Assumption \ref{assumption:general}-(1) and the finiteness of $p$, $\|\dot{\bm F}(\bm \theta_0)\bm v\|^2 = O(n)\|\bm v\|^2$, which implies $\Var(\mathbb{A})=O(na_n^2\lambda_\epsilon\lambda_w^2)\|\bm v\|^2$. Since $\E(\bm \eta) =0 $,
    \begin{equation}\label{eq:A}
        \mathbb{A} = O_p(n^{1/2}a_n \lambda_\epsilon^{1/2}\lambda_w)\|\bm v\|. 
    \end{equation}
    Now, since $\nabla^2 S_n(\bm\theta)=2\bm{\dot{F}}(\bm\theta)^T\bm\Sigma_w \bm{\dot{F}}(\bm\theta) +2\bm{\ddot{F}}(\bm\theta)^T(I\otimes \bm\Sigma_w \bm d(\bm\theta, \bm \theta_0))-2\bm{\ddot{F}}(\bm\theta)^T(I\otimes \bm\Sigma_w \bm \epsilon)$, $\mathbb{B}$ is evaluated by the following four terms.
    \begin{align*}
        \mathbb{B} & = \frac{1}{2}a_n^2 \bm v^T\nabla^2 S_n(\bm \theta_n)\bm v\\
        & = a_n^2 \bm v^T\left( \dot{\bm F}(\bm \theta_n)^T\bm\Sigma_w \dot{\bm F}(\bm \theta_n) - \dot{\bm F}(\bm \theta_0)^T\bm\Sigma_w \dot{\bm F}(\bm \theta_0)\right)\bm v \\
        & ~~~ + a_n^2 \bm v^T\dot{\bm F}(\bm \theta_0)^T\bm\Sigma_w \dot{\bm F}(\bm \theta_0)\bm v +a_n^2 \bm v^T \ddot{\bm F}(\bm \theta_n)^T\left(I\otimes\bm\Sigma_w \bm d(\bm \theta_n, \bm\theta_0) \right)\bm v \\
        & ~~~ - a_n^2 \bm v^T \ddot{\bm F}(\bm \theta_n)^T\left(I\otimes\bm\Sigma_w \bm \epsilon \right)\bm v \\
        & = \mathbb{B}_1+\mathbb{B}_2+\mathbb{B}_3+\mathbb{B}_4, \hspace{1cm} \mbox{where}\ \bm \theta_n = \bm \theta_0+a_n\bm v t.
    \end{align*}
    \begin{align}\label{eq:B1}
        \mathbb{B}_1 & = a_n^2 \bm v^T\left( \dot{\bm F}(\bm \theta_n)^T\bm\Sigma_w \dot{\bm F}(\bm \theta_n) - \dot{\bm F}(\bm \theta_0)^T\bm\Sigma_w \dot{\bm F}(\bm \theta_0)\right)\bm v \nonumber\\
        & = na_n^2\lambda_w \bm v^T\left( \frac{1}{n\lambda_w}\dot{\bm F}(\bm \theta_n)^T\bm\Sigma_w \dot{\bm F}(\bm \theta_n) - \frac{1}{n\lambda_w}\dot{\bm F}(\bm \theta_0)^T\bm\Sigma_w \dot{\bm F}(\bm \theta_0)\right)\bm v \nonumber\\
        & =  na_n^2\lambda_w \bm v^T\left(\bm\Gamma(1+o(1))\left\{ \left({\bm{\dot{F}}}(\bm\theta_0)^{T}\bm\Sigma_w {\bm{\dot{F}}}(\bm\theta_0)\right)^{-1} {\bm{\dot{F}}}(\bm\theta_n)^{T}\bm\Sigma_w {\bm{\dot{F}}}(\bm\theta_n)-I\right\}\right)\bm v \nonumber\\
        & = o(na_n^2\lambda_w)\|\bm v\|^2.
    \end{align}
    The third equality holds by Assumption \ref{assumption:general}-(3) and the last equality holds by Assumption \ref{assumption:general}-(2) since $\|\bm \theta_n - \bm\theta_0\|=a_n\|\bm v\|t\rightarrow 0$. By Assumption \ref{assumption:general}-(3) again,
    \begin{align}\label{eq:B2}
        \mathbb{B}_2 & = a_n^2 \bm v^T\dot{\bm F}(\bm \theta_0)^T\bm\Sigma_w \dot{\bm F}(\bm \theta_0)\bm v \nonumber \\
        & = na_n^2\lambda_w \bm v^T\left(\frac{1}{n\lambda_w}\dot{\bm F}(\bm \theta_0)^T\bm\Sigma_w \dot{\bm F}(\bm \theta_0)\right)\bm v \nonumber \\
        & = na_n^2\lambda_w \bm v^T\bm\Gamma\bm v (1+o(1)).
    \end{align}
    By Assumption \ref{assumption:general}-(5), $\mathbb{B}_2$ does not vanish to zero.
    \begin{align}\label{eq:B3-1}
        \mathbb{B}_3 & = a_n^2 \bm v^T \ddot{\bm F}(\bm \theta_n)^T\left(I\otimes\bm\Sigma_w \bm d(\bm \theta_n, \bm\theta_0) \right)\bm v \nonumber\\
        & = a_n^2 \bm v^T\left[\bm f_{kl}(\bm \theta_n)^T\bm\Sigma_w \bm d(\bm \theta_n, \bm \theta_0) \right]_{k,l=\{1, \ldots, p\}}\bm v.
    \end{align}
    The term with the bracket above is a matrix where the expression in the bracket equals to the element in the $k$-th row and the $l$-th column of the matrix. 
    \begin{align*}
        \left|\bm f_{kl}(\bm \theta_n)^T\bm\Sigma_w \bm d(\bm \theta_n, \bm \theta_0)\right| & \leq \|\bm f_{kl}(\bm \theta_n)\|\lambda_w\|\bm d(\bm \theta_n, \bm \theta_0)\| \\
        & = O(n^{1/2}\lambda_w) \|\bm d(\bm \theta_n, \bm \theta_0)\| \\
        & = O(n^{1/2}a_n\lambda_w)\|\bm v\|.
    \end{align*}
    The first equality follows from $\|\bm f_{kl}(\bm \theta_n)\|^2=O(n)$ by Assumption \ref{assumption:general}-(1) and the second equality holds because $\|\bm \theta_n - \bm \theta_0\|=O(a_n)\|\bm v\|$. Therefore,
    \begin{equation}\label{eq:B3-2}
        \mathbb{B}_3=O(n^{1/2}a_n^3\lambda_w)\|\bm v\|^3
    \end{equation}
    since $p$ is finite. To deal with $\mathbb{B}_4$,
    \begin{align*}
        \mathbb{B}_4 & = - a_n^2 \bm v^T \ddot{\bm F}(\bm \theta_n)^T\left(I\otimes\bm\Sigma_w \bm \epsilon \right)\bm v \\
        & = - a_n^2 \bm v^T \ddot{\bm F}(\bm \theta_n)^T\left(I\otimes\bm\Sigma_w \bm \eta \right)\bm v \\
        & =  -a_n^2 \bm v^T \left[\bm f_{kl}(\bm \theta_n)\bm\Sigma_w \bm \eta \right]_{k,l=\{1, \ldots, p\}}\bm v.
    \end{align*}
    We first show $|\bm f_{kl}(\bm\theta_n)^T\bm\Sigma_w \bm\eta|=O_p(n^{1/2}\lambda_\epsilon^{1/2}\lambda_w)$ by
    \begin{align*}
        \Var(\bm f_{kl}(\bm\theta_n)^T\bm\Sigma_w \bm \eta) & = \bm f_{kl}(\bm\theta_n)^T\bm\Sigma_w\bm\Sigma_\epsilon\bm\Sigma_w \bm f_{kl}(\bm\theta_n) \\
        & \leq \lambda_\epsilon \lambda_w^2 \|\bm f_{kl}(\bm\theta_n)\|^2  \\
        & = O(n\lambda_\epsilon \lambda_w^2).
    \end{align*}
    Thus, 
    \begin{equation}\label{eq:B4}
        \mathbb{B}_4 = O_p(n^{1/2}a_n^2\lambda_\epsilon^{1/2}\lambda_w)\|\bm v\|^2.
    \end{equation}
    By equations \eqref{eq:B1}, \eqref{eq:B2}, \eqref{eq:B3-2}, and \eqref{eq:B4}, 
    \begin{align}\label{eq:B}
        \mathbb{B} & = o(na_n^2\lambda_w)\|\bm v\|^2+na_n^2\lambda_w\bm v^T\bm\Gamma\bm v +O(n^{1/2}a_n^3\lambda_w)\|\bm v\|^3 +O_p(n^{1/2}a_n^2\lambda_\epsilon^{1/2}\lambda_w)\|\bm v\|^2 \nonumber \\
        & = na_n^2\lambda_w\bm v \bm\Gamma\bm v + o_p(na_n^2\lambda_w)\|\bm v\|^2.
    \end{align}
    The second equality holds since $\lambda_\epsilon \leq o(n)$ by Assumption \ref{assumption:general}-(5). Finally, through \eqref{eq:A} and \eqref{eq:B},
    \begin{align} \label{eq:thm1}
        S_n(\bm \theta_0+a_n \bm v) -S_n(\bm\theta_0) & =  O_p(n^{1/2}a_n \lambda_\epsilon^{1/2}\lambda_w)\|\bm v \|+na_n^2\lambda_w\bm v^T \bm\Gamma\bm v(1+o_p(1)) \nonumber \\
        & = O_p(\lambda_\epsilon\lambda_w)\|\bm v\| +\lambda_\epsilon\lambda_w\bm v^T \bm\Gamma\bm v(1+o_p(1)).
    \end{align}
    Therefore, with large enough $\|\bm v\|$, The desired result follows.
\end{proof}

\newpage

\begin{theorem}\label{thm:consistency-penalty}
For any $\varepsilon>0$ and $b_n = (\lambda_\epsilon/n)^{1/2}+c_n$, under assumptions in Lemma \ref{thm:consistency} and \ref{assumption:penalty}-(1),(2), there exists a positive constant $C$ such that
    $$P\left(\inf_{\|\bm v\|=C}  Q_n(\bm \theta_0+b_n\bm v) -  Q_n(\bm\theta_0) >0 \right) > 1-\varepsilon$$
    for large enough $n$. Therefore, with probability tending to 1, there exists a local minimizer ($\hat{\bm\theta}$) of $ Q_n (\bm \theta)$ in the ball centered at $\bm\theta_0$ with the radius $b_n \bm v$. By Assumptions \ref{assumption:general}-(5) and \ref{assumption:penalty}-(1), $b_n=o(1)$, which leads to the consistency of $\hat{\bm\theta}$.
\end{theorem}
\vspace{0.5cm}
\begin{proof}
    \begin{align}\label{eq:thm2}
    & Q_n(\bm\theta_0+b_n \bm v)- Q_n(\bm\theta_0) \nonumber \\
    & ~~~~ = \mathbb{A'+B'}+\displaystyle n \left(\sum_{i=1}^{p} p_{\tau_n}(|\theta_{0i}+b_n v_i|)-p_{\tau_n} (|\theta_{0i}|) \right) \nonumber \\
    & ~~~~  \geq \mathbb{A'+B'}+\displaystyle n \left(\sum_{i=1}^{s} p_{\tau_n}(|\theta_{0i}+b_n v_i|)-p_{\tau_n} (|\theta_{0i}|) \right) \nonumber \\
    & ~~~~ = \mathbb{A'+B'} +n\sum_{i=1}^{s}  q_{\tau_n}(|\theta_{0i}^*|)sgn(\theta_{0i}^*)b_n v_i, \hspace{0.5cm} \mbox{where }\theta_{0i}^* \mbox{ lies on a line segment }(\theta_{0i}, \theta_{0i}+b_nv_i)\nonumber \\
    & ~~~~ =  \mathbb{A'+B'} +n\sum_{i=1}^{s} \left( q_{\tau_n}(|\theta_{0i}^*|)-q_{\tau_n}(|\theta_{0i}|)+q_{\tau_n}(|\theta_{0i}|)\right)sgn(\theta_{0i}^*)b_n v_i \nonumber \\
    & ~~~~ = \mathbb{A'+B'+C+D},
    \end{align}
    where $\mathbb{A}'$ and $\mathbb{B}'$ are defined similarly to $\mathbb{A}$ and $\mathbb{B}$ in the proof of Lemma \ref{thm:consistency} with replacement of $a_n$ to $b_n$. $\mathbb{C}=n\sum_{i} (q_{\tau_n}(|\theta_{0i}^*|)-q_{\tau_n}(|\theta_{0i}|))sgn(\theta_{0i^*})b_n v_i$ and $\mathbb{D}=n\sum_{i} q_{\tau_n}(|\theta_{0i}|)sgn(\theta_{0i^*})b_n v_i$. Referring to equation \eqref{eq:thm1}
    $$\mathbb{A'+B'} = O_p(n^{1/2}b_n \lambda_\epsilon^{1/2}\lambda_w)\|\bm v \|+nb_n^2\lambda_w\bm v^T \bm\Gamma\bm v.$$
    It is enough to show that $\mathbb{C}=O(nb_n^2)\|\bm v\|$ and $\mathbb{D}=O(nb_n^2)\|\bm v\|$ since $\lambda_w \geq O(1)$. By Assumptions \ref{assumption:penalty}-(1) and (2), 
    $$|\mathbb{C}| \leq nb_n\sum_i |q_{\tau_n}(|\theta_{0i}^*|)-q_{\tau_n}(|\theta_{oi}|)||v_i|= O(nb_n^2)\|\bm v\|,$$
    $$|\mathbb{D}| \leq \displaystyle nb_n\max_{1\leq i \leq s} | q_{\tau_n}(|\theta_{0i}|)|\sum_i v_i = O(nb_nc_n)\|\bm v\| \leq O(nb_n^2)\|\bm v\|.$$
    Therefore, $nb_n^2\lambda_w \bm v^T\bm\Gamma \bm v$ dominates equation \eqref{eq:thm2} with large $\|\bm v\|$, which leads to the desired result.\\
\end{proof}

\newpage

\begin{lemma}\label{lemma:clt}[Theorem 2.2 from \cite{peligrad1997central}]\ 
Let $\bm \eta=\{\eta_1, \ldots, \eta_n\}$ be a stochastic sequence and $\bm h_n=\{h_{n,1}, \ldots, h_{n,n}\}$ be a triangular weight vector, then 
$$\bm h_n^T \bm \eta \overset{d}{\rightarrow} N(0, 1)$$
under the following conditions.
\begin{enumerate}
    \item $\bm \eta$ is a centered stochastic sequence
    \item $\sup_{n} \|\bm h_n\|_2^2 < \infty$ and $\|\bm h_n\|_{\infty} \rightarrow 0$ as $n \rightarrow \infty$.
    \item $\{\eta_i^2\}$ is uniformly integrable and $\Var(\bm h_n^T \bm \eta) = 1$
    \item For $\bm \eta$, one of the three following mixing conditions must be satisfied.
    \begin{itemize}
        \item [$-$] $\{\eta_i\}$ is a $\phi$-mixing.
        \item [$-$] $\{\eta_i\}$ is a $\rho$-mixing and $\sum_{j \in \mathcal{N}} \rho(2^j)<\infty$.
        \item [$-$] $\{\eta_i\}$ is a $\alpha$-mixing, for $\delta >0$,  $\{|\eta_i|^{2+\delta}\}$ is uniformly integrable, and $\sum_{j \in \mathcal{N}}n^{2/\delta}\alpha(n) < \infty$. 
    \end{itemize}
\end{enumerate}
\end{lemma}

\begin{proof}
We refer to \cite{peligrad1997central} for the proof.
\end{proof}
\vspace{0.5cm}
\begin{lemma}[Asymptotic normality]\label{thm:normality} 
    Under Assumption \ref{assumption:general}, $$\left(\frac{n}{\lambda_\epsilon}\right)^{1/2}\left(\hat{\bf{\bm\theta}}^{(s)}-\bm\theta_{0}\right) \overset{d}{\rightarrow} N\left(0, \bm \Gamma^{-1}\bm \Gamma_\epsilon\bm \Gamma^{-1}\right),$$
    where $\hat{\bm \theta}^{(s)}$ is a consistent estimator introduced in Lemma \ref{thm:consistency} with $S_n(\bm\theta)$.
\end{lemma}
\vspace{0.5cm}
\begin{proof}
	
    \noindent By the mean-value theorem of a vector-valued function \citep{feng2013mvt}, with $\zeta_n = (n \lambda_\epsilon\lambda_w^2)^{-1/2}$,
    \begin{align*}
    	 & \zeta_n \nabla S_n(\bm\theta_0) \\
    	 & \hspace{0.5cm} = \zeta_n \left(\nabla S_n(\hat{\bm\theta}^{(s)})+\left(\int^{1}_{0}\nabla^2 S_n\left(\bm{\hat\theta}^{(s)}+\left(\bm\theta_0-\hat{\bm\theta}^{(s)}\right)t\right)dt \right)^{T}\left(\bm\theta_0-\hat{\bm\theta}^{(s)}\right)\right)\\
    	 & \hspace{0.5cm} = \zeta_n \left(\int^{1}_{0}\nabla^2 S_n\left(\bm{\hat\theta}^{(s)}+\left(\bm\theta_0-\hat{\bm\theta}^{(s)}\right)t\right)dt \right)^{T}\left(\bm\theta_0-\hat{\bm\theta}^{(s)}\right)
    \end{align*}
    since $\nabla S_n(\hat{\bm\theta}^{(s)})=0$. We show $\zeta_n \nabla S_n(\bm\theta_0)$ follows a normal distribution asymptotically and 
    \begin{equation}\label{eq:normality-constant}
        \frac{1}{n\lambda_w}\int^{1}_{0}\nabla^2 S_n\left(\bm{\hat\theta}^{(s)}+\left(\bm\theta_0-\hat{\bm\theta}^{(s)}\right)t\right)dt  \overset{p}{\rightarrow} 2\bm\Gamma.
    \end{equation}
    For an arbitrary vector $\bm v\in \mathcal{R}^p$, 
    \begin{align*}
        \zeta_n \bm v^T\nabla S_n(\bm\theta_0) & = -2 \zeta_n \bm v^T\dot{\bm F}(\bm \theta_0)^T\bm\Sigma_w \bm \epsilon \\
        & = -2 \zeta_n \bm v^T\dot{\bm F}(\bm \theta_0)^T\bm\Sigma_w \bm \eta,
    \end{align*}
    where $\bm \eta = \bm \epsilon-\mu\bm 1$. This converges to $N(0, 4\bm v^T\bm\Gamma_\epsilon\bm v)$ since $\bm h_n^T \bm \eta \overset{d}{\rightarrow}N(0,1)$ with 
    $$\bm h_n = \left(\bm v^T \dot{\bm F}(\bm \theta_0)^T\bm\Sigma_w \bm\Sigma_\epsilon\bm\Sigma_w \dot{\bm F}(\bm \theta_0) \bm v\right)^{-1/2}\bm\Sigma_w \dot{\bm F}(\bm \theta_0)\bm v,$$
    by Lemma \ref{lemma:clt}. We first show that $\bm h_n ^T \bm \eta \overset{d}{\rightarrow} N(0,1)$ by proving the conditions given in Lemma \ref{lemma:clt} are fulfilled and then we handle the remainder term. The first condition in Lemma \ref{lemma:clt} is trivial and the fourth condition is satisfied by Assumption \ref{assumption:general}-(7). The second condition is satisfied since
    \begin{align*}
        \|\bm h_n\|^2 & =  \left(\bm v^T \dot{\bm F}(\bm \theta_0)^T\bm\Sigma_w \bm\Sigma_\epsilon\bm\Sigma_w \dot{\bm F}(\bm \theta_0) \bm v\right)^{-1} \bm v^T \dot{\bm F}(\bm \theta_0)^T\bm\Sigma_w^2 \dot{\bm F}(\bm \theta_0) \bm v \\
        & = \lambda_\epsilon^{-1}\left(\frac{1}{n\lambda_\epsilon\lambda_w^2}\bm v^T \dot{\bm F}(\bm \theta_0)^T\bm\Sigma_w \bm\Sigma_\epsilon\bm\Sigma_w \dot{\bm F}(\bm \theta_0) \bm v\right)^{-1}\frac{1}{n\lambda_w^2}\bm v^T \dot{\bm F}(\bm \theta_0)^T\bm\Sigma_w^2 \dot{\bm F}(\bm \theta_0) \bm v \\
        & \leq \lambda_\epsilon^{-1} \left(\bm v^T \bm\Gamma_\epsilon\bm v \right)^{-1} O(1) \|\bm v\|^2 ~~~~~ (\text{Assumptions \ref{assumption:general}-(1) and (3)})\\ 
        & = O(\lambda_\epsilon^{-1})\\
        & \leq O(1), ~~~~~ (\text{Assumption \ref{assumption:general}-(5)})\\
        \|\bm h_n\|_{\infty} & = \left(\bm v^T \dot{\bm F}(\bm \theta_0)^T\bm\Sigma_w \bm\Sigma_\epsilon\bm\Sigma_w \dot{\bm F}(\bm \theta_0) \bm v\right)^{-1/2}\left\|\bm\Sigma_w\dot{\bm F}(\bm\theta_0)\bm v\right\|_{\infty} \\
        & \leq \zeta_n\left(\bm v^T \bm\Gamma_\epsilon\bm v\right)^{-1/2}\left\|\bm\Sigma_w\right\|_{\infty}\left\|\dot{\bm F}(\bm\theta_0)\bm v\right\|_{\infty} \\
        & \leq O(\zeta_n) \|\bm W^T\|_{\infty}\|\bm\Sigma_n\|_{\infty}\|\bm W\|_{\infty}\left\|\dot{\bm F}(\bm\theta_0)\bm v\right\|_{\infty}\\
        & \leq O\left((n\lambda_\epsilon)^{-1/2}\right) \lambda_w^{-1}\|\bm W\|_1\|\bm W\|_{\infty} ~~~~~ (\|\dot{\bm F}(\bm\theta_0)\bm v\|_{\infty}=O(1) \mbox{ and } \|\bm\Sigma_n\|_{\infty} \leq 2)\\
        & = o(1). ~~~~~ (\text{Assumption \ref{assumption:general}-(4)})
    \end{align*}
    $\{\eta_i^2\}$ is uniformly integrable by Assumption \ref{assumption:general}-(6). The remainder of the proof is to show $\Var(\bm h_n^T \bm \eta) =1$.
    $$\Var(\bm h_n^T \bm \eta) = \left(\bm v^T \dot{\bm F}(\bm \theta_0)^T\bm\Sigma_w \bm\Sigma_\epsilon\bm\Sigma_w \dot{\bm F}(\bm \theta_0) \bm v\right)^{-1} \left(\bm v^T \dot{\bm F}(\bm \theta_0)^T\bm\Sigma_w \bm\Sigma_\epsilon\bm\Sigma_w \dot{\bm F}(\bm \theta_0) \bm v\right) = 1.$$
    Therefore, for arbitrary $\bm v$,
    \begin{align*}
        -2 \zeta_n \bm v^T\dot{\bm F}(\bm \theta_0)^T\bm\Sigma_w \bm \eta &= -2 (n \lambda_\epsilon\lambda_w^2)^{-1/2} \left(\bm v^T \dot{\bm F}(\bm \theta_0)^T\bm\Sigma_w \bm\Sigma_\epsilon\bm\Sigma_w \dot{\bm F}(\bm \theta_0) \bm v\right)^{1/2}\bm h_n^T \bm \eta \nonumber\\
        &= -2 \left(\frac{1}{n\lambda_\epsilon\lambda_w^2}\bm v^T \dot{\bm F}(\bm \theta_0)^T\bm\Sigma_w \bm\Sigma_\epsilon\bm\Sigma_w \dot{\bm F}(\bm \theta_0) \bm v\right)^{1/2}\bm h_n^T \bm \eta \nonumber \\
        &\overset{d}{\rightarrow} N(0, 4\bm v^T\bm\Gamma_\epsilon\bm v).
    \end{align*}
    The asymptotic variance in the limiting distribution comes from Assumption \ref{assumption:general}-(3). By the Cramer-Wold device, 
    \begin{equation}\label{eq:normality}
        \zeta_n \nabla S_n(\bm \theta_0) = -2 \zeta_n \dot{\bm F}(\bm \theta_0)^T\bm\Sigma_w \bm \eta \overset{d}{\rightarrow} N(0, 4\bm\Gamma_\epsilon).
    \end{equation}
    For equation \eqref{eq:normality-constant}, we need to show
    \begin{align}
        &\frac{1}{n\lambda_w}\nabla^2 S_n(\bm \theta_0) \overset{p}{\rightarrow} 2\bm\Gamma \label{eq:normality-constant1},\\
        &\frac{1}{n\lambda_w}\left(\int^{1}_{0}\nabla^2 S_n\left(\bm{\hat\theta}^{(s)}+\left(\bm\theta_0-\hat{\bm\theta}^{(s)}\right)t\right)dt -\nabla^2 S_n(\bm \theta_0)\right) \overset{p}{\rightarrow} 0.  \label{eq:normality-constant2}
    \end{align}
    For \eqref{eq:normality-constant1},
    $$\frac{1}{n\lambda_w}\nabla^2 S_n(\bm\theta_0)=\frac{2}{n\lambda_w}\bm{\dot{F}}(\bm\theta_0)^T\bm\Sigma_w \bm{\dot{F}}(\bm\theta_0)-\frac{2}{n\lambda_w}\bm{\ddot{F}}(\bm\theta_0)^T(I\otimes \bm\Sigma_w \bm \epsilon).$$
    Similar to equation \eqref{eq:B2} and \eqref{eq:B4}, the first term converges to $2\bm\Gamma$ and the second term is $O_p((\lambda_\epsilon/n)^{1/2})$, which vanishes to $o_p(1)$ by Assumption \ref{assumption:general}-(5). \eqref{eq:normality-constant2} is satisfied if
    \begin{equation}\label{eq:second_derivative}
        \displaystyle \max_{\|\bm\theta-\bm\theta_0\| \leq C a_n} \frac{1}{n\lambda_w}\left\| \nabla^2 S_n(\bm\theta)-\nabla^2 S_n(\bm \theta_0) \right\| \overset{p}{\rightarrow} 0.    
    \end{equation}
    \eqref{eq:second_derivative} can be decomposed as three following terms.
    \begin{align*}
        & \max_{\|\bm\theta-\bm\theta_0\| \leq C a_n}\frac{1}{n\lambda_w}\left\| \nabla^2 S_n(\bm\theta)-\nabla^2 S_n(\bm \theta_0)\right\| \\
        & \hspace{3cm} \leq \max_{\|\bm\theta-\bm\theta_0\| \leq C a_n} \frac{2}{n\lambda_w}\left\| \dot{\bm F}(\bm\theta)^T\bm\Sigma_w \dot{\bm F}(\bm\theta)- \dot{\bm F}(\bm\theta_0)^T\bm\Sigma_w \dot{\bm F}(\bm\theta_0)\right\|\\
        & \hspace{3.2cm} + \max_{\|\bm\theta-\bm\theta_0\| \leq C a_n}\frac{2}{n\lambda_w}\left\| \ddot{\bm F}(\bm\theta)^T\left(I \otimes\bm\Sigma_w\bm d(\bm \theta, \bm \theta_0)\right)\right\|\\
        & \hspace{3.2cm} + \max_{\|\bm\theta-\bm\theta_0\| \leq C a_n}\frac{2}{n\lambda_w}\left\|\left(\ddot{\bm F}(\bm\theta)-\ddot{\bm F}(\bm\theta_0) \right)^T \left(I \otimes\bm\Sigma_w \bm \eta\right)\right\|.
    \end{align*}
    The first part converges to 0 by a similar procedure to equation \eqref{eq:B1} and Assumption \ref{assumption:general}-(2). The second part converges to 0 because
    \begin{align*}
      \frac{2}{n\lambda_w}\left\| \ddot{\bm F}(\bm\theta)^T\left(I \otimes\bm\Sigma_w\bm d(\bm \theta, \bm \theta_0)\right)\right\|&= \frac{p}{n\lambda_w}\left\|\left[\bm f_{kl}(\bm \theta)^T\bm\Sigma_w \bm d(\bm \theta, \bm\theta_0)\right]_{k,l=\{1,\ldots,p\}}\right\|  \\
      & \leq \frac{2p}{n\lambda_w}\cdot\max_{k,l}\left|\bm f_{kl}(\bm \theta)^T\bm\Sigma_w \bm d(\bm \theta, \bm\theta_0)\right|\\
      & \leq \frac{2p}{n}\max_{k,l}\left\|\bm f_{kl}(\bm\theta) \right\|\cdot \left\| \bm d(\bm\theta, \bm\theta_0)\right\| \\
      & \leq \frac{2C_p}{n^{1/2}}\cdot \|\bm\theta-\bm\theta_0\|, \hspace{1cm} \mbox{where } C_p \mbox{ is independent with } \bm\theta \\
      & = O\left(n^{-1/2}a_n\right) = o(1).
    \end{align*}
    The results above hold if $\|\bm\theta-\bm\theta_0\|=O(a_n)$, which is still true for the maximum. For the last term,
    $$\frac{2}{n\lambda_w}\left\|\left(\ddot{\bm F}(\bm\theta)-\ddot{\bm F}(\bm\theta_0) \right)^T \left(I \otimes\bm\Sigma_w \bm \epsilon\right)\right\| = \frac{2}{n\lambda_w}\left\|\left[\left(\bm f_{kl}(\bm \theta)-\bm f_{kl}(\bm\theta_0)\right)^T\bm\Sigma_w\bm\eta\right]_{k,l=\{1,\ldots, p\}}\right\|.$$
    We evaluate $\Var((\bm f_{kl}(\bm \theta)-\bm f_{kl}(\bm\theta_0))^T\bm\Sigma_w\bm\eta)$ as follows.
    \begin{align*}
        \Var\left((\bm f_{kl}(\bm \theta)-\bm f_{kl}(\bm\theta_0))^T\bm\Sigma_w\bm\eta\right) & = (\bm f_{kl}(\bm \theta)-\bm f_{kl}(\bm\theta_0))^T\bm\Sigma_w\bm\Sigma_\epsilon\bm\Sigma_w(\bm f_{kl}(\bm \theta)-\bm f_{kl}(\bm\theta_0)) \\
        & \leq \lambda_\epsilon \lambda_w^2 \|\bm f_{kl}(\bm \theta)-\bm f_{kl}(\bm\theta_0)\|^2 \\
        & \leq \lambda_\epsilon \lambda_w^2 O\left(\|\bm\theta-\bm\theta_0\|^2\right) ~~~~~ \text{(Assumption \ref{assumption:general}-(1))} \\
        & = O\left(a_n^2 \lambda_\epsilon\lambda_w^2\right).
    \end{align*}
    Therefore, $|\left(\bm f_{kl}(\bm \theta)-\bm f_{kl}(\bm\theta_0)\right)^T\bm\Sigma_w\bm\eta|=O_p(a_n\lambda_\epsilon^{1/2}\lambda_w)$ and 
    $$\max_{\|\bm\theta-\bm\theta_0\| \leq C a_n}\frac{2}{n\lambda_w}\left\|\left(\ddot{\bm F}(\bm\theta)-\ddot{\bm F}(\bm\theta_0) \right)^T \left(I \otimes\bm\Sigma_w \bm \eta\right)\right\| =O_p\left(\frac{a_n\lambda_\epsilon^{1/2}}{n}\right) =o_p(1).$$
    Thus, we prove equation \eqref{eq:normality-constant2}. Combining results of equations \eqref{eq:normality-constant1} and \eqref{eq:normality-constant2}, we have equation \eqref{eq:normality-constant}. Recall 
    $$\zeta_n \nabla S_n(\bm\theta_0) = \zeta_n \left(\int^{1}_{0}\nabla^2 S_n\left(\bm{\hat\theta}^{(s)}+\left(\bm\theta_0-\hat{\bm\theta}^{(s)}\right)t\right)dt \right)^{T}\left(\bm\theta_0-\hat{\bm\theta}^{(s)}\right), $$
    with $\zeta_n = (n \lambda_\epsilon\lambda_w^2)^{-1/2}$. By equations \eqref{eq:normality-constant} and \eqref{eq:normality} and Slutsky's theorem,
    $$2n\lambda_w \zeta_n \bm\Gamma\left(\hat{\bm\theta}^{(s)}-\bm\theta_0\right) \overset{d}{\rightarrow} N(0, 4\bm\Gamma_\epsilon).$$
    Since $n\lambda_w \zeta_n = (n/\lambda_\epsilon)^{1/2}$, 
    $$\left(\frac{n}{\lambda_\epsilon}\right)^{1/2}\left(\hat{\bm\theta}^{(s)}-\bm\theta_0\right) \overset{d}{\rightarrow} N(0, \bm\Gamma^{-1}\bm\Gamma_\epsilon\bm\Gamma^{-1}).$$
	
\end{proof}

\newpage

\begin{theorem}[Oracle property]\label{thm:sparsity}
With $\hat{\bm \theta}$, a consistent estimator introduced in Theorem \ref{thm:consistency-penalty} using $ Q_n(\bm\theta)$, if Assumptions \ref{assumption:general} and \ref{assumption:penalty} are satisfied,
\begin{enumerate}[(i)]
    \item $P\left(\hat{\theta}_i=0\right) \rightarrow 1,$ for $i \in \{s+1, \ldots, p\}$.
    \item Also,
    $$\left(\frac{n}{\lambda_\epsilon}\right)^{1/2}\left(\hat{\bm\theta}_1-\bm\theta_{01}+\left((2\lambda_w\bm\Gamma)^{-1}\right)_{11}\bm \beta_{n,s}\right)\overset{d}{\rightarrow}N\left(0,\left( \bm\Gamma^{-1}\bm\Gamma_\epsilon\bm\Gamma^{-1}\right)_{11}\right),$$
    where $\hat{\bm \theta}_1=(\hat\theta_1, \ldots, \hat\theta_s)^T,\ \bm\theta_{01} = (\theta_{01}, \ldots, \theta_{0s})^T,\ \bm\beta_{n,s}=( q_{\tau_n}({|\theta_{01}|)sgn(\theta}_{01}),\ldots,$ $q_{\tau_n}(|\theta_{0s}|)sgn(\theta_{0s}))^{T}$ and $\bm A_{11}$ is the $s\times s$ upper-left matrix of $\bm A$.
\end{enumerate}
\end{theorem}
\vspace{0.5cm}

\begin{proof}

Proof of (i)
	
\noindent It is equivalent to show that $P\left(\hat{\theta}_i \neq0\right)\rightarrow 0$ as $n\rightarrow \infty$ for $i\in\{s+1, \ldots, p\}$. 
\begin{align*}
    P\left(\hat{\theta}_i \neq0\right) & = P\left(\hat{\theta}_i \neq0,\ |\hat{\theta}_i|>Cb_n\right)+P\left(\hat{\theta}_i \neq 0,\ |\hat{\theta}_i|\leq Cb_n\right)  \\
    & := P(\mathbb{E})+P(\mathbb{F}).
\end{align*}
For any $\varepsilon>0$ and large enough $n$, $P(\mathbb{E}) <\varepsilon/2$ by Theorem\ref{thm:consistency-penalty}. Now we show $P(\mathbb{F})<\varepsilon/2$. By the vector-valued mean value theorem \citep{feng2013mvt},
$$\zeta_n \nabla S_n(\bm\theta_0) = \zeta_n \left(\nabla S_n(\bm\theta)+\left(\int^{1}_{0}\nabla^2 S_n\left(\bm\theta+\left(\bm\theta_0-\bm\theta\right)t\right)dt \right)^{T}\left(\bm\theta_0-\bm\theta\right)\right),$$
where $\zeta_n = (n\lambda_\epsilon \lambda_w^2)^{-1/2}$.
From the proof of Lemma \ref{thm:normality}, $\zeta_n \nabla S_n(\bm\theta_0)=O_p(1)$ and with $\|\bm\theta_0-\bm\theta\|=O_p(b_n)$, the similar results of \eqref{eq:normality-constant} inform that
$$n\zeta_n\lambda_w \left(\frac{1}{n\lambda_w}\int^{1}_{0}\nabla^2 S_n\left(\bm\theta+\left(\bm\theta_0-\bm\theta\right)t\right)dt \right)^{T}\left(\bm\theta_0-\bm\theta\right) = O_p(1).$$
This leads to $\zeta_n \nabla S_n(\bm\theta)=O_p(1)$ for $\|\bm\theta-\bm\theta_0\|=O_p(b_n)$. Since $\hat{\bm\theta}$ is the local minimizer of $ Q_n(\bm\theta)$ with $\|\hat{\bm\theta}-\bm\theta_0\|=O_p(b_n)$, we attain, for $i\in\{s+1,\ldots,p\},$
$$n\zeta_n q_{\tau_n}(|\hat{\theta}_i|)=O_p(1)$$
from
$$\zeta_n \left.\frac{\partial Q_n(\bm\theta)}{\partial \theta_i}\right|_{\bm\theta=\hat{\bm\theta}}=\zeta_n \left.\frac{\partial S_n(\bm\theta)}{\partial \theta_i}\right|_{\bm\theta=\hat{\bm\theta}}+n\zeta_n q_{\tau_n}(|\hat{\theta}_i|)sgn(\hat{\theta}_i).$$
Therefore, there exists a $M>0$ such that $P\left(\left|n\zeta_n q_{\tau_n}(|\hat{\theta}_i|)\right|>M\right) <\varepsilon/2$ for large enough $n$, which implies 
$$P\left(\hat{\theta}_i \neq 0,\ |\hat{\theta}_i|\leq Cb_n,\ n\zeta_n q_{\tau_n}(|\hat{\theta}_i|)>M\right)<\frac{\varepsilon}{2}.$$
By Assumptions \ref{assumption:penalty}-(3) and (4),
$$P\left(\hat{\theta}_i \neq 0,\ |\hat{\theta}_i|\leq Cb_n,\ n\zeta_n q_{\tau_n}(|\hat{\theta}_i|)>M\right)=P\left(\hat{\theta}_i \neq 0,\ |\hat{\theta}_i|\leq Cb_n\right)$$
for large enough $n$. At last, we have $P(\mathbb{F})<\varepsilon/2$. Together with $P(\mathbb{E})<\varepsilon/2$, this implies $P(\hat{\theta}_i \neq 0) \rightarrow 0$.\\

\vspace{0.5cm}
\noindent Proof of (ii) \\

\noindent Note that
$$\zeta_n \nabla  Q_n(\hat{\bm\theta})=\zeta_n \nabla S_n(\hat{\bm\theta})+n\zeta_n   \bm q_{\tau_n}(|\hat{\bm\theta}|)sgn(\hat{\bm\theta}),$$
where $\bm q_{\tau_n}(|\hat{\bm\theta}|)sgn(\hat{\bm\theta})=( q_{\tau_n}(|\hat{\theta}_1|)sgn(\hat{\theta}_1), \ldots,  q_{\tau_n}(|\hat{\theta}_p|)sgn(\hat{\theta}_p))^T$. Since $\hat{\bm\theta}$ is a local minimizer of $ Q_n(\bm\theta)$, $\nabla  Q_n(\hat{\bm\theta})=0$, which implies
\begin{align*}
    -\zeta_n\nabla S_n(\bm\theta_0)& = \left(\frac{1}{n\lambda_w}\int^{1}_{0}\nabla^2 S_n\left(\bm\theta_0+(\hat{\bm\theta}-\bm\theta_0)t\right)dt \right)^{T}\left(n\zeta_n\lambda_w(\hat{\bm\theta}-\bm\theta_0)\right) \\
    & ~~~ +n\zeta_n \bm q_{\tau_n}(|\hat{\bm\theta}|)sgn(\hat{\bm\theta}).
\end{align*}
The left-hand side converges to $N(0, 4\bm\Gamma_\epsilon)$ and, similarly to \eqref{eq:normality-constant},
$$\frac{1}{n\lambda_w}\int^{1}_{0}\nabla^2 S_n\left(\bm\theta_0+(\hat{\bm\theta}-\bm\theta_0)t\right)dt  \overset{p}{\rightarrow} 2\bm\Gamma.$$ Thus, by the Slutsky's theorem,
    $$\left(\frac{n}{\lambda_\epsilon}\right)^{1/2}\left(2\bm\Gamma(\hat{\bm\theta}-\bm\theta_0)+\lambda_w^{-1}  \bm q_{\tau_n}(|\hat{\bm\theta}|)sgn(\hat{\bm\theta})\right)\overset{d}{\rightarrow}N\left(0, 4\bm\Gamma_\epsilon\right).$$
Slicing the first $s$ components of $\hat{\bm\theta}$, we obtain
$$\left(\frac{n}{\lambda_\epsilon}\right)^{1/2}\left(\hat{\bm\theta}_1-\bm\theta_{01}+\left((2\lambda_w\bm\Gamma)^{-1}\right)_{11}\bm \beta_{n,s}\right)\overset{d}{\rightarrow}N\left(0,\left( \bm\Gamma^{-1}\bm\Gamma_\epsilon\bm\Gamma^{-1}\right)_{11}\right).$$
\end{proof}

\newpage
\section {More simulation results}

Here, we provide tables from the simulation study conducted in Section 3 of the main article.

Tables \ref{table:add_lasso_5_estimation}- \ref{table:add_lasso_8_4_estimation} report the values of mean squared error (MSE) with standard deviation of squared error (SD) in parenthesis for the estimates from 100 repetitions of data generated from the model \begin{equation}
\label{eq:additive}
\displaystyle y_t=\frac{1}{1+\exp(-\bm x_t^{T}\bm\theta_0)}+\epsilon_t,
\end{equation}
where $\bm\theta_0=(\theta_{01}, \theta_{02}, \ldots, \theta_{0,20})^{T}$ with $\theta_{01}=1, \theta_{02}=1.2, \theta_{03}=0.6$, and the others being zero.  The first component of the covariate $\bm x$ comes from $U[-1,1]$, a uniform distribution on $[-1,1]$, and the other components of $\bm x$ are simulated from a joint normal distribution with the zero mean, the variance being 0.6 and pairwise covariance being 0.1.  For $\epsilon_t$, the AR(1) and ARMA(1,1) with the non-zero mean are considered since these processes not only represent typical time series processes but also possess the strong mixing property. The choices of the AR(1) coefficient, $\rho$, are 0.5 and 0.9 and for the ARMA process, the parameters for the AR and MA parts are fixed as 0.8 ($\rho$) and 0.4 ($\phi$), respectively. For the non-zero mean, $\mu$, the choices are 0.1 and 0.5. For the standard deviation, $\sigma=0.5$. 
The formulae to calculate MSE and SD are given in Section 3 of the main article.

\begin{table}
    \centering
    \caption{\small{Estimation results with LASSO for the equation \eqref{eq:additive} when the error process is AR(1) with $\rho=0.5$. Mean squared error values are presented with standard deviation in the parenthesis. The rows without $\rho$ or $(\rho, \phi)$ indicate that no weight matrix is used.}}
    \begin{tabular}{clrrr}
    \Xhline{3\arrayrulewidth}
    \multicolumn{5}{c}{AR(1) with $\rho=0.5$} \\ \hline
     $(\mu,\sigma)$ & Methods & \multicolumn{1}{c}{$n=50$} & \multicolumn{1}{c}{$n=100$} & \multicolumn{1}{c}{$n=200$} \\ \hline
    \multirow{8}{*}{$(0.1,0.5)$} & PMWLS & 11.09 (0.51) & 6.86 (0.54) & 4.09 (0.47) \\ \cline{2-2}
    & $\mbox{PMWLS } {\tiny (\rho=0.5)}$ & 11.22 (0.48) & 6.73 (0.53) & 3.65 (0.41) \\ \cline{2-2}
    & $\mbox{PMWLS } {\tiny (\rho=0.9)}$ & 11.67 (0.45) & 6.77 (0.53) & 3.39 (0.41) \\ \cline{2-2}
    & $\mbox{PMWLS } {\tiny (\rho=0.8,\phi=0.4)}$ & 11.96 (0.44) & 7.35 (0.54) & 3.45 (0.40) \\ \cline{2-2}
    & PWLS & 12.11 (0.46) & 7.09 (0.64) & 2.51 (0.47) \\ \cline{2-2}
    & $\mbox{PWLS } {\tiny (\rho=0.5)}$ & 12.38 (0.45) & 6.70 (0.59) & 2.81 (0.43) \\ \cline{2-2}
    & $\mbox{PWLS } {\tiny (\rho=0.9)}$ & 11.68 (0.51) & 6.72 (0.54) & 3.47 (0.39) \\ \cline{2-2}
    & $\mbox{PWLS } {\tiny (\rho=0.8,\phi=0.4)}$ & 12.23 (0.40) & 6.93 (0.53) & 3.76 (0.40) \\\hline
    \multirow{8}{*}{$(0.5,0.5)$} & PMWLS & 10.26 (0.55) & 8.02 (0.54) & 4.17 (0.46) \\ \cline{2-2}
    & $\mbox{PMWLS }{\tiny (\rho=0.5)}$ & 11.04 (0.52) & 7.56 (0.51) & 3.79 (0.44) \\ \cline{2-2}
    & $\mbox{PMWLS } {\tiny (\rho=0.9)}$ & 11.22 (0.55) & 7.34 (0.51) & 3.77 (0.45) \\ \cline{2-2}
    & $\mbox{PMWLS } {\tiny (\rho=0.8,\phi=0.4)}$ & 11.60 (0.48) & 8.23 (0.50) & 3.86 (0.44) \\ \cline{2-2}
    & PWLS & 9.41 (0.58) & 6.35 (0.64) & 2.07 (0.43)  \\ \cline{2-2}
    & $\mbox{PWLS } {\tiny (\rho=0.5)}$ & 10.92 (0.57) & 7.87 (0.61) & 2.19 (0.39)  \\ \cline{2-2}
    & $\mbox{PWLS } {\tiny (\rho=0.9)}$ & 11.36 (0.51) & 8.06 (0.49) & 4.34 (0.40) \\ \cline{2-2}
    & $\mbox{PWLS } {\tiny (\rho=0.8,\phi=0.4)}$ & 12.22 (0.45) & 8.32 (0.49) & 4.56 (0.40) \\
    \Xhline{3\arrayrulewidth}
    \multicolumn{5}{l}{\footnotesize{$\ast$ The actual MSE values are $0.01 \times$ the reported values.}}
    \end{tabular}
    \label{table:add_lasso_5_estimation}
\end{table}

\begin{table}
    \centering
    \caption{\small{Estimation results with LASSO for the equation \eqref{eq:additive} when the error process is AR(1) with $\rho=0.9$. The other configurations are identical to Table \ref{table:add_lasso_5_estimation}}.}
    \begin{tabular}{clrrr}
    \Xhline{3\arrayrulewidth}
    \multicolumn{5}{c}{AR(1) with $\rho=0.9$} \\ \hline
     $(\mu,\sigma)$ & Methods & \multicolumn{1}{c}{$n=50$} & \multicolumn{1}{c}{$n=100$} & \multicolumn{1}{c}{$n=200$} \\ \hline
    \multirow{8}{*}{$(0.1,0.5)$} & PMWLS & 8.50 (0.60) & 6.71 (0.51) & 3.64 (0.45) \\ \cline{2-2}
    & $\mbox{PMWLS } {\tiny (\rho=0.5)}$ & 6.61 (0.56) & 4.66 (0.46) & 2.03 (0.27) \\ \cline{2-2}
    & $\mbox{PMWLS } {\tiny (\rho=0.9)}$ & 7.41 (0.55) & 4.79 (0.46) & 1.75 (0.26) \\ \cline{2-2}
    & $\mbox{PMWLS } {\tiny (\rho=0.8,\phi=0.4)}$ & 7.19 (0.56) & 4.98 (0.51) & 1.87 (0.28) \\ \cline{2-2}
    & PWLS & 8.01 (0.62) & 6.34 (0.61) & 2.72 (0.47) \\ \cline{2-2}
    & $\mbox{PWLS } {\tiny (\rho=0.5)}$ & 7.40 (0.61) & 5.28 (0.54) & 1.52 (0.31) \\ \cline{2-2}
    & $\mbox{PWLS } {\tiny (\rho=0.9)}$ & 6.80 (0.53) & 4.67 (0.45) & 1.64 (0.24) \\ \cline{2-2}
    & $\mbox{PWLS } {\tiny (\rho=0.8,\phi=0.4)}$ & 7.13 (0.53) & 4.90 (0.49) & 1.90 (0.26) \\\hline
    \multirow{8}{*}{$(0.5,0.5)$} & PMWLS & 7.63 (0.58) & 6.79 (0.58) & 3.64 (0.46) \\ \cline{2-2}
    & $\mbox{PMWLS } {\tiny (\rho=0.5)}$ & 6.60 (0.57) & 3.93 (0.45) & 2.10 (0.32) \\ \cline{2-2}
    & $\mbox{PMWLS } {\tiny (\rho=0.9)}$ & 7.01 (0.57) & 3.83 (0.44) & 1.90 (0.28) \\ \cline{2-2}
    & $\mbox{PMWLS } {\tiny (\rho=0.8,\phi=0.4)}$ & 7.18 (0.55) & 3.95 (0.46) & 2.03 (0.29) \\ \cline{2-2}
    & PWLS & 7.23 (0.63) & 5.41 (0.59) & 2.07 (0.43) \\ \cline{2-2}
    & $\mbox{PWLS } {\tiny (\rho=0.5)}$ & 6.66 (0.60) & 4.58 (0.59) & 1.36 (0.32) \\ \cline{2-2}
    & $\mbox{PWLS } {\tiny (\rho=0.9)}$ & 7.34 (0.54) & 4.28 (0.42) & 2.26 (0.26) \\ \cline{2-2}
    & $\mbox{PWLS } {\tiny (\rho=0.8,\phi=0.4)}$ & 7.53 (0.54) & 5.07 (0.44) & 2.66 (0.29) \\
    \Xhline{3\arrayrulewidth}
    \multicolumn{5}{l}{\footnotesize{$\ast$ The actual MSE values are $0.01 \times$ the reported values.}}
    \end{tabular}
    \label{table:add_lasso_9_estimation}
\end{table}

\begin{table}
    \centering
    \caption{\small{Estimation results with LASSO for the equation \eqref{eq:additive} when the error process is ARMA(1,1) with $(\rho, \phi)=(0.8, 0.4)$. The other configurations are identical to Table \ref{table:add_lasso_5_estimation}.}}    
    \begin{tabular}{clrrr}
    \Xhline{3\arrayrulewidth}
    \multicolumn{5}{c}{ARMA(1,1) with $\rho=0.8,\phi=0.4$} \\ \hline
    $(\mu,\sigma)$ & Methods & \multicolumn{1}{c}{$n=50$} & \multicolumn{1}{c}{$n=100$} & \multicolumn{1}{c}{$n=200$} \\ \hline
    \multirow{8}{*}{$(0.1,0.5)$} & PMWLS & 6.55 (0.54) & 4.11 (0.48) & 2.47 (0.40) \\ \cline{2-2}
    & $\mbox{PMWLS } {\tiny (\rho=0.5)}$ & 5.13 (0.52) & 2.64 (0.34) & 1.52 (0.26) \\ \cline{2-2}
    & $\mbox{PMWLS } {\tiny (\rho=0.9)}$ & 5.12 (0.51) & 2.61 (0.34) & 1.42 (0.24) \\ \cline{2-2}
    & $\mbox{PMWLS } {\tiny (\rho=0.8,\phi=0.4)}$ & 5.45 (0.51) & 2.58 (0.35) & 1.36 (0.24) \\ \cline{2-2}
    & PWLS & 6.20 (0.63) & 3.05 (0.50) & 1.62 (0.38) \\ \cline{2-2}
    & $\mbox{PWLS } {\tiny (\rho=0.5)}$ & 5.52 (0.57) & 2.12 (0.37) & 0.96 (0.23) \\ \cline{2-2}
    & $\mbox{PWLS } {\tiny (\rho=0.9)}$ & 5.14 (0.52) & 2.68 (0.32) & 1.37 (0.20) \\ \cline{2-2}
    & $\mbox{PWLS } {\tiny (\rho=0.8,\phi=0.4)}$ & 5.35 (0.54) & 2.73 (0.36) & 1.61 (0.24) \\\hline
    \multirow{8}{*}{$(0.5,0.5)$} & PMWLS & 7.37 (0.59) & 3.95 (0.50) & 2.66 (0.39) \\ \cline{2-2}
    & $\mbox{PMWLS } {\tiny (\rho=0.5)}$ & 5.35 (0.51) & 2.57 (0.36) & 1.29 (0.24) \\ \cline{2-2}
    & $\mbox{PMWLS } {\tiny (\rho=0.9)}$ & 5.86 (0.50) & 2.49 (0.37) & 1.40 (0.23) \\ \cline{2-2}
    & $\mbox{PMWLS } {\tiny (\rho=0.8,\phi=0.4)}$ & 6.04 (0.52) & 2.47 (0.36) & 1.74 (0.27) \\ \cline{2-2}
    & PWLS & 6.13 (0.61) & 1.67 (0.41) & 1.35 (0.34) \\ \cline{2-2}
    & $\mbox{PWLS } {\tiny (\rho=0.5)}$ & 5.73 (0.57) & 1.87 (0.38) & 0.80 (0.22)  \\ \cline{2-2}
    & $\mbox{PWLS } {\tiny (\rho=0.9)}$ & 6.41 (0.49) & 2.96 (0.35) & 1.94 (0.24) \\ \cline{2-2}
    & $\mbox{PWLS } {\tiny (\rho=0.8,\phi=0.4)}$ & 6.81 (0.50) & 3.18 (0.33) & 2.03 (0.27) \\
    \Xhline{3\arrayrulewidth}
    \multicolumn{5}{l}{\footnotesize{$\ast$ The actual MSE values are $0.01 \times$ the reported values.}}
    \end{tabular}
    \label{table:add_lasso_8_4_estimation}
\end{table}

Tables \ref{table:add_lasso_5_selection}-\ref{table:add_lasso_8_4_selection} demonstrate selection results of PMWLS and PWLS methods with the LASSO penalty. True positive (TP) counts the number of significant estimates among the significant true parameters and true negative (TN) counts the number of insignificant estimates among the insignificant true parameters. 

\begin{table}
    \centering
    \caption{\small{Selection results with LASSO for the equation \eqref{eq:additive} when the error process is AR(1) with $\rho=0.5$.}}
    \begin{tabular}{clrrrrrrr}
    \Xhline{3\arrayrulewidth}
    \multicolumn{9}{c}{AR(1) with $\rho=0.5$} \\ \hline
    \multirow{2}{*}{$(\mu,\sigma)$} & \multirow{2}{*}{Methods} & \multicolumn{3}{c}{TP} && \multicolumn{3}{c}{TN} \\ \cline{3-5} \cline{7-9} 
    & & \multicolumn{1}{c}{50} & \multicolumn{1}{c}{100} & \multicolumn{1}{c}{200} && \multicolumn{1}{c}{50} & \multicolumn{1}{c}{100} & \multicolumn{1}{c}{200}\\ \hline
    \multirow{8}{*}{$(0.1,0.5)$} & PMWLS & 0.75 & 1.85 & 2.61 & & 16.87 & 16.65 & 16.38 \\ \cline{2-2}
    & $\mbox{PMWLS } {\tiny (\rho=0.5)}$ & 0.69 & 1.89 & 2.76 & & 16.85 & 16.72 & 16.70 \\ \cline{2-2}
    & $\mbox{PMWLS } {\tiny (\rho=0.9)}$ & 0.59 & 1.90 & 2.79 & & 16.86 & 16.74 & 16.55 \\ \cline{2-2}
    & $\mbox{PMWLS } {\tiny (\rho=0.8,\phi=0.4)}$ & 0.52 & 1.71 & 2.75 & & 16.85 & 16.80 & 16.54  \\ \cline{2-2}
    & PWLS & 0.42 & 1.66 & 2.71 & & 16.95 & 16.82 & 16.66 \\ \cline{2-2}
    & $\mbox{PWLS } {\tiny (\rho=0.5)}$ & 0.36 & 1.78 & 2.74 & & 16.91 & 16.79 & 16.84 \\ \cline{2-2}
    & $\mbox{PWLS } {\tiny (\rho=0.9)}$ & 0.58 & 1.86 & 2.75 & & 16.79 & 16.73 & 16.62 \\ \cline{2-2}
    & $\mbox{PWLS } {\tiny (\rho=0.8,\phi=0.4)}$ & 0.41 & 1.75 & 2.69 & & 16.89 & 16.76 & 16.56 \\\hline
    \multirow{8}{*}{$(0.5,0.5)$} & PMWLS & 0.94 & 1.53 & 2.55 & & 16.78 & 16.72 & 16.51 \\ \cline{2-2}
    & $\mbox{PMWLS } {\tiny (\rho=0.5)}$ & 0.75 & 1.71 & 2.58 & & 16.83 & 16.77 & 16.59 \\ \cline{2-2}
    & $\mbox{PMWLS } {\tiny (\rho=0.9)}$ & 0.70 & 1.73 & 2.58 & & 16.78 & 16.76 & 16.61 \\ \cline{2-2}
    & $\mbox{PMWLS } {\tiny (\rho=0.8,\phi=0.4)}$ & 0.64 & 1.50 & 2.60 & & 16.92 & 16.84 & 16.51 \\ \cline{2-2}
    & PWLS & 1.06 & 1.81 & 2.77 & & 16.87 & 16.75 & 16.76 \\ \cline{2-2}
    & $\mbox{PWLS } {\tiny (\rho=0.5)}$ & 0.77 & 1.42 & 2.77 & & 16.78 & 16.88 & 16.78 \\ \cline{2-2}
    & $\mbox{PWLS } {\tiny (\rho=0.9)}$ & 0.66 & 1.58 & 2.56 & & 16.89 & 16.89 & 16.74 \\ \cline{2-2}
    & $\mbox{PWLS } {\tiny (\rho=0.8,\phi=0.4)}$ & 0.46 & 1.52 & 2.51 & & 16.91 & 16.88 & 16.75 \\
    \Xhline{3\arrayrulewidth}
    \end{tabular}
    \label{table:add_lasso_5_selection}
\end{table}
	
\begin{table}
    \centering
    \caption{\small{Selection results with LASSO for the equation \eqref{eq:additive} when the error process is AR(1) with $\rho= 0.9$.}}
    \begin{tabular}{clrrrrrrr}
    \Xhline{3\arrayrulewidth}
    \multicolumn{9}{c}{AR(1) with $\rho=0.9$} \\ \hline
    \multirow{2}{*}{$(\mu,\sigma)$} & \multirow{2}{*}{Methods} & \multicolumn{3}{c}{TP} && \multicolumn{3}{c}{TN} \\ \cline{3-5} \cline{7-9} 
    & & \multicolumn{1}{c}{50} & \multicolumn{1}{c}{100} & \multicolumn{1}{c}{200} && \multicolumn{1}{c}{50} & \multicolumn{1}{c}{100} & \multicolumn{1}{c}{200}\\ \hline
    \multirow{8}{*}{$(0.1,0.5)$} & PMWLS & 1.38 & 1.89 & 2.62 & & 16.52 & 16.54 & 16.28 \\ \cline{2-2}
    & $\mbox{PMWLS } {\tiny (\rho=0.5)}$ & 1.85 & 2.36 & 2.96 & & 16.85 & 16.94 & 16.82 \\ \cline{2-2}
    & $\mbox{PMWLS } {\tiny (\rho=0.9)}$ & 1.67 & 2.31 & 2.95 & & 16.88 & 16.94 & 16.80 \\ \cline{2-2}
    & $\mbox{PMWLS } {\tiny (\rho=0.8,\phi=0.4)}$ & 1.75 & 2.21 & 2.94 & & 16.85 & 16.89 & 16.78  \\ \cline{2-2}
    & PWLS & 1.36 & 1.84 & 2.73 & & 16.89 & 16.80 & 16.58 \\ \cline{2-2}
    & $\mbox{PWLS } {\tiny (\rho=0.5)}$ & 1.59 & 2.15 & 2.90 & & 16.85 & 16.98 & 16.93 \\ \cline{2-2}
    & $\mbox{PWLS } {\tiny (\rho=0.9)}$ & 1.84 & 2.32 & 2.95 & & 16.90 & 16.90 & 16.83 \\ \cline{2-2}
    & $\mbox{PWLS } {\tiny (\rho=0.8,\phi=0.4)}$ & 1.79 & 2.24 & 2.94 & & 16.86 & 16.91 & 16.79 \\\hline
    \multirow{8}{*}{$(0.5,0.5)$} & PMWLS & 1.67 & 1.88 & 2.63 & & 16.45 & 16.53 & 16.30 \\ \cline{2-2}
    & $\mbox{PMWLS } {\tiny (\rho=0.5)}$ & 1.83 & 2.47 & 2.88 & & 16.73 & 16.83 & 16.80 \\ \cline{2-2}
    & $\mbox{PMWLS } {\tiny (\rho=0.9)}$ & 1.73 & 2.48 & 2.92 & & 16.80 & 16.90 & 16.94 \\ \cline{2-2}
    & $\mbox{PMWLS } {\tiny (\rho=0.8,\phi=0.4)}$ & 1.71 & 2.44 & 2.92 & & 16.78 & 16.81 & 16.85 \\ \cline{2-2}
    & PWLS & 1.59 & 2.02 & 2.73 & & 16.76 & 16.83 & 16.65 \\ \cline{2-2}
    & $\mbox{PWLS } {\tiny (\rho=0.5)}$ & 1.79 & 2.17 & 2.89 & & 16.79 & 16.88 & 16.94 \\ \cline{2-2}
    & $\mbox{PWLS } {\tiny (\rho=0.9)}$ & 1.69 & 2.44 & 2.94 & & 16.88 & 16.97 & 16.95 \\ \cline{2-2}
    & $\mbox{PWLS } {\tiny (\rho=0.8,\phi=0.4)}$ & 1.68 & 2.26 & 2.89 & & 16.85 & 16.95 & 16.90 \\
    \Xhline{3\arrayrulewidth}
    \end{tabular}
    \label{table:add_lasso_9_selection}
\end{table}

\begin{table}
    \centering
    \caption{\small{Selection results with LASSO for the equation \eqref{eq:additive} when the error process is ARMA(1,1) with $(\rho, \phi)=(0.8, 0.4)$.}}
    \begin{tabular}{clrrrrrrr}
    \Xhline{3\arrayrulewidth}
    \multicolumn{9}{c}{ARMA(1,1) with $\rho=0.8,\phi=0.4$} \\ \hline
    \multirow{2}{*}{$(\mu,\sigma)$} & \multirow{2}{*}{Methods} & \multicolumn{3}{c}{TP} && \multicolumn{3}{c}{TN} \\ \cline{3-5} \cline{7-9} 
    & & \multicolumn{1}{c}{50} & \multicolumn{1}{c}{100} & \multicolumn{1}{c}{200} && \multicolumn{1}{c}{50} & \multicolumn{1}{c}{100} & \multicolumn{1}{c}{200}\\ \hline
    \multirow{8}{*}{$(0.1,0.5)$} & PMWLS & 1.96 & 2.50 & 2.89 & & 16.63 & 16.36 & 15.99 \\ \cline{2-2}
    & $\mbox{PMWLS } {\tiny (\rho=0.5)}$ & 2.29 & 2.82 & 3.00 & & 16.74 & 16.74 & 16.59 \\ \cline{2-2}
    & $\mbox{PMWLS } {\tiny (\rho=0.9)}$ & 2.34 & 2.79 & 3.00 & & 16.81 & 16.76 & 16.60 \\ \cline{2-2}
    & $\mbox{PMWLS } {\tiny (\rho=0.8,\phi=0.4)}$ & 2.22 & 2.82 & 3.00 & & 16.77 & 16.69 & 16.48 \\ \cline{2-2}
    & PWLS & 1.85 & 2.59 & 2.93 & & 16.80 & 16.55 & 16.45 \\ \cline{2-2}
    & $\mbox{PWLS } {\tiny (\rho=0.5)}$ & 2.11 & 2.83 & 3.00 & & 16.77 & 16.85 & 16.87 \\ \cline{2-2}
    & $\mbox{PWLS } {\tiny (\rho=0.9)}$ & 2.32 & 2.83 & 3.00 & & 16.78 & 16.86 & 16.77 \\ \cline{2-2}
    & $\mbox{PWLS } {\tiny (\rho=0.8,\phi=0.4)}$ & 2.20 & 2.78 & 3.00 & & 16.72 & 16.76 & 16.64 \\\hline
    \multirow{8}{*}{$(0.5,0.5)$} & PMWLS & 1.75 & 2.58 & 2.95 & & 16.45 & 16.16 & 16.03 \\ \cline{2-2}
    & $\mbox{PMWLS } {\tiny (\rho=0.5)}$ & 2.22 & 2.82 & 3.00 & & 16.67 & 16.72 & 16.47 \\ \cline{2-2}
    & $\mbox{PMWLS } {\tiny (\rho=0.9)}$ & 2.10 & 2.81 & 3.00 & & 16.84 & 16.68 & 16.72 \\ \cline{2-2}
    & $\mbox{PMWLS } {\tiny (\rho=0.8,\phi=0.4)}$ & 2.04 & 2.81 & 2.98 & & 16.75 & 16.68 & 16.57 \\ \cline{2-2}
    & PWLS & 1.88 & 2.81 & 2.96 & & 16.71 & 16.63 & 16.58 \\ \cline{2-2}
    & $\mbox{PWLS } {\tiny (\rho=0.5)}$ & 2.02 & 2.81 & 3.00 & & 16.80 & 16.88 & 16.87 \\ \cline{2-2}
    & $\mbox{PWLS } {\tiny (\rho=0.9)}$ & 2.00 & 2.77 & 2.99 & & 16.88 & 16.87 & 16.89 \\ \cline{2-2}
    & $\mbox{PWLS } {\tiny (\rho=0.8,\phi=0.4)}$ & 1.95 & 2.77 & 2.97 & & 16.88 & 16.86 & 16.77 \\
    \Xhline{3\arrayrulewidth}
    \end{tabular}
    \label{table:add_lasso_8_4_selection}
\end{table}

Tables \ref{table:multi_add_estimation} and \ref{table:multi_add_selection} provides the estimation and selection results using nonlinear multiplicative model: 
\begin{equation}
\label{eq:multiplicative}
\displaystyle y_t=\frac{1}{1+\exp(-\bm x_t^{T}\bm\theta_0)}\times\epsilon_t.
\end{equation}
For $\epsilon_t$, the exponentiated AR processes or an ARMA process are considered since the $\epsilon_{t}$'s in the equation~\eqref{eq:multiplicative} are allowed to have only positive values. The AR and ARMA coefficients and the parameter setting of $\bm\theta$ are the same as the one in the model \eqref{eq:additive}. We transformed the model in the log scale and apply our approach. For comparison, an `Additive' method is considered, where the estimator is calculated as if the data are from a nonlinear additive model without log transformation. For this simulation, we provide the results using the SCAD penalty. 

\begin{table}
    \centering
    \caption{\small{Estimation results with SCAD for the data from the equation \eqref{eq:multiplicative}. The Model column refers to the exponentiated error process for the data generation. The other configurations are identical to Table \ref{table:add_lasso_5_estimation}}.}
    \begin{tabular}{cclrrr}
    \Xhline{3\arrayrulewidth}
    Model & $(\mu,\sigma)$ & Methods & \multicolumn{1}{c}{$n=50$} & \multicolumn{1}{c}{$n=100$} & \multicolumn{1}{c}{$n=200$} \\ \hline
    & \multirow{2}{*}{$(0.1,0.5)$} & PMWLS & 0.88 (0.42) & 0.32 (0.26) & 0.15 (0.17) \\ \cline{3-3}
    \multirow{1}{*}{AR(1)}& & Additive & 6.59 (1.07) & 4.25 (0.72) & 2.67 (0.49) \\\cline{2-6}
    \multirow{1}{*}{$\rho=0.5$}& \multirow{2}{*}{$(0.5,0.5)$} & PMWLS & 0.68 (0.37) & 0.30 (0.25) & 0.14 (0.16) \\ \cline{3-3}
    & & Additive & 79.04 (3.22) & 85.64 (2.64) & 48.82 (1.46) \\
    \Xhline{3\arrayrulewidth}
    & \multirow{2}{*}{$(0.1,0.5)$} & PMWLS & 0.61 (0.35) & 0.24 (0.22) & 0.13 (0.16) \\ \cline{3-3}
    \multirow{1}{*}{AR(1)}& & Additive & 16.30 (1.68) & 10.87 (1.30) & 3.98 (0.66) \\\cline{2-6}
    \multirow{1}{*}{$\rho=0.9$}& \multirow{2}{*}{$(0.5,0.5)$} & PMWLS & 0.55 (0.33) & 0.28 (0.24) & 0.13 (0.16) \\ \cline{3-3}
    & & Additive & 92.78 (3.47) & 68.63 (2.54) & 47.81 (1.93) \\
    \Xhline{3\arrayrulewidth}
    & \multirow{2}{*}{$(0.1,0.5)$} & PMWLS & 0.37 (0.27) & 0.14 (0.17) & 0.09 (0.14) \\ \cline{3-3}
    \multirow{1}{*}{ARMA(1,1)}& & Additive & 8.27 (1.16) & 3.44 (0.68) & 1.51 (0.37)  \\\cline{2-6}
    \multirow{1}{*}{$(\rho, \phi)=(0.8,0.4)$}& \multirow{2}{*}{$(0.5,0.5)$} & PMWLS & 0.44 (0.30) & 0.15 (0.18) & 0.07 (0.12) \\ \cline{3-3}
    & & Additive & 66.88 (2.77) & 41.68 (1.91) & 31.44 (1.04) \\
    \Xhline{3\arrayrulewidth}
    \end{tabular}
    \label{table:multi_add_estimation}
\end{table}

\begin{table}
    \centering
    \caption{\small{Selection results with SCAD for the data from the equation \eqref{eq:multiplicative}. The other configurations refer to Table \ref{table:multi_add_estimation}}.}
    \begin{tabular}{cclrrrrrrr}
    \Xhline{3\arrayrulewidth}
    \multirow{2}{*}{Model} &\multirow{2}{*}{$(\mu,\sigma)$} & \multirow{2}{*}{Methods} & \multicolumn{3}{c}{TP} && \multicolumn{3}{c}{TN} \\ \cline{4-6} \cline{8-10} 
    && & \multicolumn{1}{c}{n=50} & \multicolumn{1}{c}{n=100} & \multicolumn{1}{c}{200} && \multicolumn{1}{c}{50} & \multicolumn{1}{c}{100} & \multicolumn{1}{c}{200}\\ \hline
     & \multirow{2}{*}{$(0.1,0.5)$} & PMWLS & 2.88 & 2.99 & 3.00 & & 16.83 & 16.92 & 16.94 \\ \cline{3-3}
    \multirow{1}{*}{AR(1)}& & Additive & 2.54 & 2.72 & 2.99 & & 16.99 & 17.00 & 17.00 \\\cline{2-10}
    \multirow{1}{*}{$\rho=0.5$}& \multirow{2}{*}{$(0.5,0.5)$} & PMWLS & 2.94 & 2.99 & 3.00 & & 16.84 & 16.94 & 17.00 \\ \cline{3-3}
    & & Additive & 2.26 & 2.61 & 2.89 & & 16.99 & 17.00 & 16.99 \\
    \Xhline{3\arrayrulewidth}
    & \multirow{2}{*}{$(0.1,0.5)$} & PMWLS & 2.94 & 3.00 & 3.00 & & 16.85 & 16.92 & 16.96 \\ \cline{3-3}
    \multirow{1}{*}{AR(1)}& & Additive & 2.62 & 2.90 & 2.96 & & 16.99 & 17.00 & 17.00 \\\cline{2-10}
    \multirow{1}{*}{$\rho=0.9$}& \multirow{2}{*}{$(0.5,0.5)$} & PMWLS & 2.97 & 2.99 & 3.00 & & 16.74 & 16.90 & 16.97 \\ \cline{3-3}
    & & Additive & 2.40 & 2.68 & 2.82 & & 16.99 & 17.00 & 16.99 \\
    \Xhline{3\arrayrulewidth}
    & \multirow{2}{*}{$(0.1,0.5)$} & PMWLS & 2.99 & 3.00 & 3.00 & & 16.81 & 16.96 & 16.92 \\ \cline{3-3}
    \multirow{1}{*}{ARMA(1,1)}& & Additive & 2.87 & 2.98 & 3.00 & & 16.93 & 16.99 & 16.99 \\\cline{2-10}
    \multirow{1}{*}{$(\rho, \phi)=(0.8,0.4)$}& \multirow{2}{*}{$(0.5,0.5)$} & PMWLS & 2.98 & 3.00 & 3.00 & & 16.70 & 16.94 & 16.95 \\ \cline{3-3}
    & & Additive & 2.64 & 2.93 & 2.97 & & 16.95 & 16.99 & 17.00 \\
    \Xhline{3\arrayrulewidth}
    \end{tabular}
    \label{table:multi_add_selection}
\end{table}

\newpage
\bibliographystyle{chicago}
\bibliography{references}